# GWTC-4.0: Constraints on the Cosmic Expansion Rate and Modified Gravitational-wave Propagation

A. G. Abac,[1] I. Abouelfettouh,[2] F. Acernese,[3,4] K. Ackley,[5] C. Adamcewicz,[6] S. Adhicary,[7] D. Adhikari,[8,9]
N. Adhikari,[10] R. X. Adhikari,[11] V. K. Adkins,[12] S. Afroz,[13] A. Agapito,[14] D. Agarwal,[15] M. Agathos,[16] N. Aggarwal,[17]
S. Aggarwal,[18] O. D. Aguiar,[19] I.-L. Ahrend,[20] L. Aiello,[21,22] A. Ain,[23] P. Ajith,[24] T. Akutsu,[25,26] S. Albanesi,[27,28]
W. Ali,[29,30] S. Al-Kershi,[8,9] C. Allénè,[31] A. Allocca,[32,4] S. Al-Shammari,[33] P. A. Altin,[34] S. Alvarez-Lopez,[35]
W. Amar,[31] O. Amarasinghe,[33] A. Amato,[36,37] F. Amicucci,[38,39] C. Amra,[40] A. Ananyeva,[11] S. B. Anderson,[11]
W. G. Anderson,[11] M. Andia,[41] M. Ando,[42] M. Andrés-Carcasona,[43] T. Andrić,[44,45,8,9] J. Anglin,[46] S. Ansoldi,[47,48]
J. M. Antelis,[49] S. Antier,[41] M. Aoumi,[50] E. Z. Appavuravther,[51,52] S. Appert,[11] S. K. Apple,[53] K. Arai,[11] A. Araya,[42]
M. C. Araya,[11] M. Arca Sedda,[44,45] J. S. Areeda,[54] N. Aritomi,[2] F. Armato,[29,30] S. Armstrong,[55] N. Arnaud,[56]
M. Arogeti,[57] S. M. Aronson,[12] K. G. Arun,[58] G. Ashton,[59] Y. Aso,[25,60] L. Asprea,[28] M. Assiduo,[61,62]
S. Assis de Souza Melo,[63] S. M. Aston,[64] P. Astone,[38] F. Attadio,[39,38] F. Aubin,[65] K. AultONeal,[66] G. Avallone,[67]
E. A. Avila,[49] S. Babak,[20] C. Badger,[68] S. Bae,[69] S. Bagnasco,[28] L. Baiotti,[70] R. Bajpai,[71] T. Baka,[72,37] A. M. Baker,[6]
K. A. Baker,[73] T. Baker,[74] G. Baldi,[75,76] N. Baldicchi,[77,51] M. Ball,[78] G. Ballardin,[63] S. W. Ballmer,[79] S. Banagiri,[6]
B. Banerjee,[44] D. Bankar,[80] T. M. Baptiste,[12] P. Baral,[10] M. Baratti,[81,82] J. C. Barayoga,[11] B. C. Barish,[11] D. Barker,[2]
N. Barman,[80] P. Barneo,[83,84,85] F. Barone,[86,4] B. Barr,[87] L. Barsotti,[35] M. Barsuglia,[20] D. Barta,[88] A. M. Bartoletti,[89]
M. A. Barton,[87] I. Bartos,[46] A. Basalaev,[8,9] R. Bassiri,[90] A. Basti,[82,81] M. Bawaj,[77,51] P. Baxi,[91] J. C. Bayley,[87]
A. C. Baylor,[10] P. A. Baynard II,[57] M. Bazzan,[92,93] V. M. Bedakihale,[94] F. Beirnaert,[95] M. Bejger,[96] D. Belardinelli,[22]
A. S. Bell,[87] D. S. Bellie,[97] L. Bellizzi,[81,82] W. Benoit,[18] I. Bentara,[56] J. D. Bentley,[98] M. Ben Yaala,[55] S. Bera,[99,100]
F. Bergamin,[33] B. K. Berger,[90] S. Bernuzzi,[27] M. Beroiz,[11] C. P. L. Berry,[87] D. Bersanetti,[29] T. Bertheas,[101]
A. Bertolini,[37,36] J. Betzwieser,[64] D. Beveridge,[73] G. Bevilacqua,[102] N. Bevins,[103] R. Bhandare,[104] R. Bhatt,[11]
D. Bhattacharjee,[105,106] S. Bhattacharyya,[107] S. Bhaumik,[46] V. Biancalana,[102] A. Bianchi,[37,108] I. A. Bilenko,[109]
M. Bilicki,[110] G. Billingsley,[11] A. Binetti,[111] S. Bini,[11,75,76] C. Binu,[112] S. Biot,[113] O. Birnholtz,[114] S. Biscoveanu,[97]
A. Bisht,[9] M. Bitossi,[63,81] M.-A. Bizouard,[115] S. Blaber,[116] J. K. Blackburn,[11] L. A. Blagg,[78] C. D. Blair,[73,64]
D. G. Blair,[73] N. Bode,[8,9] N. Boettner,[98] G. Boileau,[115] M. Boldrini,[38] G. N. Bolingbroke,[117] A. Bolliand,[118,40]
L. D. Bonavena,[46] R. Bondarescu,[83] F. Bondu,[119] E. Bonilla,[90] M. S. Bonilla,[54] A. Bonino,[120] R. Bonnand,[31,118]
A. Borchers,[8,9] S. Borhanian,[7] V. Boschi,[81] S. Bose,[121] V. Bossilkov,[64] Y. Bothra,[37,108] A. Boudon,[56] L. Bourg,[57]
M. Boyle,[122] A. Bozzi,[63] C. Bradaschia,[81] P. R. Brady,[10] A. Branch,[64] M. Branchesi,[44,45] I. Braun,[105] T. Briant,[123]
A. Brillet,[115] M. Brinkmann,[8,9] P. Brockill,[10] E. Brockmueller,[8,9] A. F. Brooks,[11] B. C. Brown,[46] D. D. Brown,[117]
M. L. Brozzetti,[77,51] S. Brunett,[11] G. Bruno,[15] R. Bruntz,[124] J. Bryant,[120] Y. Bu,[125] F. Bucci,[62] J. Buchanan,[124]
O. Bulashenko,[83,84] T. Bulik,[126] H. J. Bulten,[37] A. Buonanno,[127,1] K. Burtnyk,[2] R. Buscicchio,[128,129] D. Buskulic,[31]
C. Buy,[101] R. L. Byer,[90] G. S. Cabourn Davies,[74] R. Cabrita,[15] V. Cáceres-Barbosa,[7] L. Cadonati,[57] G. Cagnoli,[130]
C. Cahillane,[79] A. Calafat,[99] T. A. Callister,[131] E. Calloni,[32,4] S. R. Callos,[78] M. Canepa,[30,29] G. Caneva Santoro,[43]
K. C. Cannon,[42] H. Cao,[35] L. A. Capistran,[132] E. Capocasa,[20] E. Capote,[2,11] G. Capurri,[82,81] G. Carapella,[67,133]
F. Carbognani,[63] M. Carlassara,[8,9] J. B. Carlin,[125] T. K. Carlson,[134] M. F. Carney,[105] M. Carpinelli,[128,63]
G. Carrillo,[78] J. J. Carter,[8,9] G. Carullo,[120,135] A. Casallas-Lagos,[136] J. Casanueva Diaz,[63] C. Casentini,[137,22]
S. Y. Castro-Lucas,[138] S. Caudill,[134] M. Cavaglià,[106] R. Cavalieri,[63] A. Ceja,[54] G. Cella,[81] P. Cerdá-Durán,[139,140]
E. Cesarini,[22] N. Chabbra,[34] W. Chaibi,[115] A. Chakraborty,[13] P. Chakraborty,[8,9] S. Chakraborty,[104]
S. Chalathadka Subrahmanya,[98] J. C. L. Chan,[141] M. Chan,[116] K. Chang,[142] S. Chao,[143,142] P. Charlton,[144]
E. Chassande-Mottin,[20] C. Chatterjee,[145] Debarati Chatterjee,[80] Deep Chatterjee,[35] M. Chaturvedi,[104] S. Chaty,[20]
K. Chatziioannou,[11] A. Chen,[146] A. H.-Y. Chen,[147] D. Chen,[148] H. Chen,[143] H. Y. Chen,[149] S. Chen,[145] Yanbei Chen,[150]
Yitian Chen,[122] H. P. Cheng,[151] P. Chessa,[77,51] H. T. Cheung,[91] S. Y. Cheung,[6] F. Chiadini,[152,133] G. Chiarini,[8,9,93]
A. Chiba,[153] A. Chincarini,[29] M. L. Chiofalo,[82,81] A. Chiummo,[4,63] C. Chou,[147] S. Choudhary,[73] N. Christensen,[115,154]
S. S. Y. Chua,[34] G. Ciani,[75,76] P. Ciecielag,[96] M. Cieślar,[126] M. Cifaldi,[22] B. Cirok,[155] F. Clara,[2] J. A. Clark,[11,57]
T. A. Clarke,[6] P. Clearwater,[156] S. Clesse,[113] F. Cleva,[115,118] E. Coccia,[44,45,43] E. Codazzo,[157,158] P.-F. Cohadon,[123]
S. Colace,[30] E. Colangeli,[74] M. Colleoni,[99] C. G. Collette,[159] J. Collins,[64] S. Colloms,[87] A. Colombo,[160,129]
C. M. Compton,[2] G. Connolly,[78] L. Conti,[93] T. R. Corbitt,[12] I. Cordero-Carrión,[161] S. Corezzi,[77,51] N. J. Cornish,[162]
I. Coronado,[163] A. Corsi,[164] R. Cottingham,[64] M. W. Coughlin,[18] A. Couineaux,[38] P. Couvares,[11,57] D. M. Coward,[73]
R. Coyne,[165] A. Cozzumbo,[44] J. D. E. Creighton,[10] T. D. Creighton,[166] P. Cremonese,[99] S. Crook,[64] R. Crouch,[2]
J. Csizmazia,[2] J. R. Cudell,[167] T. J. Cullen,[11] A. Cumming,[87] E. Cuoco,[168,169] M. Cusinato,[139] L. V. Da Conceição,[170]
T. Dal Canton,[41] S. Dal Pra,[171] G. Dálya,[101] B. D'Angelo,[29] S. Danilishin,[36,37] S. D'Antonio,[38] K. Danzmann,[9,8,9]
K. E. Darroch,[124] L. P. Dartez,[64] R. Das,[107] A. Dasgupta,[94] V. Dattilo,[63] A. Daumas,[20] N. Davari,[172,173] I. Dave,[104]
A. Davenport,[138] M. Davier,[41] T. F. Davies,[73] D. Davis,[11] L. Davis,[73] M. C. Davis,[18] P. Davis,[174,175] E. J. Daw,[176] M. Dax,[1]
J. De Bolle,[95] M. Deenadayalan,[80] J. Degallaix,[177] M. De Laurentis,[32,4] F. De Lillo,[23] S. Della Torre,[129]
W. Del Pozzo,[82,81] A. Demagny,[31] F. De Marco,[39,38] G. Demasi,[178,62] F. De Matteis,[21,22] N. Demos,[35] T. Dent,[179]

Corresponding author: LSC P&P Committee, via LVK Publications as proxy

lvc.publications@ligo.org

A. Depasse,[15] N. DePergola,[103] R. De Pietri,[180,181] R. De Rosa,[32,4] C. De Rossi,[63] M. Desai,[35] R. DeSalvo,[182] A. DeSimone,[183] R. De Simone,[152,133] A. Dhani,[1] R. Diab,[46] M. C. Díaz,[166] M. Di Cesare,[32,4] G. Dideron,[184] T. Dietrich,[1] L. Di Fiore,[4] C. Di Fronzo,[73] M. Di Giovanni,[39,38] T. Di Girolamo,[32,4] D. Diksha,[37,36] J. Ding,[20,185] S. Di Pace,[39,38] I. Di Palma,[39,38] D. Di Piero,[186,48] F. Di Renzo,[56] Divyajyoti,[33] A. Dmitriev,[120] J. P. Docherty,[87] Z. Doctor,[97] N. Doerksen,[170] E. Dohmen,[2] A. Doke,[134] A. Domiciano De Souza,[187] L. D'Onofrio,[38] F. Donovan,[35] K. L. Dooley,[33] T. Dooney,[72] S. Doravari,[80] O. Dorosh,[188] W. J. D. Doyle,[124] M. Drago,[39,38] J. C. Driggers,[2] L. Dunn,[125] U. Dupletsa,[44] P.-A. Duverne,[20] D. D'Urso,[172,157] P. Dutta Roy,[46] H. Duval,[189] S. E. Dwyer,[2] C. Eassa,[2] M. Ebersold,[190,31] T. Eckhardt,[98] G. Eddolls,[79] A. Effler,[64] J. Eichholz,[34] H. Einsle,[115] M. Eisenmann,[25] M. Emma,[59] K. Endo,[153] R. Enficiaud,[1] L. Errico,[32,4] R. Espinosa,[166] M. Esposito,[4,32] R. C. Essick,[191] H. Estellés,[1] T. Etzel,[11] M. Evans,[35] T. Evstafyeva,[184] B. E. Ewing,[7] J. M. Ezquiaga,[141] F. Fabrizi,[61,62] V. Fafone,[21,22] S. Fairhurst,[33] A. M. Farah,[131] B. Farr,[78] W. M. Farr,[192,193] G. Favaro,[92] M. Favata,[194] M. Fays,[167] M. Fazio,[55] J. Feicht,[11] M. M. Fejer,[90] R. Felicetti,[186,48] E. Fenyvesi,[88,195] J. Fernandes,[196] T. Fernandes,[197,139] D. Fernando,[112] S. Ferraiuolo,[198,39,38] T. A. Ferreira,[12] F. Fidecaro,[82,81] P. Figura,[96] A. Fiori,[81,82] I. Fiori,[63] M. Fishbach,[191] R. P. Fisher,[124] R. Fittipaldi,[199,133] V. Fiumara,[200,133] R. Flaminio,[31] S. M. Fleischer,[201] L. S. Fleming,[202] E. Floden,[18] H. Fong,[116] J. A. Font,[139,140] F. Fontinele-Nunes,[18] C. Foo,[1] B. Fornal,[203] K. Franceschetti,[180] F. Frappez,[31] S. Frasca,[39,38] F. Frasconi,[81] J. P. Freed,[66] Z. Frei,[204] A. Freise,[37,108] O. Freitas,[197,139] R. Frey,[78] W. Frischhertz,[64] P. Fritschel,[35] V. V. Frolov,[64] G. G. Fronzé,[28] M. Fuentes-Garcia,[11] S. Fujii,[205] T. Fujimori,[206] P. Fulda,[46] M. Fyffe,[64] B. Gadre,[72] J. R. Gair,[1] S. Galaudage,[187] V. Galdi,[207] R. Gamba,[7] A. Gamboa,[1] S. Gamoji,[182] D. Ganapathy,[208] A. Ganguly,[80] B. Garaventa,[29] J. García-Bellido,[209] C. García-Quirós,[190] J. W. Gardner,[34] K. A. Gardner,[116] S. Garg,[42] J. Gargiulo,[63] X. Garrido,[41] A. Garron,[99] F. Garufi,[32,4] P. A. Garver,[90] C. Gasbarra,[21,22] B. Gateley,[2] F. Gautier,[210] V. Gayathri,[10] T. Gayer,[79] G. Gemme,[29] A. Gennai,[81] V. Gennari,[101] J. George,[104] R. George,[149] O. Gerberding,[98] L. Gergely,[155] Archisman Ghosh,[95] Sayantan Ghosh,[196] Shaon Ghosh,[194] Shrobana Ghosh,[8,9] Suprovo Ghosh,[211] Tathagata Ghosh,[80] J. A. Giaime,[12,64] K. D. Giardina,[64] D. R. Gibson,[202] C. Gier,[55] S. Gkaitatzis,[82,81] J. Glanzer,[11] F. Glotin,[41] J. Godfrey,[78] R. V. Godley,[8,9] P. Godwin,[11] A. S. Goettel,[33] E. Goetz,[116] J. Golomb,[11] S. Gomez Lopez,[39,38] B. Goncharov,[44] G. González,[12] P. Goodarzi,[212] S. Goode,[6] A. W. Goodwin-Jones,[15] M. Gosselin,[63] R. Gouaty,[31] D. W. Gould,[34] K. Govorkova,[35] A. Grado,[77,51] V. Graham,[87] A. E. Granados,[18] M. Granata,[177] V. Granata,[213,133] S. Gras,[35] P. Grassia,[11] J. Graves,[57] C. Gray,[2] R. Gray,[87] G. Greco,[51] A. C. Green,[37,108] L. Green,[214] S. M. Green,[74] S. R. Green,[215] C. Greenberg,[134] A. M. Gretarsson,[66] H. K. Griffin,[18] D. Griffith,[11] H. L. Griggs,[57] G. Grignani,[77,51] C. Grimaud,[31] H. Grote,[33] S. Grunewald,[1] D. Guerra,[139] D. Guetta,[216] G. M. Guidi,[61,62] A. R. Guimaraes,[12] H. K. Gulati,[94] F. Gulminelli,[174,175] H. Guo,[146] W. Guo,[73] Y. Guo,[37,36] Anuradha Gupta,[217] I. Gupta,[7] N. C. Gupta,[94] S. K. Gupta,[46] V. Gupta,[18] N. Gupte,[1] J. Gurs,[98] N. Gutierrez,[177] N. Guttman,[6] F. Guzman,[132] D. Haba,[218] M. Haberland,[1] S. Haino,[219] E. D. Hall,[35] E. Z. Hamilton,[99] G. Hammond,[87] M. Haney,[37] J. Hanks,[2] C. Hanna,[7] M. D. Hannam,[33] O. A. Hannuksela,[220] A. G. Hanselman,[131] H. Hansen,[2] J. Hanson,[64] S. Hanumasagar,[57] R. Harada,[42] A. R. Hardison,[183] S. Harikumar,[188] K. Haris,[37,72] I. Harley-Trochimczyk,[132] T. Harmark,[135] J. Harms,[44,45] G. M. Harry,[221] I. W. Harry,[74] J. Hart,[105] B. Haskell,[96,222,223] C. J. Haster,[214] K. Haughian,[87] H. Hayakawa,[50] K. Hayama,[224] M. C. Heintze,[64] J. Heinze,[120] J. Heinzel,[35] H. Heitmann,[115] F. Hellman,[208] A. F. Helmling-Cornell,[78] G. Hemming,[63] O. Henderson-Sapir,[117] M. Hendry,[87] I. S. Heng,[87] M. H. Hennig,[87] C. Henshaw,[57] M. Heurs,[8,9] A. L. Hewitt,[225,226] J. Heynen,[15] J. Heyns,[35] S. Higginbotham,[33] S. Hild,[36,37] S. Hill,[87] Y. Himemoto,[227] N. Hirata,[25] C. Hirose,[228] D. Hofman,[177] B. E. Hogan,[66] N. A. Holland,[37,108] I. J. Hollows,[176] D. E. Holz,[131] L. Honet,[113] D. J. Horton-Bailey,[208] J. Hough,[87] S. Hourihane,[11] N. T. Howard,[145] E. J. Howell,[73] C. G. Hoy,[74] C. A. Hrishikesh,[21] P. Hsi,[35] H.-F. Hsieh,[143] H.-Y. Hsieh,[143] C. Hsiung,[229] S.-H. Hsu,[147] W.-F. Hsu,[111] Q. Hu,[87] H. Y. Huang,[142] Y. Huang,[7] Y. T. Huang,[79] A. D. Huddart,[230] B. Hughey,[66] V. Hui,[31] S. Husa,[99] R. Huxford,[7] L. Iampieri,[39,38] G. A. Iandolo,[36] M. Ianni,[22,21] G. Iannone,[133] J. Iascau,[78] K. Ide,[231] R. Iden,[218] A. Ierardi,[44,45] S. Ikeda,[148] H. Imafuku,[42] Y. Inoue,[142] G. Iorio,[92] P. Iosif,[186,48] M. H. Iqbal,[34] J. Irwin,[87] R. Ishikawa,[231] M. Isi,[192,193] K. S. Isleif,[232] Y. Itoh,[206,233] M. Iwaya,[205] B. R. Iyer,[24] C. Jacquet,[101] P.-E. Jacquet,[123] T. Jacquot,[41] S. J. Jadhav,[234] S. P. Jadhav,[156] M. Jain,[134] T. Jain,[225] A. L. James,[11] K. Jani,[145] J. Janquart,[15] N. N. Janthalur,[234] S. Jaraba,[235] P. Jaranowski,[236] R. Jaume,[99] W. Javed,[33] A. Jennings,[2] M. Jensen,[2] W. Jia,[35] J. Jiang,[151] H.-B. Jin,[237,238] G. R. Johns,[124] N. A. Johnson,[46] M. C. Johnston,[214] R. Johnston,[87] N. Johny,[8,9] D. H. Jones,[34] D. I. Jones,[211] R. Jones,[87] H. E. Jose,[78] P. Joshi,[7] S. K. Joshi,[80] G. Joubert,[56] J. Ju,[239] L. Ju,[73] K. Jung,[240] J. Junker,[34] V. Juste,[113] H. B. Kabagoz,[64,35] T. Kajita,[241] I. Kaku,[206] V. Kalogera,[97] M. Kalomenopoulos,[214] M. Kamiizumi,[50] N. Kanda,[233,206] S. Kandhasamy,[80] G. Kang,[242] N. C. Kannachel,[6] J. B. Kanner,[11] S. A. KantiMahanty,[18] S. J. Kapadia,[80] D. P. Kapasi,[54] M. Karthikeyan,[134] M. Kasprzack,[11] H. Kato,[153] T. Kato,[205] E. Katsavounidis,[35] W. Katzman,[64] R. Kaushik,[104] K. Kawabe,[2] R. Kawamoto,[206] D. Keitel,[99] L. J. Kemperman,[117] J. Kennington,[7] F. A. Kerkow,[18] R. Kesharwani,[80] J. S. Key,[243] R. Khadela,[8,9] S. Khadka,[90] S. S. Khadkikar,[7] F. Y. Khalili,[109] F. Khan,[8,9] T. Khanam,[164] M. Khursheed,[104] N. M. Khusid,[192,193] W. Kiendrebeogo,[115,244] N. Kijbunchoo,[117] C. Kim,[245] J. C. Kim,[246] K. Kim,[247] M. H. Kim,[239] S. Kim,[248] Y.-M. Kim,[247] C. Kimball,[97] K. Kimes,[54] M. Kinnear,[33] J. S. Kissel,[2] S. Klimenko,[46] A. M. Knee,[116] E. J. Knox,[78] N. Knust,[8,9] K. Kobayashi,[205] S. M. Koehlenbeck,[90] G. Koekoek,[37,36] K. Kohri,[249,250] K. Kokeyama,[33,251] S. Koley,[44,167] P. Kolitsidou,[120] A. E. Koloniari,[252] K. Komori,[42] A. K. H. Kong,[143] A. Kontos,[253] L. M. Koponen,[120] M. Korobko,[98] X. Kou,[18] A. Koushik,[23] N. Kouvatsos,[68] M. Kovalam,[73] T. Koyama,[153] D. B. Kozak,[11] S. L. Kranzhoff,[36,37] V. Kringel,[8,9] N. V. Krishnendu,[120] S. Kroker,[254] A. Królak,[255,188] K. Kruska,[8,9] J. Kubisz,[256] G. Kuehn,[8,9] S. Kulkarni,[217] A. Kulur Ramamohan,[34] Achal Kumar,[46] Anil Kumar,[234] Praveen Kumar,[179] Prayush Kumar,[24] Rahul Kumar,[2] Rakesh Kumar,[94] J. Kume,[257,258,42] K. Kuns,[35] N. Kuntimaddi,[33] S. Kuroyanagi,[209,259] S. Kuwahara,[42] K. Kwak,[240] K. Kwan,[34] S. Kwon,[42] G. Lacaille,[87] D. Laghi,[190,101] A. H. Laity,[165] E. Lalande,[260] M. Lalleman,[23] P. C. Lalremruati,[261] M. Landry,[2] B. B. Lane,[35] R. N. Lang,[35] J. Lange,[149] R. Langgin,[214] B. Lantz,[90] I. La Rosa,[99] J. Larsen,[201] A. Lartaux-Vollard,[41] P. D. Lasky,[6] J. Lawrence,[166] M. Laxen,[64] C. Lazarte,[139] A. Lazzarini,[11] C. Lazzaro,[158,157]

P. Leaci,[39,38] L. Leali,[18] Y. K. Lecoeuche,[116] H. M. Lee,[262] H. W. Lee,[263] J. Lee,[79] K. Lee,[239] R.-K. Lee,[143] R. Lee,[35]
Sungho Lee,[247] Sunjae Lee,[239] Y. Lee,[142] I. N. Legred,[11] J. Lehmann,[8,9] L. Lehner,[184] M. Le Jean,[177,118] A. Lemaître,[264]
M. Lenti,[62,178] M. Leonardi,[75,76,265] M. Lequime,[40] N. Leroy,[41] M. Lesovsky,[11] N. Letendre,[31] M. Lethuillier,[56]
Y. Levin,[6] K. Leyde,[74] A. K. Y. Li,[11] K. L. Li,[266] T. G. F. Li,[111] X. Li,[150] Y. Li,[97] Z. Li,[87] A. Lihos,[124] E. T. Lin,[143] F. Lin,[142]
L. C.-C. Lin,[266] Y.-C. Lin,[143] C. Lindsay,[202] S. D. Linker,[182] A. Liu,[220] G. C. Liu,[229] Jian Liu,[73] F. Llamas Villarreal,[166]
J. Llobera-Querol,[99] R. K. L. Lo,[141] J.-P. Locquet,[111] S. C. G. Loggins,[267] M. R. Loizou,[134] L. T. London,[68] A. Longo,[61,62]
D. Lopez,[167] M. Lopez Portilla,[72] M. Lorenzini,[21,22] A. Lorenzo-Medina,[179] V. Loriette,[41] M. Lormand,[64]
G. Losurdo,[268,81] E. Lotti,[134] T. P. Lott IV,[57] J. D. Lough,[8,9] H. A. Loughlin,[35] C. O. Lousto,[112] N. Low,[125] N. Lu,[34]
L. Lucchesi,[81] H. Lück,[9,8,9] D. Lumaca,[22] A. P. Lundgren,[269,270] A. W. Lussier,[260] R. Macas,[74] M. MacInnis,[35]
D. M. Macleod,[33] I. A. O. MacMillan,[11] A. Macquet,[41] K. Maeda,[153] S. Maenaut,[111] S. S. Magare,[80] R. M. Magee,[11]
E. Maggio,[1] R. Maggiore,[37,108] M. Magnozzi,[29,30] M. Mahesh,[98] M. Maini,[165] S. Majhi,[80] E. Majorana,[39,38]
C. N. Makarem,[11] D. Malakar,[106] J. A. Malaquias-Reis,[19] U. Mali,[191] S. Maliakal,[11] A. Malik,[104] L. Mallick,[170,191]
A.-K. Malz,[59] N. Man,[115] M. Mancarella,[100] V. Mandic,[18] V. Mangano,[172,157] B. Mannix,[78] G. L. Mansell,[79]
M. Manske,[10] M. Mantovani,[63] M. Mapelli,[92,93,271] C. Marinelli,[102] F. Marion,[31] A. S. Markosyan,[90] A. Markowitz,[11]
E. Maros,[11] S. Marsat,[101] F. Martelli,[61,62] I. W. Martin,[87] R. M. Martin,[194] B. B. Martinez,[132] D. A. Martinez,[54]
M. Martinez,[43,272] V. Martinez,[130] A. Martini,[75,76] J. C. Martins,[19] D. V. Martynov,[120] E. J. Marx,[35] L. Massaro,[36,37]
A. Masserot,[31] M. Masso-Reid,[87] S. Mastrogiovanni,[38] T. Matcovich,[51] M. Matiushechkina,[8,9] L. Maurin,[210]
N. Mavalvala,[35] N. Maxwell,[2] G. McCarrol,[64] R. McCarthy,[2] D. E. McClelland,[34] S. McCormick,[64] L. McCuller,[11]
S. McEachin,[124] C. McElhenny,[124] G. I. McGhee,[87] J. McGinn,[87] K. B. M. McGowan,[145] J. McIver,[116] A. McLeod,[73]
I. McMahon,[190] T. McRae,[34] R. McTeague,[87] D. Meacher,[10] B. N. Meagher,[79] R. Mechum,[112] Q. Meijer,[72] A. Melatos,[125]
C. S. Menoni,[138] F. Mera,[2] R. A. Mercer,[10] L. Mereni,[177] K. Merfeld,[164] E. L. Merilh,[64] J. R. Mérou,[99] J. D. Merritt,[78]
M. Merzougui,[115] C. Messick,[10] B. Mestichelli,[44] M. Meyer-Conde,[273] F. Meylahn,[8,9] A. Mhaske,[80] A. Miani,[75,76]
H. Miao,[274] C. Michel,[177] Y. Michimura,[42] H. Middleton,[120] D. P. Mihaylov,[105] A. L. Miller,[37,72] S. J. Miller,[11]
M. Millhouse,[57] E. Milotti,[186,48] V. Milotti,[92] Y. Minenkov,[22] E. M. Minihan,[66] Ll. M. Mir,[43] L. Mirasola,[157,158]
M. Miravet-Tenés,[139] C.-A. Miritescu,[43] A. Mishra,[24] C. Mishra,[107] T. Mishra,[46] A. L. Mitchell,[37,108] J. G. Mitchell,[66]
S. Mitra,[80] V. P. Mitrofanov,[109] K. Mitsuhashi,[25] R. Mittleman,[35] O. Miyakawa,[50] S. Miyoki,[50] A. Miyoko,[66] G. Mo,[35]
L. Mobilia,[61,62] S. R. P. Mohapatra,[11] S. R. Mohite,[7] M. Molina-Ruiz,[208] M. Mondin,[182] M. Montani,[61,62] C. J. Moore,[225]
D. Moraru,[2] A. More,[80] S. More,[80] C. Moreno,[136] E. A. Moreno,[35] G. Moreno,[2] A. Moreso Serra,[83] S. Morisaki,[42,205]
Y. Moriwaki,[153] G. Morras,[209] A. Moscatello,[92] M. Mould,[35] B. Mours,[65] C. M. Mow-Lowry,[37,108] L. Muccillo,[178,62]
F. Muciaccia,[39,38] D. Mukherjee,[120] Samanwaya Mukherjee,[24] Soma Mukherjee,[166] Subroto Mukherjee,[94]
Suvodip Mukherjee,[13] N. Mukund,[35] A. Mullavey,[64] H. Mullock,[116] J. Mundi,[221] C. L. Mungioli,[73] M. Murakoshi,[231]
P. G. Murray,[87] D. Nabari,[75,76] S. L. Nadji,[8,9] A. Nagar,[28,275] N. Nagarajan,[87] K. Nakagaki,[50] K. Nakamura,[25]
H. Nakano,[276] M. Nakano,[11] D. Nanadoumgar-Lacroze,[43] D. Nandi,[12] V. Napolano,[63] P. Narayan,[217] I. Nardecchia,[22]
T. Narikawa,[205] H. Narola,[72] L. Naticchioni,[38] R. K. Nayak,[261] L. Negri,[72] A. Nela,[87] C. Nelle,[78] A. Nelson,[132]
T. J. N. Nelson,[64] M. Nery,[8,9] A. Neunzert,[2] S. Ng,[54] L. Nguyen Quynh,[277] S. A. Nichols,[12] A. B. Nielsen,[278]
Y. Nishino,[25,42] A. Nishizawa,[279] S. Nissanke,[280,37] W. Niu,[7] F. Nocera,[63] J. Noller,[281] M. Norman,[33] C. North,[33]
J. Novak,[118,235,282] R. Nowicki,[145] J. F. Nuño Siles,[209] L. K. Nuttall,[74] K. Obayashi,[231] J. Oberling,[2] J. O'Dell,[230]
E. Oelker,[35] M. Oertel,[235,118,283,282] G. Oganesyan,[44,45] T. O'Hanlon,[64] M. Ohashi,[50] F. Ohme,[8,9] R. Oliveri,[118,283,282]
R. Omer,[18] B. O'Neal,[124] M. Onishi,[153] K. Oohara,[284] B. O'Reilly,[64] M. Orselli,[51,77] R. O'Shaughnessy,[112] S. O'Shea,[87]
S. Oshino,[50] C. Osthelder,[11] I. Ota,[12] D. J. Ottaway,[117] A. Ouzriat,[56] H. Overmier,[64] B. J. Owen,[285] R. Ozaki,[231]
A. E. Pace,[7] R. Pagano,[12] M. A. Page,[25] A. Pai,[196] L. Paiella,[44] A. Pal,[286] S. Pal,[261] M. A. Palaia,[81,82] M. Pálfi,[204]
P. P. Palma,[39,21,22] C. Palomba,[38] P. Palud,[20] H. Pan,[143] J. Pan,[73] K. C. Pan,[143] P. K. Panda,[234] Shiksha Pandey,[7]
Swadha Pandey,[35] P. T. H. Pang,[37,72] F. Pannarale,[39,38] K. A. Pannone,[54] B. C. Pant,[104] F. H. Panther,[73] M. Panzeri,[61,62]
F. Paoletti,[81] A. Paolone,[38,287] A. Papadopoulos,[87] E. E. Papalexakis,[212] L. Papalini,[81,82] G. Papigkiotis,[252] A. Paquis,[41]
A. Parisi,[77,51] B.-J. Park,[247] J. Park,[288] W. Parker,[64] G. Pascale,[8,9] D. Pascucci,[95] A. Pasqualetti,[63] R. Passaquieti,[82,81]
L. Passenger,[6] D. Passuello,[81] O. Patane,[2] A. V. Patel,[142] D. Pathak,[80] A. Patra,[33] B. Patricelli,[82,81] B. G. Patterson,[33]
K. Paul,[107] S. Paul,[78] E. Payne,[11] T. Pearce,[33] M. Pedraza,[11] A. Pele,[11] F. E. Peña Arellano,[289] X. Peng,[120] Y. Peng,[57]
S. Penn,[290] M. D. Penuliar,[54] A. Perego,[75,76] Z. Pereira,[134] C. Périgois,[291,93,92] G. Perna,[92] A. Perreca,[75,76,44] J. Perret,[20]
S. Perriès,[56] J. W. Perry,[37,108] D. Pesios,[252] S. Peters,[167] S. Petracca,[207] C. Petrillo,[77] H. P. Pfeiffer,[1] H. Pham,[64]
K. A. Pham,[18] K. S. Phukon,[120] H. Phurailatpam,[220] M. Piarulli,[101] L. Piccari,[39,38] O. J. Piccinni,[34] M. Pichot,[115]
M. Piendibene,[82,81] F. Piergiovanni,[61,62] L. Pierini,[38] G. Pierra,[38] V. Pierro,[292,133] M. Pietrzak,[96] M. Pillas,[167] F. Pilo,[81]
L. Pinard,[177] I. M. Pinto,[292,133,293,32] M. Pinto,[63] B. J. Piotrzkowski,[10] M. Pirello,[2] M. D. Pitkin,[225,87] A. Placidi,[51]
E. Placidi,[39,38] M. L. Planas,[99] W. Plastino,[213,22] C. Plunkett,[35] R. Poggiani,[82,81] E. Polini,[35] J. Pomper,[81,82] L. Pompili,[1]
J. Poon,[220] E. Porcelli,[37] E. K. Porter,[20] C. Posnansky,[7] R. Poulton,[63] J. Powell,[156] G. S. Prabhu,[80] M. Pracchia,[167]
B. K. Pradhan,[80] T. Pradier,[65] A. K. Prajapati,[94] K. Prasai,[294] R. Prasanna,[234] P. Prasia,[80] G. Pratten,[120]
G. Principe,[186,48] G. A. Prodi,[75,76] P. Prosperi,[81] P. Prosposito,[21,22] A. C. Providence,[66] A. Puecher,[1] J. Pullin,[12]
P. Puppo,[38] M. Pürrer,[165] H. Qi,[16] J. Qin,[34] G. Quéméner,[175,118] V. Quetschke,[166] P. J. Quinonez,[66] N. Qutob,[57]
R. Rading,[232] P. Raffai,[295] I. Rainho,[139] S. Raja,[104] C. Rajan,[104] B. Rajbhandari,[112] K. E. Ramirez,[64] F. A. Ramis Vidal,[99]
M. Ramos Arevalo,[166] A. Ramos-Buades,[99,37] S. Ranjan,[57] K. Ransom,[64] P. Rapagnani,[39,38] B. Ratto,[66]
A. Ravichandran,[134] A. Ray,[97] V. Raymond,[33] M. Razzano,[82,81] J. Read,[54] T. Regimbau,[31] S. Reid,[55] C. Reissel,[35]
D. H. Reitze,[11] A. I. Renzini,[11,128] B. Revenu,[296,41] A. Revilla Peña,[83] R. Reyes,[182] L. Ricca,[15] F. Ricci,[39,38] M. Ricci,[38,39]
A. Ricciardone,[82,81] J. Rice,[79] J. W. Richardson,[212] M. L. Richardson,[117] A. Rijal,[66] K. Riles,[91] H. K. Riley,[33]
S. Rinaldi,[271] J. Rittmeyer,[98] C. Robertson,[230] F. Robinet,[41] M. Robinson,[2] A. Rocchi,[22] L. Rolland,[31] J. G. Rollins,[297]
R. Romano,[3,4] A. Romero,[31] I. M. Romero-Shaw,[225] J. H. Romie,[64] S. Ronchini,[7] T. J. Roocke,[117] L. Rosa,[4,32]

T. J. Rosauer,[212] C. A. Rose,[57] D. Rosińska,[126] M. P. Ross,[53] M. Rossello-Sastre,[99] S. Rowan,[87] S. K. Roy,[192,193] S. Roy,[15] D. Rozza,[128,129] P. Ruggi,[63] N. Ruhama,[240] E. Ruiz Morales,[298,209] K. Ruiz-Rocha,[145] S. Sachdev,[57] T. Sadecki,[2] P. Saffarieh,[37,108] S. Safi-Harb,[170] M. R. Sah,[13] S. Saha,[143] T. Sainrat,[65] S. Sajith Menon,[216,39,38] K. Sakai,[299] Y. Sakai,[273] M. Sakellariadou,[68] S. Sakon,[7] O. S. Salafia,[160,129,128] F. Salces-Carcoba,[11] L. Salconi,[63] M. Saleem,[149] F. Salemi,[39,38] M. Sallé,[37] S. U. Salunkhe,[80] S. Salvador,[175,174] A. Salvarese,[149] A. Samajdar,[72,37] A. Sanchez,[2] E. J. Sanchez,[11] L. E. Sanchez,[11] N. Sanchis-Gual,[139] J. R. Sanders,[183] E. M. Sänger,[1] F. Santoliquido,[44,45] F. Sarandrea,[28] T. R. Saravanan,[80] N. Sarin,[6] P. Sarkar,[8,9] A. Sasli,[252] P. Sassi,[51,77] B. Sassolas,[177] B. S. Sathyaprakash,[7,33] R. Sato,[228] S. Sato,[153] Yukino Sato,[153] Yu Sato,[153] O. Sauter,[46] R. L. Savage,[2] T. Sawada,[50] H. L. Sawant,[80] S. Sayah,[177] V. Scacco,[21,22] D. Schaetzl,[11] M. Scheel,[150] A. Schiebelbein,[191] M. G. Schiworski,[79] P. Schmidt,[120] S. Schmidt,[72] R. Schnabel,[98] M. Schneewind,[8,9] R. M. S. Schofield,[78] K. Schouteden,[111] B. W. Schulte,[8,9] B. F. Schutz,[33,8,9] E. Schwartz,[300] M. Scialpi,[301] J. Scott,[87] S. M. Scott,[34] R. M. Sedas,[64] T. C. Seetharamu,[87] M. Seglar-Arroyo,[43] Y. Sekiguchi,[302] D. Sellers,[64] N. Sembo,[206] A. S. Sengupta,[303] E. G. Seo,[87] J. W. Seo,[111] V. Sequino,[32,4] M. Serra,[38] A. Sevrin,[189] T. Shaffer,[2] U. S. Shah,[57] M. A. Shaikh,[262] L. Shao,[304] A. K. Sharma,[99] Preeti Sharma,[12] Prianka Sharma,[104] Ritwik Sharma,[18] S. Sharma Chaudhary,[106] P. Shawhan,[127] N. S. Shcheblanov,[305,264] E. Sheridan,[145] Z.-H. Shi,[143] M. Shikauchi,[42] R. Shimomura,[306] H. Shinkai,[306] S. Shirke,[80] D. H. Shoemaker,[35] D. M. Shoemaker,[149] R. W. Short,[2] S. ShyamSundar,[104] A. Sider,[159] H. Siegel,[192,193] D. Sigg,[2] L. Silenzi,[36,37] L. Silvestri,[39,171] M. Simmonds,[117] L. P. Singer,[307] Amitesh Singh,[217] Anika Singh,[11] D. Singh,[208] N. Singh,[99] S. Singh,[218,60] A. M. Sintes,[99] V. Sipala,[172,157] V. Skliris,[33] B. J. J. Slagmolen,[34] D. A. Slater,[201] T. J. Slaven-Blair,[73] J. Smetana,[120] J. R. Smith,[54] L. Smith,[87,186,48] R. J. E. Smith,[6] W. J. Smith,[145] S. Soares de Albuquerque Filho,[61] M. Soares-Santos,[190] K. Somiya,[218] I. Song,[143] S. Soni,[35] V. Sordini,[56] F. Sorrentino,[29] H. Sotani,[308] F. Spada,[81] V. Spagnuolo,[37] A. P. Spencer,[87] P. Spinicelli,[63] A. K. Srivastava,[94] F. Stachurski,[87] C. J. Stark,[124] D. A. Steer,[309] N. Steinle,[170] J. Steinlechner,[36,37] S. Steinlechner,[36,37] N. Stergioulas,[252] P. Stevens,[41] S. P. Stevenson,[156] M. StPierre,[165] M. D. Strong,[12] A. Strunk,[2] A. L. Stuver,[103,*] M. Suchenek,[96] S. Sudhagar,[96] Y. Sudo,[231] N. Sueltmann,[98] L. Suleiman,[54] K. D. Sullivan,[12] J. Sun,[242] L. Sun,[34] S. Sunil,[94] J. Suresh,[115] B. J. Sutton,[68] P. J. Sutton,[33] K. Suzuki,[218] M. Suzuki,[205] B. L. Swinkels,[37] A. Syx,[118] M. J. Szczepańczyk,[310] P. Szewczyk,[126] M. Tacca,[37] H. Tagoshi,[205] K. Takada,[205] H. Takahashi,[273] R. Takahashi,[25] A. Takamori,[42] S. Takano,[311] H. Takeda,[312,313] K. Takeshita,[218] I. Takimoto Schmiegelow,[44,45] M. Takou-Ayaoh,[79] C. Talbot,[131] M. Tamaki,[205] N. Tamanini,[101] D. Tanabe,[142] K. Tanaka,[50] S. J. Tanaka,[231] S. Tanioka,[33] D. B. Tanner,[46] W. Tanner,[8,9] L. Tao,[212] R. D. Tapia,[7] E. N. Tapia San Martín,[37] C. Taranto,[21,22] A. Taruya,[314] J. D. Tasson,[154] J. G. Tau,[112] D. Tellez,[54] R. Tenorio,[99] H. Themann,[182] A. Theodoropoulos,[139] M. P. Thirugnanasambandam,[80] L. M. Thomas,[11] M. Thomas,[64] P. Thomas,[2] J. E. Thompson,[211] S. R. Thondapu,[104] K. A. Thorne,[64] E. Thrane,[6] J. Tissino,[44,45] A. Tiwari,[80] Pawan Tiwari,[44] Praveer Tiwari,[196] S. Tiwari,[190] V. Tiwari,[120] M. R. Todd,[79] M. Toffano,[92] A. M. Toivonen,[18] K. Toland,[87] A. E. Tolley,[74] T. Tomaru,[25] V. Tommasini,[11] T. Tomura,[50] H. Tong,[6] C. Tong-Yu,[142] A. Torres-Forné,[139,140] C. I. Torrie,[11] I. Tosta e Melo,[315] E. Tournefier,[31] M. Trad Nery,[115] K. Tran,[124] A. Trapananti,[52,51] R. Travaglini,[169] F. Travasso,[52,51] G. Traylor,[64] M. Trevor,[127] M. C. Tringali,[63] A. Tripathee,[91] G. Troian,[186,48] A. Trovato,[186,48] L. Trozzo,[4] R. J. Trudeau,[11] T. Tsang,[33] S. Tsuchida,[316] L. Tsukada,[214] K. Turbang,[189,23] M. Turconi,[115] C. Turski,[95] H. Ubach,[83,84] N. Uchikata,[205] T. Uchiyama,[50] R. P. Udall,[11] T. Uehara,[317] K. Ueno,[42] V. Undheim,[278] L. E. Uronen,[220] T. Ushiba,[50] M. Vacatello,[81,82] H. Vahlbruch,[8,9] N. Vaidya,[11] G. Vajente,[11] A. Vajpeyi,[6] J. Valencia,[99] M. Valentini,[108,37] S. A. Vallejo-Peña,[318] S. Vallero,[28] V. Valsan,[10] M. van Dael,[37,319] E. Van den Bossche,[189] J. F. J. van den Brand,[36,108,37] C. Van Den Broeck,[72,37] M. van der Sluys,[37,72] A. Van de Walle,[41] J. van Dongen,[37,108] K. Vandra,[103] M. VanDyke,[121] H. van Haevermaet,[23] J. V. van Heijningen,[37,108] P. Van Hove,[65] J. Vanier,[260] M. VanKeuren,[105] J. Vanosky,[2] N. van Remortel,[23] M. Vardaro,[36,37] A. F. Vargas,[125] V. Varma,[134] A. N. Vazquez,[90] A. Vecchio,[120] G. Vedovato,[93] J. Veitch,[87] P. J. Veitch,[117] S. Venikoudis,[15] R. C. Venterea,[18] P. Verdier,[56] M. Vereecken,[15] D. Verkindt,[31] B. Verma,[134] Y. Verma,[104] S. M. Vermeulen,[11] F. Vetrano,[61] A. Veutro,[38,39] A. Viceré,[61,62] S. Vidyant,[79] A. D. Viets,[89] A. Vijaykumar,[191] A. Vilkha,[112] N. Villanueva Espinosa,[139] V. Villa-Ortega,[179] E. T. Vincent,[57] J.-Y. Vinet,[115] S. Viret,[56] S. Vitale,[35] H. Vocca,[77,51] D. Voigt,[98] E. R. G. von Reis,[2] J. S. A. von Wrangel,[8,9] W. E. Vossius,[232] L. Vujeva,[141] S. P. Vyatchanin,[109] J. Wack,[11] L. E. Wade,[105] M. Wade,[105] K. J. Wagner,[112] L. Wallace,[11] E. J. Wang,[90] H. Wang,[218] J. Z. Wang,[91] W. H. Wang,[166] Y. F. Wang,[1] G. Waratkar,[196] J. Warner,[2] M. Was,[31] T. Washimi,[25] N. Y. Washington,[11] D. Watarai,[42] B. Weaver,[2] S. A. Webster,[87] N. L. Weickhardt,[98] M. Weinert,[8,9] A. J. Weinstein,[11] R. Weiss,[35] L. Wen,[73] K. Wette,[34] J. T. Whelan,[112] B. F. Whiting,[46] C. Whittle,[11] E. G. Wickens,[74] D. Wilken,[8,9,9] A. T. Wilkin,[212] B. M. Williams,[121] D. Williams,[87] M. J. Williams,[74] N. S. Williams,[1] J. L. Willis,[11] B. Willke,[9,8,9] M. Wils,[111] L. Wilson,[105] C. W. Winborn,[106] J. Winterflood,[73] C. C. Wipf,[11] G. Woan,[87] J. Woehler,[36,37] N. E. Wolfe,[35] H. T. Wong,[142] I. C. F. Wong,[220,111] K. Wong,[191] T. Wouters,[72,37] J. L. Wright,[2] M. Wright,[87,72] B. Wu,[79] C. Wu,[143] D. S. Wu,[8,9] H. Wu,[143] K. Wu,[121] Q. Wu,[53] Y. Wu,[97] Z. Wu,[101] E. Wuchner,[54] D. M. Wysocki,[10] V. A. Xu,[208] Y. Xu,[99] N. Yadav,[28] H. Yamamoto,[11] K. Yamamoto,[153] T. S. Yamamoto,[42] T. Yamamoto,[50] R. Yamazaki,[231] T. Yan,[120] K. Z. Yang,[18] Y. Yang,[147] Z. Yarbrough,[12] J. Yebana,[99] S.-W. Yeh,[143] A. B. Yelikar,[145] X. Yin,[35] J. Yokoyama,[320,42] T. Yokozawa,[50] S. Yuan,[73] H. Yuzurihara,[50] M. Zanolin,[66] M. Zeeshan,[112] T. Zelenova,[63] J.-P. Zendri,[93] M. Zeoli,[15] M. Zerrad,[40] M. Zevin,[97] L. Zhang,[11] N. Zhang,[57] R. Zhang,[151] T. Zhang,[120] C. Zhao,[73] Yue Zhao,[163] Yuhang Zhao,[20] Z.-C. Zhao,[321] Y. Zheng,[106] H. Zhong,[18] H. Zhou,[79] H. O. Zhu,[73] Z.-H. Zhu,[321,322] A. B. Zimmerman,[149] L. Zimmermann,[56] M. E. Zucker,[35,11] J. Zweizig,[11]

The LIGO Scientific Collaboration, the Virgo Collaboration, and the KAGRA Collaboration

[1]*Max Planck Institute for Gravitational Physics (Albert Einstein Institute), D-14476 Potsdam, Germany*

[2]*LIGO Hanford Observatory, Richland, WA 99352, USA*
[3]*Dipartimento di Farmacia, Università di Salerno, I-84084 Fisciano, Salerno, Italy*
[4]*INFN, Sezione di Napoli, I-80126 Napoli, Italy*
[5]*University of Warwick, Coventry CV4 7AL, United Kingdom*
[6]*OzGrav, School of Physics & Astronomy, Monash University, Clayton 3800, Victoria, Australia*
[7]*The Pennsylvania State University, University Park, PA 16802, USA*
[8]*Max Planck Institute for Gravitational Physics (Albert Einstein Institute), D-30167 Hannover, Germany*
[9]*Leibniz Universität Hannover, D-30167 Hannover, Germany*
[10]*University of Wisconsin-Milwaukee, Milwaukee, WI 53201, USA*
[11]*LIGO Laboratory, California Institute of Technology, Pasadena, CA 91125, USA*
[12]*Louisiana State University, Baton Rouge, LA 70803, USA*
[13]*Tata Institute of Fundamental Research, Mumbai 400005, India*
[14]*Centre de Physique Théorique, Aix-Marseille Université, Campus de Luminy, 163 Av. de Luminy, 13009 Marseille, France*
[15]*Université catholique de Louvain, B-1348 Louvain-la-Neuve, Belgium*
[16]*Queen Mary University of London, London E1 4NS, United Kingdom*
[17]*University of California, Davis, Davis, CA 95616, USA*
[18]*University of Minnesota, Minneapolis, MN 55455, USA*
[19]*Instituto Nacional de Pesquisas Espaciais, 12227-010 São José dos Campos, São Paulo, Brazil*
[20]*Université Paris Cité, CNRS, Astroparticule et Cosmologie, F-75013 Paris, France*
[21]*Università di Roma Tor Vergata, I-00133 Roma, Italy*
[22]*INFN, Sezione di Roma Tor Vergata, I-00133 Roma, Italy*
[23]*Universiteit Antwerpen, 2000 Antwerpen, Belgium*
[24]*International Centre for Theoretical Sciences, Tata Institute of Fundamental Research, Bengaluru 560089, India*
[25]*Gravitational Wave Science Project, National Astronomical Observatory of Japan, 2-21-1 Osawa, Mitaka City, Tokyo 181-8588, Japan*
[26]*Advanced Technology Center, National Astronomical Observatory of Japan, 2-21-1 Osawa, Mitaka City, Tokyo 181-8588, Japan*
[27]*Theoretisch-Physikalisches Institut, Friedrich-Schiller-Universität Jena, D-07743 Jena, Germany*
[28]*INFN Sezione di Torino, I-10125 Torino, Italy*
[29]*INFN, Sezione di Genova, I-16146 Genova, Italy*
[30]*Dipartimento di Fisica, Università degli Studi di Genova, I-16146 Genova, Italy*
[31]*Univ. Savoie Mont Blanc, CNRS, Laboratoire d'Annecy de Physique des Particules - IN2P3, F-74000 Annecy, France*
[32]*Università di Napoli "Federico II", I-80126 Napoli, Italy*
[33]*Cardiff University, Cardiff CF24 3AA, United Kingdom*
[34]*OzGrav, Australian National University, Canberra, Australian Capital Territory 0200, Australia*
[35]*LIGO Laboratory, Massachusetts Institute of Technology, Cambridge, MA 02139, USA*
[36]*Maastricht University, 6200 MD Maastricht, Netherlands*
[37]*Nikhef, 1098 XG Amsterdam, Netherlands*
[38]*INFN, Sezione di Roma, I-00185 Roma, Italy*
[39]*Università di Roma "La Sapienza", I-00185 Roma, Italy*
[40]*Aix Marseille Univ, CNRS, Centrale Med, Institut Fresnel, F-13013 Marseille, France*
[41]*Université Paris-Saclay, CNRS/IN2P3, IJCLab, 91405 Orsay, France*
[42]*University of Tokyo, Tokyo, 113-0033, Japan*
[43]*Institut de Física d'Altes Energies (IFAE), The Barcelona Institute of Science and Technology, Campus UAB, E-08193 Bellaterra (Barcelona), Spain*
[44]*Gran Sasso Science Institute (GSSI), I-67100 L'Aquila, Italy*
[45]*INFN, Laboratori Nazionali del Gran Sasso, I-67100 Assergi, Italy*
[46]*University of Florida, Gainesville, FL 32611, USA*
[47]*Dipartimento di Scienze Matematiche, Informatiche e Fisiche, Università di Udine, I-33100 Udine, Italy*
[48]*INFN, Sezione di Trieste, I-34127 Trieste, Italy*
[49]*Tecnologico de Monterrey, Escuela de Ingeniería y Ciencias, 64849 Monterrey, Nuevo León, Mexico*
[50]*Institute for Cosmic Ray Research, KAGRA Observatory, The University of Tokyo, 238 Higashi-Mozumi, Kamioka-cho, Hida City, Gifu 506-1205, Japan*
[51]*INFN, Sezione di Perugia, I-06123 Perugia, Italy*
[52]*Università di Camerino, I-62032 Camerino, Italy*
[53]*University of Washington, Seattle, WA 98195, USA*
[54]*California State University Fullerton, Fullerton, CA 92831, USA*
[55]*SUPA, University of Strathclyde, Glasgow G1 1XQ, United Kingdom*
[56]*Université Claude Bernard Lyon 1, CNRS, IP2I Lyon / IN2P3, UMR 5822, F-69622 Villeurbanne, France*
[57]*Georgia Institute of Technology, Atlanta, GA 30332, USA*
[58]*Chennai Mathematical Institute, Chennai 603103, India*

[59]Royal Holloway, University of London, London TW20 0EX, United Kingdom
[60]Astronomical course, The Graduate University for Advanced Studies (SOKENDAI), 2-21-1 Osawa, Mitaka City, Tokyo 181-8588, Japan
[61]Università degli Studi di Urbino "Carlo Bo", I-61029 Urbino, Italy
[62]INFN, Sezione di Firenze, I-50019 Sesto Fiorentino, Firenze, Italy
[63]European Gravitational Observatory (EGO), I-56021 Cascina, Pisa, Italy
[64]LIGO Livingston Observatory, Livingston, LA 70754, USA
[65]Université de Strasbourg, CNRS, IPHC UMR 7178, F-67000 Strasbourg, France
[66]Embry-Riddle Aeronautical University, Prescott, AZ 86301, USA
[67]Dipartimento di Fisica "E.R. Caianiello", Università di Salerno, I-84084 Fisciano, Salerno, Italy
[68]King's College London, University of London, London WC2R 2LS, United Kingdom
[69]Korea Institute of Science and Technology Information, Daejeon 34141, Republic of Korea
[70]International College, Osaka University, 1-1 Machikaneyama-cho, Toyonaka City, Osaka 560-0043, Japan
[71]Accelerator Laboratory, High Energy Accelerator Research Organization (KEK), 1-1 Oho, Tsukuba City, Ibaraki 305-0801, Japan
[72]Institute for Gravitational and Subatomic Physics (GRASP), Utrecht University, 3584 CC Utrecht, Netherlands
[73]OzGrav, University of Western Australia, Crawley, Western Australia 6009, Australia
[74]University of Portsmouth, Portsmouth, PO1 3FX, United Kingdom
[75]Università di Trento, Dipartimento di Fisica, I-38123 Povo, Trento, Italy
[76]INFN, Trento Institute for Fundamental Physics and Applications, I-38123 Povo, Trento, Italy
[77]Università di Perugia, I-06123 Perugia, Italy
[78]University of Oregon, Eugene, OR 97403, USA
[79]Syracuse University, Syracuse, NY 13244, USA
[80]Inter-University Centre for Astronomy and Astrophysics, Pune 411007, India
[81]INFN, Sezione di Pisa, I-56127 Pisa, Italy
[82]Università di Pisa, I-56127 Pisa, Italy
[83]Institut de Ciències del Cosmos (ICCUB), Universitat de Barcelona (UB), c. Martí i Franquès, 1, 08028 Barcelona, Spain
[84]Departament de Física Quàntica i Astrofísica (FQA), Universitat de Barcelona (UB), c. Martí i Franqués, 1, 08028 Barcelona, Spain
[85]Institut d'Estudis Espacials de Catalunya, c. Gran Capità, 2-4, 08034 Barcelona, Spain
[86]Dipartimento di Medicina, Chirurgia e Odontoiatria "Scuola Medica Salernitana", Università di Salerno, I-84081 Baronissi, Salerno, Italy
[87]IGR, University of Glasgow, Glasgow G12 8QQ, United Kingdom
[88]HUN-REN Wigner Research Centre for Physics, H-1121 Budapest, Hungary
[89]Concordia University Wisconsin, Mequon, WI 53097, USA
[90]Stanford University, Stanford, CA 94305, USA
[91]University of Michigan, Ann Arbor, MI 48109, USA
[92]Università di Padova, Dipartimento di Fisica e Astronomia, I-35131 Padova, Italy
[93]INFN, Sezione di Padova, I-35131 Padova, Italy
[94]Institute for Plasma Research, Bhat, Gandhinagar 382428, India
[95]Universiteit Gent, B-9000 Gent, Belgium
[96]Nicolaus Copernicus Astronomical Center, Polish Academy of Sciences, 00-716, Warsaw, Poland
[97]Northwestern University, Evanston, IL 60208, USA
[98]Universität Hamburg, D-22761 Hamburg, Germany
[99]IAC3–IEEC, Universitat de les Illes Balears, E-07122 Palma de Mallorca, Spain
[100]Aix-Marseille Université, Université de Toulon, CNRS, CPT, Marseille, France
[101]Laboratoire des 2 Infinis - Toulouse (L2IT-IN2P3), F-31062 Toulouse Cedex 9, France
[102]Università di Siena, Dipartimento di Scienze Fisiche, della Terra e dell'Ambiente, I-53100 Siena, Italy
[103]Villanova University, Villanova, PA 19085, USA
[104]RRCAT, Indore, Madhya Pradesh 452013, India
[105]Kenyon College, Gambier, OH 43022, USA
[106]Missouri University of Science and Technology, Rolla, MO 65409, USA
[107]Indian Institute of Technology Madras, Chennai 600036, India
[108]Department of Physics and Astronomy, Vrije Universiteit Amsterdam, 1081 HV Amsterdam, Netherlands
[109]Lomonosov Moscow State University, Moscow 119991, Russia
[110]Center for Theoretical Physics of the Polish Academy of Sciences, al. Lotnik'ow 32/46, 02-668, Warsaw, Poland
[111]Katholieke Universiteit Leuven, Oude Markt 13, 3000 Leuven, Belgium
[112]Rochester Institute of Technology, Rochester, NY 14623, USA
[113]Université libre de Bruxelles, 1050 Bruxelles, Belgium
[114]Bar-Ilan University, Ramat Gan, 5290002, Israel
[115]Université Côte d'Azur, Observatoire de la Côte d'Azur, CNRS, Artemis, F-06304 Nice, France

[116]University of British Columbia, Vancouver, BC V6T 1Z4, Canada
[117]OzGrav, University of Adelaide, Adelaide, South Australia 5005, Australia
[118]Centre national de la recherche scientifique, 75016 Paris, France
[119]Univ Rennes, CNRS, Institut FOTON - UMR 6082, F-35000 Rennes, France
[120]University of Birmingham, Birmingham B15 2TT, United Kingdom
[121]Washington State University, Pullman, WA 99164, USA
[122]Cornell University, Ithaca, NY 14850, USA
[123]Laboratoire Kastler Brossel, Sorbonne Université, CNRS, ENS-Université PSL, Collège de France, F-75005 Paris, France
[124]Christopher Newport University, Newport News, VA 23606, USA
[125]OzGrav, University of Melbourne, Parkville, Victoria 3010, Australia
[126]Astronomical Observatory Warsaw University, 00-478 Warsaw, Poland
[127]University of Maryland, College Park, MD 20742, USA
[128]Università degli Studi di Milano-Bicocca, I-20126 Milano, Italy
[129]INFN, Sezione di Milano-Bicocca, I-20126 Milano, Italy
[130]Université de Lyon, Université Claude Bernard Lyon 1, CNRS, Institut Lumière Matière, F-69622 Villeurbanne, France
[131]University of Chicago, Chicago, IL 60637, USA
[132]University of Arizona, Tucson, AZ 85721, USA
[133]INFN, Sezione di Napoli, Gruppo Collegato di Salerno, I-80126 Napoli, Italy
[134]University of Massachusetts Dartmouth, North Dartmouth, MA 02747, USA
[135]Niels Bohr Institute, Copenhagen University, 2100 København, Denmark
[136]Universidad de Guadalajara, 44430 Guadalajara, Jalisco, Mexico
[137]Istituto di Astrofisica e Planetologia Spaziali di Roma, 00133 Roma, Italy
[138]Colorado State University, Fort Collins, CO 80523, USA
[139]Departamento de Astronomía y Astrofísica, Universitat de València, E-46100 Burjassot, València, Spain
[140]Observatori Astronòmic, Universitat de València, E-46980 Paterna, València, Spain
[141]Niels Bohr Institute, University of Copenhagen, 2100 Kóbenhavn, Denmark
[142]National Central University, Taoyuan City 320317, Taiwan
[143]National Tsing Hua University, Hsinchu City 30013, Taiwan
[144]OzGrav, Charles Sturt University, Wagga Wagga, New South Wales 2678, Australia
[145]Vanderbilt University, Nashville, TN 37235, USA
[146]University of the Chinese Academy of Sciences / International Centre for Theoretical Physics Asia-Pacific, Bejing 100049, China
[147]Department of Electrophysics, National Yang Ming Chiao Tung University, 101 Univ. Street, Hsinchu, Taiwan
[148]Kamioka Branch, National Astronomical Observatory of Japan, 238 Higashi-Mozumi, Kamioka-cho, Hida City, Gifu 506-1205, Japan
[149]University of Texas, Austin, TX 78712, USA
[150]CaRT, California Institute of Technology, Pasadena, CA 91125, USA
[151]Northeastern University, Boston, MA 02115, USA
[152]Dipartimento di Ingegneria Industriale (DIIN), Università di Salerno, I-84084 Fisciano, Salerno, Italy
[153]Faculty of Science, University of Toyama, 3190 Gofuku, Toyama City, Toyama 930-8555, Japan
[154]Carleton College, Northfield, MN 55057, USA
[155]University of Szeged, Dóm tér 9, Szeged 6720, Hungary
[156]OzGrav, Swinburne University of Technology, Hawthorn VIC 3122, Australia
[157]INFN Cagliari, Physics Department, Università degli Studi di Cagliari, Cagliari 09042, Italy
[158]Università degli Studi di Cagliari, Via Università 40, 09124 Cagliari, Italy
[159]Université Libre de Bruxelles, Brussels 1050, Belgium
[160]INAF, Osservatorio Astronomico di Brera sede di Merate, I-23807 Merate, Lecco, Italy
[161]Departamento de Matemáticas, Universitat de València, E-46100 Burjassot, València, Spain
[162]Montana State University, Bozeman, MT 59717, USA
[163]The University of Utah, Salt Lake City, UT 84112, USA
[164]Johns Hopkins University, Baltimore, MD 21218, USA
[165]University of Rhode Island, Kingston, RI 02881, USA
[166]The University of Texas Rio Grande Valley, Brownsville, TX 78520, USA
[167]Université de Liège, B-4000 Liège, Belgium
[168]DIFA- Alma Mater Studiorum Università di Bologna, Via Zamboni, 33 - 40126 Bologna, Italy
[169]Istituto Nazionale Di Fisica Nucleare - Sezione di Bologna, viale Carlo Berti Pichat 6/2 - 40127 Bologna, Italy
[170]University of Manitoba, Winnipeg, MB R3T 2N2, Canada
[171]INFN-CNAF - Bologna, Viale Carlo Berti Pichat, 6/2, 40127 Bologna BO, Italy
[172]Università degli Studi di Sassari, I-07100 Sassari, Italy

[173]INFN, Laboratori Nazionali del Sud, I-95125 Catania, Italy
[174]Université de Normandie, ENSICAEN, UNICAEN, CNRS/IN2P3, LPC Caen, F-14000 Caen, France
[175]Laboratoire de Physique Corpusculaire Caen, 6 boulevard du maréchal Juin, F-14050 Caen, France
[176]The University of Sheffield, Sheffield S10 2TN, United Kingdom
[177]Université Claude Bernard Lyon 1, CNRS, Laboratoire des Matériaux Avancés (LMA), IP2I Lyon / IN2P3, UMR 5822, F-69622 Villeurbanne, France
[178]Università di Firenze, Sesto Fiorentino I-50019, Italy
[179]IGFAE, Universidade de Santiago de Compostela, E-15782 Santiago de Compostela, Spain
[180]Dipartimento di Scienze Matematiche, Fisiche e Informatiche, Università di Parma, I-43124 Parma, Italy
[181]INFN, Sezione di Milano Bicocca, Gruppo Collegato di Parma, I-43124 Parma, Italy
[182]California State University, Los Angeles, Los Angeles, CA 90032, USA
[183]Marquette University, Milwaukee, WI 53233, USA
[184]Perimeter Institute, Waterloo, ON N2L 2Y5, Canada
[185]Corps des Mines, Mines Paris, Université PSL, 60 Bd Saint-Michel, 75272 Paris, France
[186]Dipartimento di Fisica, Università di Trieste, I-34127 Trieste, Italy
[187]Université Côte d'Azur, Observatoire de la Côte d'Azur, CNRS, Lagrange, F-06304 Nice, France
[188]National Center for Nuclear Research, 05-400 Świerk-Otwock, Poland
[189]Vrije Universiteit Brussel, 1050 Brussel, Belgium
[190]University of Zurich, Winterthurerstrasse 190, 8057 Zurich, Switzerland
[191]Canadian Institute for Theoretical Astrophysics, University of Toronto, Toronto, ON M5S 3H8, Canada
[192]Stony Brook University, Stony Brook, NY 11794, USA
[193]Center for Computational Astrophysics, Flatiron Institute, New York, NY 10010, USA
[194]Montclair State University, Montclair, NJ 07043, USA
[195]HUN-REN Institute for Nuclear Research, H-4026 Debrecen, Hungary
[196]Indian Institute of Technology Bombay, Powai, Mumbai 400 076, India
[197]Centro de Física das Universidades do Minho e do Porto, Universidade do Minho, PT-4710-057 Braga, Portugal
[198]Aix Marseille Univ, CNRS/IN2P3, CPPM, Marseille, France
[199]CNR-SPIN, I-84084 Fisciano, Salerno, Italy
[200]Scuola di Ingegneria, Università della Basilicata, I-85100 Potenza, Italy
[201]Western Washington University, Bellingham, WA 98225, USA
[202]SUPA, University of the West of Scotland, Paisley PA1 2BE, United Kingdom
[203]Barry University, Miami Shores, FL 33168, USA
[204]Eötvös University, Budapest 1117, Hungary
[205]Institute for Cosmic Ray Research, KAGRA Observatory, The University of Tokyo, 5-1-5 Kashiwa-no-Ha, Kashiwa City, Chiba 277-8582, Japan
[206]Department of Physics, Graduate School of Science, Osaka Metropolitan University, 3-3-138 Sugimoto-cho, Sumiyoshi-ku, Osaka City, Osaka 558-8585, Japan
[207]University of Sannio at Benevento, I-82100 Benevento, Italy and INFN, Sezione di Napoli, I-80100 Napoli, Italy
[208]University of California, Berkeley, CA 94720, USA
[209]Instituto de Fisica Teorica UAM-CSIC, Universidad Autonoma de Madrid, 28049 Madrid, Spain
[210]Laboratoire d'Acoustique de l'Université du Mans, UMR CNRS 6613, F-72085 Le Mans, France
[211]University of Southampton, Southampton SO17 1BJ, United Kingdom
[212]University of California, Riverside, Riverside, CA 92521, USA
[213]Dipartimento di Ingegneria Industriale, Elettronica e Meccanica, Università degli Studi Roma Tre, I-00146 Roma, Italy
[214]University of Nevada, Las Vegas, Las Vegas, NV 89154, USA
[215]University of Nottingham NG7 2RD, UK
[216]Ariel University, Ramat HaGolan St 65, Ari'el, Israel
[217]The University of Mississippi, University, MS 38677, USA
[218]Graduate School of Science, Institute of Science Tokyo, 2-12-1 Ookayama, Meguro-ku, Tokyo 152-8551, Japan
[219]Institute of Physics, Academia Sinica, 128 Sec. 2, Academia Rd., Nankang, Taipei 11529, Taiwan
[220]The Chinese University of Hong Kong, Shatin, NT, Hong Kong
[221]American University, Washington, DC 20016, USA
[222]Dipartimento di Fisica, Università degli studi di Milano, Via Celoria 16, I-20133, Milano, Italy
[223]INFN, sezione di Milano, Via Celoria 16, I-20133, Milano, Italy
[224]Department of Applied Physics, Fukuoka University, 8-19-1 Nanakuma, Jonan, Fukuoka City, Fukuoka 814-0180, Japan
[225]University of Cambridge, Cambridge CB2 1TN, United Kingdom
[226]University of Lancaster, Lancaster LA1 4YW, United Kingdom
[227]College of Industrial Technology, Nihon University, 1-2-1 Izumi, Narashino City, Chiba 275-8575, Japan
[228]Faculty of Engineering, Niigata University, 8050 Ikarashi-2-no-cho, Nishi-ku, Niigata City, Niigata 950-2181, Japan

[229]*Department of Physics, Tamkang University, No. 151, Yingzhuan Rd., Danshui Dist., New Taipei City 25137, Taiwan*
[230]*Rutherford Appleton Laboratory, Didcot OX11 0DE, United Kingdom*
[231]*Department of Physical Sciences, Aoyama Gakuin University, 5-10-1 Fuchinobe, Sagamihara City, Kanagawa 252-5258, Japan*
[232]*Helmut Schmidt University, D-22043 Hamburg, Germany*
[233]*Nambu Yoichiro Institute of Theoretical and Experimental Physics (NITEP), Osaka Metropolitan University, 3-3-138 Sugimoto-cho, Sumiyoshi-ku, Osaka City, Osaka 558-8585, Japan*
[234]*Directorate of Construction, Services & Estate Management, Mumbai 400094, India*
[235]*Observatoire Astronomique de Strasbourg, 11 Rue de l'Université, 67000 Strasbourg, France*
[236]*Faculty of Physics, University of Białystok, 15-245 Białystok, Poland*
[237]*National Astronomical Observatories, Chinese Academic of Sciences, 20A Datun Road, Chaoyang District, Beijing, China*
[238]*School of Astronomy and Space Science, University of Chinese Academy of Sciences, 20A Datun Road, Chaoyang District, Beijing, China*
[239]*Sungkyunkwan University, Seoul 03063, Republic of Korea*
[240]*Department of Physics, Ulsan National Institute of Science and Technology (UNIST), 50 UNIST-gil, Ulju-gun, Ulsan 44919, Republic of Korea*
[241]*Institute for Cosmic Ray Research, The University of Tokyo, 5-1-5 Kashiwa-no-Ha, Kashiwa City, Chiba 277-8582, Japan*
[242]*Chung-Ang University, Seoul 06974, Republic of Korea*
[243]*University of Washington Bothell, Bothell, WA 98011, USA*
[244]*Laboratoire de Physique et de Chimie de l'Environnement, Université Joseph KI-ZERBO, 9GH2+3V5, Ouagadougou, Burkina Faso*
[245]*Ewha Womans University, Seoul 03760, Republic of Korea*
[246]*National Institute for Mathematical Sciences, Daejeon 34047, Republic of Korea*
[247]*Korea Astronomy and Space Science Institute, Daejeon 34055, Republic of Korea*
[248]*Department of Astronomy and Space Science, Chungnam National University, 9 Daehak-ro, Yuseong-gu, Daejeon 34134, Republic of Korea*
[249]*Institute of Particle and Nuclear Studies (IPNS), High Energy Accelerator Research Organization (KEK), 1-1 Oho, Tsukuba City, Ibaraki 305-0801, Japan*
[250]*Division of Science, National Astronomical Observatory of Japan, 2-21-1 Osawa, Mitaka City, Tokyo 181-8588, Japan*
[251]*Nagoya University, Nagoya, 464-8601, Japan*
[252]*Department of Physics, Aristotle University of Thessaloniki, 54124 Thessaloniki, Greece*
[253]*Bard College, Annandale-On-Hudson, NY 12504, USA*
[254]*Technical University of Braunschweig, D-38106 Braunschweig, Germany*
[255]*Institute of Mathematics, Polish Academy of Sciences, 00656 Warsaw, Poland*
[256]*Astronomical Observatory, Jagiellonian University, 31-007 Cracow, Poland*
[257]*Department of Physics and Astronomy, University of Padova, Via Marzolo, 8-35151 Padova, Italy*
[258]*Sezione di Padova, Istituto Nazionale di Fisica Nucleare (INFN), Via Marzolo, 8-35131 Padova, Italy*
[259]*Department of Physics, Nagoya University, ES building, Furocho, Chikusa-ku, Nagoya, Aichi 464-8602, Japan*
[260]*Université de Montréal/Polytechnique, Montreal, Quebec H3T 1J4, Canada*
[261]*Indian Institute of Science Education and Research, Kolkata, Mohanpur, West Bengal 741252, India*
[262]*Seoul National University, Seoul 08826, Republic of Korea*
[263]*Department of Computer Simulation, Inje University, 197 Inje-ro, Gimhae, Gyeongsangnam-do 50834, Republic of Korea*
[264]*NAVIER, École des Ponts, Univ Gustave Eiffel, CNRS, Marne-la-Vallée, France*
[265]*Gravitational Wave Science Project, National Astronomical Observatory of Japan (NAOJ), Mitaka City, Tokyo 181-8588, Japan*
[266]*Department of Physics, National Cheng Kung University, No.1, University Road, Tainan City 701, Taiwan*
[267]*St. Thomas University, Miami Gardens, FL 33054, USA*
[268]*Scuola Normale Superiore, I-56126 Pisa, Italy*
[269]*Institució Catalana de Recerca i Estudis Avançats, E-08010 Barcelona, Spain*
[270]*Institut de Física d'Altes Energies, E-08193 Barcelona, Spain*
[271]*Institut fuer Theoretische Astrophysik, Zentrum fuer Astronomie Heidelberg, Universitaet Heidelberg, Albert Ueberle Str. 2, 69120 Heidelberg, Germany*
[272]*Institucio Catalana de Recerca i Estudis Avançats (ICREA), Passeig de Lluís Companys, 23, 08010 Barcelona, Spain*
[273]*Research Center for Space Science, Advanced Research Laboratories, Tokyo City University, 3-3-1 Ushikubo-Nishi, Tsuzuki-Ku, Yokohama, Kanagawa 224-8551, Japan*
[274]*Tsinghua University, Beijing 100084, China*
[275]*Institut des Hautes Etudes Scientifiques, F-91440 Bures-sur-Yvette, France*
[276]*Faculty of Law, Ryukoku University, 67 Fukakusa Tsukamoto-cho, Fushimi-ku, Kyoto City, Kyoto 612-8577, Japan*
[277]*Phenikaa Institute for Advanced Study (PIAS), Phenikaa University, Yen Nghia, Ha Dong, Hanoi, Vietnam*
[278]*University of Stavanger, 4021 Stavanger, Norway*
[279]*Physics Program, Graduate School of Advanced Science and Engineering, Hiroshima University, 1-3-1 Kagamiyama, Higashihiroshima City, Hiroshima 739-8526, Japan*
[280]*GRAPPA, Anton Pannekoek Institute for Astronomy and Institute for High-Energy Physics, University of Amsterdam, 1098 XH Amsterdam, Netherlands*
[281]*University College London, London WC1E 6BT, United Kingdom*
[282]*Observatoire de Paris, 75014 Paris, France*

[283]Laboratoire Univers et Théories, Observatoire de Paris, 92190 Meudon, France
[284]Graduate School of Science and Technology, Niigata University, 8050 Ikarashi-2-no-cho, Nishi-ku, Niigata City, Niigata 950-2181, Japan
[285]University of Maryland, Baltimore County, Baltimore, MD 21250, USA
[286]CSIR-Central Glass and Ceramic Research Institute, Kolkata, West Bengal 700032, India
[287]Consiglio Nazionale delle Ricerche - Istituto dei Sistemi Complessi, I-00185 Roma, Italy
[288]Department of Astronomy, Yonsei University, 50 Yonsei-Ro, Seodaemun-Gu, Seoul 03722, Republic of Korea
[289]Department of Physics, University of Guadalajara, Av. Revolucion 1500, Colonia Olimpica C.P. 44430, Guadalajara, Jalisco, Mexico
[290]Hobart and William Smith Colleges, Geneva, NY 14456, USA
[291]INAF, Osservatorio Astronomico di Padova, I-35122 Padova, Italy
[292]Dipartimento di Ingegneria, Università del Sannio, I-82100 Benevento, Italy
[293]Museo Storico della Fisica e Centro Studi e Ricerche "Enrico Fermi", I-00184 Roma, Italy
[294]Kennesaw State University, Kennesaw, GA 30144, USA
[295]Eotvos University, Budapest 1117, Hungary
[296]Subatech, CNRS/IN2P3 - IMT Atlantique - Nantes Université, 4 rue Alfred Kastler BP 20722 44307 Nantes CÉDEX 03, France
[297]LIGO Laboratory, California Institrenziniute of Technology, Pasadena, CA 91125, USA
[298]Departamento de Física - ETSIDI, Universidad Politécnica de Madrid, 28012 Madrid, Spain
[299]Department of Electronic Control Engineering, National Institute of Technology, Nagaoka College, 888 Nishikatakai, Nagaoka City, Niigata 940-8532, Japan
[300]Trinity College, Hartford, CT 06106, USA
[301]Dipartimento di Fisica e Scienze della Terra, Università Degli Studi di Ferrara, Via Saragat, 1, 44121 Ferrara FE, Italy
[302]Faculty of Science, Toho University, 2-2-1 Miyama, Funabashi City, Chiba 274-8510, Japan
[303]Indian Institute of Technology, Palaj, Gandhinagar, Gujarat 382355, India
[304]Kavli Institute for Astronomy and Astrophysics, Peking University, Yiheyuan Road 5, Haidian District, Beijing 100871, China
[305]Laboratoire MSME, Cité Descartes, 5 Boulevard Descartes, Champs-sur-Marne, 77454 Marne-la-Vallée Cedex 2, France
[306]Faculty of Information Science and Technology, Osaka Institute of Technology, 1-79-1 Kitayama, Hirakata City, Osaka 573-0196, Japan
[307]NASA Goddard Space Flight Center, Greenbelt, MD 20771, USA
[308]Faculty of Science and Technology, Kochi University, 2-5-1 Akebono-cho, Kochi-shi, Kochi 780-8520, Japan
[309]Laboratoire de Physique de l'École Normale Supérieure, ENS, (CNRS, Université PSL, Sorbonne Université, Université Paris Cité), F-75005 Paris, France
[310]Faculty of Physics, University of Warsaw, Ludwika Pasteura 5, 02-093 Warszawa, Poland
[311]Laser Interferometry and Gravitational Wave Astronomy, Max Planck Institute for Gravitational Physics, Callinstrasse 38, 30167 Hannover, Germany
[312]The Hakubi Center for Advanced Research, Kyoto University, Yoshida-honmachi, Sakyou-ku, Kyoto City, Kyoto 606-8501, Japan
[313]Department of Physics, Kyoto University, Kita-Shirakawa Oiwake-cho, Sakyou-ku, Kyoto City, Kyoto 606-8502, Japan
[314]Yukawa Institute for Theoretical Physics (YITP), Kyoto University, Kita-Shirakawa Oiwake-cho, Sakyou-ku, Kyoto City, Kyoto 606-8502, Japan
[315]University of Catania, Department of Physics and Astronomy, Via S. Sofia, 64, 95123 Catania CT, Italy
[316]National Institute of Technology, Fukui College, Geshi-cho, Sabae-shi, Fukui 916-8507, Japan
[317]Department of Communications Engineering, National Defense Academy of Japan, 1-10-20 Hashirimizu, Yokosuka City, Kanagawa 239-8686, Japan
[318]Universidad de Antioquia, Medellín, Colombia
[319]Eindhoven University of Technology, 5600 MB Eindhoven, Netherlands
[320]Kavli Institute for the Physics and Mathematics of the Universe (Kavli IPMU), WPI, The University of Tokyo, 5-1-5 Kashiwa-no-Ha, Kashiwa City, Chiba 277-8583, Japan
[321]Department of Astronomy, Beijing Normal University, Xinjiekouwai Street 19, Haidian District, Beijing 100875, China
[322]School of Physics and Technology, Wuhan University, Bayi Road 299, Wuchang District, Wuhan, Hubei, 430072, China

## ABSTRACT

We analyze data from 142 of the 218 gravitational-wave (GW) sources in the fourth LIGO–Virgo–KAGRA Collaboration (LVK) Gravitational-Wave Transient Catalog (GWTC-4.0) to estimate the Hubble constant $H_0$ jointly with the population properties of merging compact binaries. We measure the luminosity distance and redshifted masses of GW sources directly; in contrast, we infer GW source redshifts statistically through i) location of features in the compact object mass spectrum and merger rate evolution, and ii) identifying potential host galaxies in the GW localization volume. Probing the relationship between source luminosity distances and redshifts obtained in this way yields constraints on cosmological parameters. We also constrain parameterized deviations from general relativity which affect GW propagation, specifically those modifying the dependence of a GW signal on the source luminosity distance. Assuming our fiducial model for the source-frame mass distribution and using GW candidates detected up to the end of the fourth observing run (O4a), together with the GLADE+ all-sky galaxy catalog, we estimate $H_0 = 76.6^{+13.0}_{-9.5}$ $(76.6^{+25.2}_{-14.0})\,\mathrm{km\,s^{-1}\,Mpc^{-1}}$. This value is reported as a median with 68.3% (90%) symmetric credible interval, and includes combination with the $H_0$ measurement from GW170817 and its electromagnetic counterpart. Using a parametrization of modified GW propagation in terms of the magnitude parameter $\Xi_0$, we estimate $\Xi_0 = 1.2^{+0.8}_{-0.4}\,(1.2^{+2.4}_{-0.5})$, where $\Xi_0 = 1$ recovers the behavior of general relativity.

## 1. INTRODUCTION

Obtaining independent measurements of the Hubble constant ($H_0$) is a major focus of gravitational-wave (GW) cosmology, driven by the existing discrepancy between early Universe measurements from the cosmic microwave background (CMB) radiation and local measurements from standardizable sources such as Type Ia supernovae (SNe Ia). Measurements of $H_0$ made by the Planck Collaboration in the Planck 2018 Data Release (Aghanim et al. 2020) and the Supernovae H0 for the Equation of State (SH0ES) project with the recalibration of supernovae by Large Magellanic Cloud Cepheids (Riess et al. 2022) have now reached an $\sim 8\%$ discrepancy with $\gtrsim 5\sigma$ credibility, although other local measurements, including alternative methods of calibrating the distance ladder, suggest a smaller tension (e.g., Di Valentino & Brout 2024).

The possibility of using GW detections to infer cosmological parameters, such as $H_0$, was first proposed by Schutz (1986). GWs from compact binary coalescences (CBCs) serve as *standard sirens* (Holz & Hughes 2005), providing a self-calibrated measure of luminosity distance that is independent of traditional methods such as the cosmic distance ladder. If combined with redshift information, GWs can be used as probes of the luminosity distance-redshift relation, which depends on the cosmological model and its parameters. In this way GW sources may help to resolve the $H_0$ discrepancy, and can also provide insights into possible new physics beyond the standard Lambda cold dark matter ($\Lambda$CDM) cosmological model (Bull et al. 2016; Perivolaropoulos & Skara 2022; Abdalla et al. 2022; Di Valentino et al. 2025).

However, the redshift of a CBC source cannot be determined from the GW signal itself due to its degeneracy with the binary source masses (Krolak & Schutz 1987). Several methods have been proposed to break this degeneracy. If a counterpart in the electromagnetic (EM) spectrum can be uniquely associated to the GW event, the redshift of the galaxy host can be determined via astronomical photometry or spectroscopy (Holz & Hughes 2005; Dalal et al. 2006; Nissanke et al. 2010, 2013a; Abbott et al. 2017a; Chen et al. 2018; Feeney et al. 2019): we will refer to such an event as a *bright* siren. The only bright siren observed to date is the binary neutron star (BNS) merger GW170817 (Abbott et al. 2017b), which, combined with coincident EM transients associated with the host galaxy NGC 4993 (Abbott et al. 2017c), provided the first bright standard siren measurement of $H_0$ (Abbott et al. 2017a). While we are waiting for the next bright siren event, the steady increase of detections from binary black hole (BBH), neutron star–black hole binary (NSBH) and other BNS candidates without confident EM counterparts has driven forward other methods to measure $H_0$.

One approach relies on the presence of features in the mass spectrum of binary compact objects to break the mass-redshift degeneracy (Chernoff & Finn 1993; Markovic 1993; Taylor et al. 2012; Farr et al. 2019; You et al. 2021; Mastrogiovanni et al. 2021; Ezquiaga & Holz 2021, 2022), a method we will refer to as the *spectral* siren method (also sometimes called the *population* method). By making some assumptions about the source-frame mass distribution of CBCs, the cosmological parameters are sampled together with a set of population parameters describing the source-frame mass distribution and the CBC merger rate (distributions of other CBC parameters, such as spins, may be included). This method has been applied in Abbott et al. (2023a) to the BBH candidates reported in the Gravitational-Wave Transient Catalog (GWTC) 3.0 (Abbott et al. 2021b).

A second approach consists of supplementing the spectral siren method with additional redshift information from galaxy surveys (Schutz 1986; MacLeod & Hogan 2008; Del Pozzo 2012; Nishizawa 2017; Fishbach et al. 2019; Soares-Santos et al. 2019; Gray et al. 2020; Palmese et al. 2020; Abbott et al. 2021a; Finke et al. 2021a; Abbott et al. 2023a; Gair et al. 2023; Borghi et al. 2024; Bom et al. 2024). We will refer to this as the *dark* siren method (also called *galaxy catalog* method, or *galaxy host identification* method) Alternative approaches to infer the source redshift, which we will not consider in this work, take advantage of the cross-correlation between the spatial distribution of GWs and galaxies (Camera & Nishizawa 2013; Oguri 2016; Mukherjee et al. 2020, 2021b, 2024; Afroz & Mukherjee 2024; Fonseca et al. 2023; Zazzera et al. 2025; Ferri et al. 2024; Pedrotti et al. 2025), the adoption of theoretical priors on the merger-redshift distributions (Ding et al. 2019; Ye & Fishbach 2021), and the use of tidal distortions of neutron stars (NSs) (Messenger & Read 2012; Del Pozzo et al. 2017; Chatterjee et al. 2021).

In previous LIGO–Virgo–KAGRA Collaboration (LVK) analyses, it was possible to apply the dark siren method to GWs only by fixing the population parameters to some fiducial values, due to the computational challenges of sampling the cosmological and population parameter space together with highly structured redshift information coming from a galaxy catalog. However, as shown in Abbott et al. (2023a), this made the results strongly dependent on the

* Deceased, September 2024.

assumed BBH source mass distribution parameters. These challenges have recently been overcome in the latest version of the codes used by the LVK, `gwcosmo 3.0` (Gray et al. 2020, 2022, 2023) and `icarogw 2.0` (Mastrogiovanni et al. 2023, 2024), from hereon simply referred to as `gwcosmo` and `icarogw`. Both codes now implement the dark siren method allowing marginalization over the GW population parameters, while incorporating galaxy catalog information. By applying this new method to the full set of publicly available LVK GW observations, we are able to obtain cosmological constraints that are more robust to the systematic uncertainties introduced by the population assumptions (Mastrogiovanni et al. 2021; Abbott et al. 2023a).

The fourth observing run (O4) of the LVK network of detectors began on 2023 May 24 at 15:00:00 UTC, and included the two Laser Interferometer Gravitational-Wave Observatory (LIGO; Aasi et al. 2015) detectors in observing mode after several upgrades that improved their sensitivity (Ganapathy et al. 2023; Jia et al. 2024; Capote et al. 2025), while the Virgo (Acernese et al. 2015) and KAGRA (Akutsu et al. 2021) detectors did not join the observing run in order to continue commissioning (Abac et al. 2025b). The first part of the fourth observing run (O4a) ended on 2024 January 16 at 16:00:00 UTC, and the accompanying version of the Gravitational Wave Transient Catalog 4.0, hereafter referred to as GWTC-4.0 (Abac et al. 2025b,c,d), contains all the candidates reported in previous observing runs, which include the first observing run (O1; Abbott et al. 2016), the second observing run (O2; Abbott et al. 2019a), and the third observing run (O3; Abbott et al. 2021b, 2023b, 2024), in addition to the latest observations from O4a, for a total of 218 candidates. See Abac et al. (2025b) for a general introduction to GWTC-4.0, and the articles presented in the GWTC-4.0 Focus Issue (Abac et al. 2025e) for other aspects of this data set.

In this paper we present an updated estimate of $H_0$ using the full population of BNS, NSBH, and BBH candidates reported in GWTC-4.0. We select candidates for inclusion in the analysis based on a false alarm rate (FAR) of less than 0.25 per year to reduce contamination from noise events. This allows us to combine the bright siren event GW170817 with an additional 141 GW detections used as dark sirens to obtain our final estimate of $H_0$.

In addition, we present constraints on deviations from general relativity (GR) that affect the propagation of GWs and which can be parametrized in terms of a modified GW–EM luminosity-distance ratio (Belgacem et al. 2018a; Ezquiaga 2021; Mancarella et al. 2022; Leyde et al. 2022; Mastrogiovanni et al. 2023; Chen et al. 2024a). These constraints test the hypothesis that gravity behaves differently from GR on cosmological scales, leading to a mistaken inference of a dark energy component (see Clifton et al. 2012 for a comprehensive review of modified gravity models).

The remainder of this paper is organized as follows. In Section 2 we present the spectral and dark siren statistical methods adopted in this study to infer the cosmological and population parameters. In Section 3 we detail the properties of the GW candidates and the galaxy catalog used. In Section 4 we present the results of our analysis and the tests made to check its robustness against systematic errors, while in Section 5 we discuss how our results compare with the literature and the limitations of our analysis. In Section 6 we present our conclusions.

Throughout this paper, unless otherwise stated, we assume a flat-$\Lambda$CDM cosmology and the best-fit Planck-2015 value of $\Omega_{\rm m} = 0.3065$ for the fractional matter density in the current epoch (Ade et al. 2016).

## 2. METHODS

### 2.1. *Dark Sirens Statistical Framework*

To infer cosmology and population–level properties of GW sources from the observed event catalog, we employ a hierarchical Bayesian framework (Mandel et al. 2019; Vitale et al. 2020). The observed sample is modeled as resulting from an inhomogeneous Poisson process in the presence of selection effects, assuming statistically independent and non-overlapping events. Each event in the catalog is described by detector–frame parameters $\boldsymbol{\theta}^{\rm det}$, which include the detector–frame masses and GW luminosity distance, $\boldsymbol{\theta}^{\rm det} \ni \{m_1^{\rm det}, m_2^{\rm det}, D_{\rm L}^{\rm GW}\}$ (where $m_1^{\rm det} \geq m_2^{\rm det}$). For each event, labeled by the index $i$, individual parameter constraints are given in the form of samples from the posterior probability $p\big(\boldsymbol{\theta}_i^{\rm det}|d_i\big)$ for the parameters $\boldsymbol{\theta}_i^{\rm det}$ given the observed data $d_i$. These are assumed to be obtained with a parameter estimation prior that we denote $\pi_{\rm PE}(\boldsymbol{\theta}^{\rm det})$. The event parameters are drawn from a distribution which is modeled as a function of source–frame quantities $\boldsymbol{\theta}$, which include the source–frame masses and redshift, $\boldsymbol{\theta} \ni \{m_1, m_2, z\}$. The population distribution $p_{\rm pop}(\boldsymbol{\theta}|\boldsymbol{\Lambda})$ is described parametrically by a set of *hyperparameters* $\boldsymbol{\Lambda}$ (sometimes simply referred as *parameters*). We infer the cosmological hyperparameters, denoted here as $\boldsymbol{\Lambda}_c$, in addition to the population hyperparameters. As population properties are modeled in source–frame, while GW observations provide information on detector–frame quantities, evaluating the population function implies assuming a cosmology. We therefore write the source–frame variables as functions of the detector–frame ones and of the parameters $\boldsymbol{\Lambda}_{\rm c}$, $\boldsymbol{\theta}_i = \boldsymbol{\theta}_i(\boldsymbol{\theta}_i^{\rm det}, \boldsymbol{\Lambda}_{\rm c})$.

The posterior probability on the parameters $\{\boldsymbol{\Lambda}, \boldsymbol{\Lambda}_c\}$ given the ensemble of GW strain data $\{d\}$ from $N_{\rm det}$ detections can be written as (Loredo 2004; Mandel et al. 2019; Vitale et al. 2020):

$$
\begin{aligned}
&p\,(\boldsymbol{\Lambda}, \boldsymbol{\Lambda}_{\rm c}|\{d\}, N_{\rm det}) \propto \pi(\boldsymbol{\Lambda})\,\pi(\boldsymbol{\Lambda}_{\rm c}) \\
&\times \xi(\boldsymbol{\Lambda}, \boldsymbol{\Lambda}_{\rm c})^{-N_{\rm det}} \prod_{i=1}^{N_{\rm det}} \int {\rm d}\boldsymbol{\theta}_i^{\rm det}\, \frac{p\big(\boldsymbol{\theta}_i^{\rm det}|d_i\big)}{\pi_{\rm PE}\big(\boldsymbol{\theta}_i^{\rm det}\big)} \\
&\times \left[\left|\frac{{\rm d}\boldsymbol{\theta}_i^{\rm det}(\boldsymbol{\theta}_i, \boldsymbol{\Lambda}_{\rm c})}{{\rm d}\boldsymbol{\theta}_i}\right|^{-1} p_{\rm pop}(\boldsymbol{\theta}_i|\boldsymbol{\Lambda})\right]_{\boldsymbol{\theta}_i=\boldsymbol{\theta}_i(\boldsymbol{\theta}_i^{\rm det}, \boldsymbol{\Lambda}_{\rm c})},
\end{aligned}
\tag{1}
$$

where $\pi(\cdot)$ denotes a prior, $\left|\mathrm{d}\boldsymbol{\theta}_i^{\mathrm{det}}(\boldsymbol{\theta}_i,\boldsymbol{\Lambda}_{\mathrm{c}})/\mathrm{d}\boldsymbol{\theta}_i\right|$ is the Jacobian of the transformation from source to detector frame, and

$$\xi(\boldsymbol{\Lambda},\boldsymbol{\Lambda}_{\mathrm{c}}) = \int \mathrm{d}\boldsymbol{\theta}^{\mathrm{det}}\, P(\mathrm{det}|\boldsymbol{\theta}^{\mathrm{det}}) \times \left[\left|\frac{\mathrm{d}\boldsymbol{\theta}^{\mathrm{det}}(\boldsymbol{\theta},\boldsymbol{\Lambda}_{\mathrm{c}})}{\mathrm{d}\boldsymbol{\theta}}\right|^{-1} p_{\mathrm{pop}}(\boldsymbol{\theta}|\boldsymbol{\Lambda})\right]_{\boldsymbol{\theta}=\boldsymbol{\theta}(\boldsymbol{\theta}^{\mathrm{det}},\boldsymbol{\Lambda}_{\mathrm{c}})} \tag{2}$$

is the expected fraction of detected events in the population. This term corrects for selection effects, namely the fact that the detectors observe a fraction of the real underlying population described by $p_{\mathrm{pop}}(\boldsymbol{\theta}|\boldsymbol{\Lambda})$. Here, $P(\mathrm{det}|\boldsymbol{\theta}^{\mathrm{det}}) \in [0,1]$ is the probability of detecting an event with parameters $\boldsymbol{\theta}^{\mathrm{det}}$. This function must be evaluated by matching the detection criterion used to obtain the observed catalog (Essick & Fishbach 2024). Finally, Equation (1) assumes marginalization over the overall total number of mergers in the observing time, $N$, with a scale–invariant prior $\propto 1/N$ (Mandel et al. 2019).

The population distribution in Equations (1) and (2) inherits an explicit dependence on the cosmological parameters stemming from the conversion from detector to source frame. This property allows constraints on cosmological parameters. Specifically, we can relate detector–and source–frame quantities using

$$z = z(D_{\mathrm{L}}^{\mathrm{GW}};\boldsymbol{\Lambda}_{\mathrm{c}})\,, \tag{3}$$

$$m_{1,2} = \frac{m_{1,2}^{\mathrm{det}}}{1+z(D_{\mathrm{L}}^{\mathrm{GW}};\boldsymbol{\Lambda}_{\mathrm{c}})}\,. \tag{4}$$

The redshift is obtained from the luminosity distance for given cosmological parameters via the inversion of the distance–redshift relation (see Section 2.4.1 for details). In the presence of features in the source–frame mass distribution as modeled in $p_{\mathrm{pop}}$, the above relation between the source–frame mass and the redshifted mass can be used to probe cosmology even in the absence of an explicit EM counterpart, which corresponds to the *spectral* method (Chernoff & Finn 1993; Taylor et al. 2012; Taylor & Gair 2012; Farr et al. 2019). The Jacobian transformation from source to detector frame also introduces a dependence on the cosmological parameters. Explicitly,

$$\left|\frac{\mathrm{d}\boldsymbol{\theta}^{\mathrm{det}}(\boldsymbol{\theta},\boldsymbol{\Lambda}_{\mathrm{c}})}{\mathrm{d}\boldsymbol{\theta}_i}\right| = (1+z)^2\,\frac{\mathrm{d}D_{\mathrm{L}}^{\mathrm{GW}}(z,\boldsymbol{\Lambda}_{\mathrm{c}})}{\mathrm{d}z}\,. \tag{5}$$

The explicit expression for the luminosity distance needed to compute the Jacobian is given in Equations (19), (22), and (23) below for the scenarios considered in this paper.

In addition to the mass distribution, an informative population prior on the redshift of an event can be constructed by using a galaxy catalog, corresponding to the *dark siren* method. We describe the construction of the redshift prior in detail in Section 2.2. Any consistent inference must account for selection effects via Equation (2), where the detection probability is a function of any variable $\boldsymbol{\theta}^{\mathrm{det}}$ determining the GW waveform. As a consequence, any analysis based on the construction of a redshift prior from a galaxy catalog also requires assumptions on the mass distribution to compute the selection effects. This implies one has to marginalize over the parameters of the mass distribution to obtain consistent and unbiased results. Also note that both the individual–event likelihood and detection probability $P(\mathrm{det}|\boldsymbol{\theta}^{\mathrm{det}})$ depend on more parameters than just mass and redshift, e.g., inclination angles (namely, the angle between the orbital angular momentum of the binary and the observer's line-of-sight) and spins of the compact objects. Neglecting those additional parameters corresponds to implicitly assuming that their astrophysical distribution $p_{\mathrm{pop}}$ coincides with the prior used in the individual–event parameter estimation. In particular, in absence of a specific model, spins are assumed to have a distribution uniform in magnitude and isotropic in orientation (Abac et al. 2025d).

The two pipelines used in our analysis, `icarogw` and `gwcosmo` (Mastrogiovanni et al. 2024; Gray et al. 2023), adopt different strategies to evaluate the posterior in Equation (1). Detailed technical descriptions are provided in Appendix A.

### 2.2. *Construction of Redshift Priors*

In this Section, we detail the construction of population priors on redshift. We give here a general overview of the method, and we refer the reader to Mastrogiovanni et al. (2023) and Gray et al. (2023) for more specifics.

We split the source–frame parameters $\boldsymbol{\theta}$ as $\boldsymbol{\theta} \equiv \{z,\Omega,\bar{\boldsymbol{\theta}}\}$ where $\bar{\boldsymbol{\theta}}$ denotes all source–frame waveform parameters other than redshift and sky position $\Omega$. We write the population distribution appearing in Equation (1) in terms of the source–frame merger rate as (Mastrogiovanni et al. 2023; Gray et al. 2023; Mastrogiovanni et al. 2024)

$$p_{\mathrm{pop}}(\boldsymbol{\theta}_i|\boldsymbol{\Lambda}) \propto p_{\mathrm{pop}}(\bar{\boldsymbol{\theta}}_i|\boldsymbol{\Lambda})\frac{\psi(z|\boldsymbol{\Lambda})}{1+z} \times \left[\frac{\mathrm{d}N^{\mathrm{eff}}_{\mathrm{gal,cat}}}{\mathrm{d}z\mathrm{d}\Omega} + \frac{\mathrm{d}N^{\mathrm{eff}}_{\mathrm{gal,out}}}{\mathrm{d}z\mathrm{d}\Omega}\right]. \tag{6}$$

In the above Equation, $\psi(z|\boldsymbol{\Lambda})$ parametrizes the redshift dependence of the CBC merger rate and the factor of $(1+z)^{-1}$ accounts for the conversion of time intervals from source to observer frame. The terms in square brackets represent the contributions to the redshift prior from galaxies within the catalog (first term), and a model for unobserved 'out-of-catalog' galaxies (second term). We will discuss the details of these two terms next.

*In–catalog part,* $\mathrm{d}N^{\mathrm{eff}}_{\mathrm{gal,cat}}/(\mathrm{d}z\mathrm{d}\Omega)$—This term is built starting from the galaxies in the catalog. The sky is divided in equal–size pixels, labeled with their central coordinates $\Omega$, of area $\Delta\Omega$, with the `healpix` pixelization algorithm (Górski et al. 2005; Zonca et al. 2019). Inside each pixel, we select all galaxies with apparent magnitude brighter than the median inside the pixel, denoted as $m_{\mathrm{thr}}(\Omega)$. To compute this median threshold, we adopt `nside` = 32 in the `healpix` scheme. However, as described in Section 3.2, a higher resolution is used to pixelize the galaxy catalog used in the analysis. The choice of a coarser resolution to compute the skymap of median thresholds ensures robustness against small-number statistics with the numbers of galaxies (Gray et al. 2022, 2023). This median threshold can depend on the sky position if the galaxy catalog is compiled from multiple surveys and does not have uniform coverage. For each pixel, a redshift prior is constructed as a weighted sum of the posterior distributions for the true redshift $z$ given observed redshifts $z^j_{\rm obs}$ for the selected galaxies $j = 1, \ldots, N_{\rm gal}(\Omega)$ in the pixel, each denoted by $p(z|z^j_{\rm obs}, \sigma^j_{\rm z,obs}, \mathbf{\Lambda}_{\rm c})$. The in–catalog term is then obtained as

$$\frac{{\rm d}N^{\rm eff}_{\rm gal,cat}}{{\rm d}z{\rm d}\Omega} = \frac{1}{\Delta\Omega}\sum_{j}^{N_{\rm gal}(\Omega)} w_j(\epsilon, M_j) \times p(z|z^j_{\rm obs}, \sigma^j_{\rm z,obs}, \mathbf{\Lambda}_{\rm c})\,\delta(\Omega - \Omega_j)\,, \tag{7}$$

where $M_j$ is the absolute magnitude of a galaxy in a specific band. We assume negligible uncertainties on the sky position, and define the weights (Gray et al. 2020)

$$w_j(\epsilon, M_j) = \left|\frac{L_j}{L_*}\right|^{\epsilon} = 10^{-0.4\epsilon(M_j - M_*)}\,, \tag{8}$$

where $L_*$ and $M_*$ are the reference luminosity and corresponding magnitude at the knee of the luminosity function, respectively. We assume the luminosity function to be given by the Schechter function (Schechter 1976), described in more detail in Appendix B.

In Equation (7), we weight each galaxy by Equation (8), namely by the absolute luminosity in a specific band, $L_j$, raised to a power $\epsilon$ which we treat as a fixed parameter. In particular, we consider the cases $\epsilon = 0$, corresponding to equal probability for all galaxies to host CBCs, which we will refer to as *no–weighting* case, and $\epsilon = 1$ corresponding to a linear weight of galaxies by their luminosity, which we will refer to as *luminosity–weighting* case. It is known that luminosity in specific magnitude bands correlates with galaxy properties such as stellar mass or star formation rate, for example. Luminosity–weighting reflects an assumption that such galaxy properties may also correlate with likelihood to host CBC mergers, see Gray et al. (2020); Pálfi et al. (2025) for more extended discussions.

The absolute magnitude $M_j$ is obtained for each galaxy from the measured apparent magnitude $m_j$ via $M_j = m_j + 5 - 5\log D_{\rm L}(z, \mathbf{\Lambda}_{\rm c}) - K_{\rm corr}$ (with $D_{\rm L}$ expressed in pc), where the *K-correction* term $K_{\rm corr}$ accounts for the shifting of the observed spectrum for galaxy at redshift $z$.[1,2] K–corrections are computed following Kochanek et al. (2001). The conversion from apparent to absolute magnitude depends in principle on the computation of a distance, hence on cosmology. However, the overall dependence on $H_0$ in such conversion cancels out in the ratio $L/L_*$, as $L_*$ shares the same scaling with $H_0$. This leaves in principle a residual dependence on other parameters of the distance–redshift relation such as $\Omega_{\rm m}$ in our dark siren analyses, which we consider fixed. Under this assumption, the difference $M_j - M_*$ in Equation (8) can be considered only a function of $m_j$ and $z$.

Each redshift measurement is assumed to be described by a Gaussian distribution with mean $z^j_{\rm obs}$ and standard deviation $\sigma^j_{\rm z,obs}$ (see Palmese et al. 2020; Turski et al. 2023, for the impact of using more generalized distributions). Specifically, we test both the assumption that the Gaussian distribution models directly the redshift posterior probability, in which case $p(z|z^j_{\rm obs}, \sigma^j_{\rm z,obs}, \mathbf{\Lambda}_{\rm c}) = \mathcal{N}(z|z^j_{\rm obs}, \sigma^j_{\rm z,obs})$, and that it models the likelihood instead. In the second case, we obtain the posterior as

$$p(z|z^j_{\rm obs}, \sigma^j_{\rm z,obs}, \mathbf{\Lambda}_{\rm c}) = \frac{\mathcal{N}(z^j_{\rm obs}|z, \sigma^j_{\rm z,obs})\,\pi_V(z, \mathbf{\Lambda}_{\rm c})}{\int {\rm d}z\,\mathcal{N}(z^j_{\rm obs}|z, \sigma^j_{\rm z,obs})\,\pi_V(z, \mathbf{\Lambda}_{\rm c})}\,, \tag{9}$$

where the likelihood $\mathcal{N}(z^j_{\rm obs}|z, \sigma^j_{\rm z,obs})$ is multiplied by a volumetric prior $\pi_V(z, \mathbf{\Lambda}_{\rm c})$ (representing our prior knowledge for the true galaxies' redshift in absence of measurements) to obtain the posterior. We choose the prior as uniform in comoving volume, that is

$$\pi_V(z, \mathbf{\Lambda}_{\rm c}) \propto \frac{{\rm d}V_{\rm c}(z, \mathbf{\Lambda}_{\rm c})}{{\rm d}z{\rm d}\Omega}\,, \tag{10}$$

with ${\rm d}V_{\rm c}(z, \mathbf{\Lambda}_{\rm c})/({\rm d}z{\rm d}\Omega)$ being the differential comoving volume element:

$$\frac{{\rm d}V_{\rm c}}{{\rm d}z{\rm d}\Omega}(z, \mathbf{\Lambda}_{\rm c}) = \frac{c\,D_{\rm L}^2}{H_0\,(1+z)^2\,E(z)}\,, \tag{11}$$

where $E(z)$ is the expansion rate defined below in Section 2.4.1. We verified that both assumptions lead to negligible differences.

*Out–of–catalog part,* ${\rm d}N^{\rm eff}_{\rm gal,out}/({\rm d}z{\rm d}\Omega)$—This term models the contributions from galaxies that are missed by the survey due to magnitude limits. It requires some prior assumption on the number and distribution of missing galaxies in luminosity, redshift, and sky position. We assume that galaxies are uniformly distributed in comoving volume and solid angle, and that their absolute magnitude $M$ follows a redshift–independent Schechter function ${\rm Sch}(M;\lambda)$ with parameters $\lambda = \{\alpha, \phi_*, M_*\}$ between lower and upper ends $M_{\rm min}$ and $M_{\rm max}$. Here $\alpha$ is the faint-end slope of the Schecter function, $\phi_*$ is the overall amplitude, and $M^*$ was introduced in Equation 8. We note that these parameters take different values in different luminosity bands. See Appendix B for details. The number of missing galaxies per unit

[1] We use the symbol $m$ for both source-frame masses and galaxy apparent magnitudes, clarifying its meaning when necessary.

[2] We have assumed that the apparent magnitude is known with a negligible uncertainty, as we have used galaxies in the $K$–band that are significantly bright compared to the flux limits. This assumption may have to be revisited when using fainter galaxies closer to the flux limits.

redshift, solid angle, and absolute magnitude is estimated as

$$\frac{\mathrm{d}N_{\rm gal,out}}{\mathrm{d}z\mathrm{d}\Omega\mathrm{d}M} = \frac{\mathrm{d}V_{\rm c}}{\mathrm{d}z\mathrm{d}\Omega}(z,\Omega)\,\mathrm{Sch}(M;\lambda)\,p_{\rm miss}(z,\Omega,M)\,, \tag{12}$$

where $\mathrm{d}V_{\rm c}/(\mathrm{d}z\mathrm{d}\Omega)$ is the comoving volume element, and $p_{\rm miss}(z,\Omega,M) = \Theta\left(M - M_{\rm thr}\left[z, m_{\rm thr}(\Omega)\right]\right)$ is the probability of missing a galaxy. The latter is modeled as a Heaviside step function following the assumption that a galaxy is included in the in–catalog part if its apparent magnitude $m$ is smaller (i.e., it is brighter) than the threshold $m_{\rm thr}(\Omega)$.

An effective out–of–catalog term can be obtained integrating Equation (12) with a luminosity weight $\propto 10^{-0.4\epsilon(M-M_*)}$ (analog to Equation (8)) over the absolute magnitude $M$ between the faint end of the Schechter function $M_{\rm max}$ and the threshold $M_{\rm thr}(z, m_{\rm thr}(\Omega))$. We provide details in Appendix B. One obtains

$$\frac{\mathrm{d}N^{\rm eff}_{\rm gal,out}}{\mathrm{d}z\mathrm{d}\Omega}(z,\Omega) = \frac{\mathrm{d}V_{\rm c}}{\mathrm{d}z\mathrm{d}\Omega}(z,\Omega)\,\phi_* \int_{x_{\rm thr}}^{x_{\rm max}} \mathrm{d}x\, x^{\alpha+\epsilon} e^{-x}\,, \tag{13}$$

where

$$x_{\rm thr} = 10^{0.4[M_* - M_{\rm thr}(z, m_{\rm thr}(\Omega))]}, \tag{14}$$

$$x_{\rm max} = 10^{0.4(M_* - M_{\rm max})}. \tag{15}$$

We note that the out–of–catalog part is independent of $H_0$. In the luminosity–weighting case ($\epsilon = 1$), the probability of galaxies to host a GW candidate reaches its maximum at the knee $M_*$ of the luminosity function. As long as $M_{\rm max}$ is sufficiently fainter than $M_*$, there is little sensitivity of our results to $M_{\rm max}$. For the no–weighting case ($\epsilon = 0$), choice of arbitrarily faint $M_{\rm max}$ would lead to a large increase in the number of galaxies that could potentially host GW events. Such faint galaxies cannot be seen out to large redshifts due to the flux limit of the survey, which can drive up the incompleteness and subsequently result in the redshift prior to be completely dominated by the out–of–catalog term (Bera et al. 2020).

### 2.3. *Population Models*

We construct CBC rate models from independent redshift and source mass distributions, while we assume the CBC spins to be isotropically distributed with uniform distribution in the spin magnitudes. Specifically, the term $\psi(z|\mathbf{\Lambda})$ in Equation (6), describing the merger rate evolution as a function of the redshift, is modeled with a Madau–Dickinson parametrization (Madau & Dickinson 2014), which is characterized by parameters $\{\gamma, \kappa, z_{\rm p}\} \in \mathbf{\Lambda}$, where $\gamma$ and $\kappa$ are the power–law slopes respectively before and after the redshift turning point, $z_{\rm p}$, between the two power–law regimes. Explicitly,

$$\psi\left(z|\gamma,\kappa,z_{\rm p}\right) = \left[1 + \left(1+z_{\rm p}\right)^{-\gamma-\kappa}\right] \times \frac{(1+z)^{\gamma}}{1 + \left[(1+z)/\left(1+z_{\rm p}\right)\right]^{\gamma+\kappa}}\,. \tag{16}$$

This parametrization is more complex than the one adopted in studies that focus solely on GW population properties, where usually it takes the form of simple power–laws, $\psi(z) \propto (1+z)^{\gamma}$ (Abbott et al. 2023c; Abac et al. 2025f). This choice is motivated by the fact that, when varying the cosmology, a GW event at given distance can be associated with a redshift which is significantly higher than the one corresponding to the fiducial cosmology. The model in Equation (16) ensures that the merger rate decays after a peak at $z = z_{\rm p}$, consistently with astrophysical expectations. The Madau–Dickinson distribution is typically used to describe the cosmic star formation rate, while the CBC merger rate is then obtained by convolving with a time-delay distribution. In practice, this is equivalent to using the same functional form with different values of $\gamma$ and $\kappa$, and by adopting wide priors on these parameters we effectively account for a broad range of possible delay times.

In this study we consider three different models for the distribution of primary mass, $p\left(m_1|\mathbf{\Lambda}\right)$, which enters the term $p_{\rm pop}$ in Equation (6). These models are denoted as: POWER LAW + PEAK (PLP), MULTI PEAK (MLTP), and FULLPOP-4.0. These are phenomenological parametric models defined in terms of relatively simple functional forms that contain features motivated by either astrophysical expectations or previous GW observations. These models are constructed as superpositions of truncated Gaussian and power–law distributions with different parameters (described in Appendix C), and they are suited for the BBH spectrum description only with the exception of the FULLPOP-4.0 model (see below). In this work we consider these mass models as redshift-independent; see Mukherjee (2022); Karathanasis et al. (2023); Rinaldi et al. (2024) for investigations into their possible evolution. We will comment upon this further in Section 5.2. Figure 1 shows a sketch of the typical form of these models, with the different mass features that characterize them highlighted. We now briefly describe these models (see Appendix C and Abac et al. 2025f for more details).

The PLP mass model (Talbot & Thrane 2018) has been used for the analysis of previous GW catalogs (Abbott et al. 2023c,a). It is based on a power–law distribution with a smooth low–mass cutoff. In addition to this power–law component, the model includes a Gaussian peak to capture an excess of events at intermediate masses, and a high–mass cutoff. This model is described by eight population parameters.

The MLTP mass model is an extension of the PLP model, originally introduced in Abbott et al. (2021c). Like the PLP model, it features a power–law distribution for the primary mass spectrum with a smooth low–mass cutoff and includes a Gaussian peak to capture an excess at intermediate masses. The distinguishing feature of the MLTP model is the inclusion of a second Gaussian peak, making it a combination of a power–law and two Gaussian components. This model is similar to the "BROKEN POWER LAW + 2 PEAKS" model adopted in Abac et al. (2025f), except our model has one power law instead of two. This model is characterized by eleven population parameters.

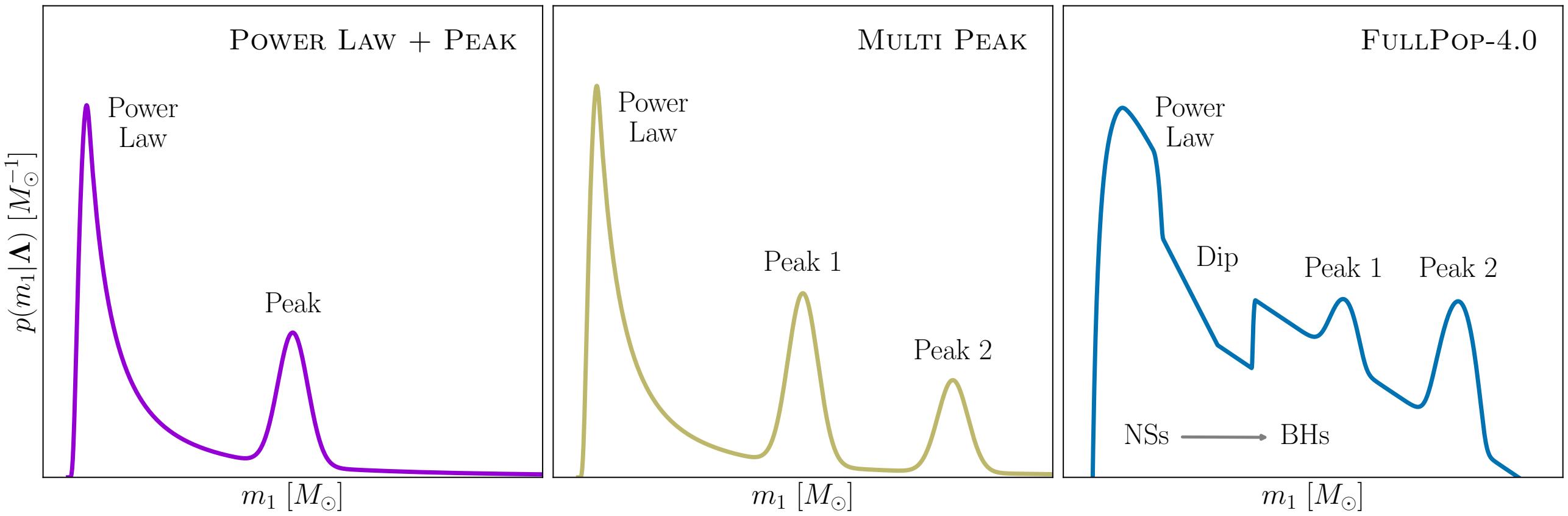

**Figure 1.** Qualitative graphical representation of the three source-frame mass models considered in this paper and described in Section 2.3 and Appendix C. The mass distribution models displayed in the first two panels represent the mass ranges of BHs, while the third panel includes both BHs NSs. The mass ranges shown are not to scale.

In the PLP and MLTP models the full mass distributions are factorized as

$$p(m_1, m_2|\mathbf{\Lambda}) = p(m_1|\mathbf{\Lambda})\, S_{\rm h}(m_1|\mathbf{\Lambda}) \times p(m_2|m_1, \mathbf{\Lambda})\, S_{\rm h}(m_2|\mathbf{\Lambda}), \quad (17)$$

where $p\,(m_2|m_1, \mathbf{\Lambda})$ is the distribution of the secondary mass component conditioned on the primary mass and $S_{\rm h}(m|\mathbf{\Lambda})$ is a smoothing function defined in Appendix C. This is modeled assuming that the mass ratio $q = m_2/m_1$ follows a power–law distribution.

The FullPop-4.0 model is a generalization of the previous mass models, extending the distribution to encompass the full mass spectrum of CBCs, including BNS, NSBH, and BBH mergers. It is designed to cover a wide mass range, from a few to several hundred solar masses (Fishbach et al. 2020; Farah et al. 2022; Mali & Essick 2025). The model combines a first power–law component for the low–mass region (representing NS-containing events) with a smooth low–mass cut–off, and a second power–law component for the BBH mass distribution, which includes two Gaussian peaks. A dip function is introduced at the junction between the two power–law regimes, aiming to model the apparent mass gap between NSs and BHs. The parameters governing this dip are treated as population parameters. This model is characterized by nineteen parameters.

By modeling the full population of compact objects in a unified framework, the FullPop-4.0 model allows us to include a broader set of GW events in our analysis, offering greater sensitivity to features in the mass spectrum and enabling tighter constraints on cosmological parameters. Another major distinction from the PLP and MLTP mass models lies in the parametrization of the secondary mass. Instead of modeling $m_2$ as a power–law conditioned on $m_1$, as in Equation (17), the FullPop-4.0 model assumes that the distribution of $m_2$ is given by $p(m_2|\mathbf{\Lambda})$ and employs a pairing function $f(m_1, m_2|\mathbf{\Lambda})$ enforcing the condition $m_1 \geq m_2$ and allowing for further flexibility for the secondary mass (Fishbach & Holz 2020). Therefore, in this case, we have

$$p(m_1, m_2|\mathbf{\Lambda}) \propto\; p_{\rm S}(m_1|\mathbf{\Lambda})\, p_{\rm S}(m_2|\mathbf{\Lambda}) f(m_1, m_2|\mathbf{\Lambda})\,, \quad (18)$$

where $p_{\rm S}(m|\mathbf{\Lambda})$ is defined in terms of $p(m|\mathbf{\Lambda})$ and the smoothing functions defined in Appendix C.

The equations which describe our three population models can be found in Appendix C. For more details, see also Abac et al. (2025f). In Section 4 we compare our analysis obtained using *single–population* models (the PLP and MLTP models which are valid for BBH candidates only) to that obtained using a *multi-population* model (BNS + NSBH + BBH candidates), i.e., the FullPop-4.0 model.

### 2.4. *Cosmological Models*

#### 2.4.1. *Background Evolution*

Under the assumptions of homogeneity and isotropy, the luminosity distance can be computed based on the Friedmann–Lemaître-Robertson–Walker (FLRW) metric as

$$D_{\rm L} = \frac{c(1+z)}{H_0} \int_0^z \frac{{\rm d}z'}{E(z')}\,, \quad (19)$$

where $E(z) = H(z)/H_0$ is the dimensionless expansion rate of the Universe. This depends on the cosmological model assumed and can be computed using the Friedmann equations. In this paper, we restrict our focus to a flat–$\Lambda$CDM model. Under this assumption, $E(z)$ is given by

$$E(z) = \left[\Omega_{\rm m}(1+z)^3 + \Omega_\Lambda\right]^{1/2}\,. \quad (20)$$

Here, $\Omega_{\rm m}$ is the fractional energy density in matter components today (cold dark matter + baryonic matter), and we have ignored the radiation energy density which is negligible at the redshifts of our interest. Under this approximation, the dark energy density fraction today is $\Omega_\Lambda = 1 - \Omega_{\rm m}$.

More generally, the cosmic expansion history can be extended to include dark energy with a constant equation–of–state parameter $w_0 \neq -1$. If $w_0$ is a constant, the dark energy density evolves with redshift as $\sim(1+z)^{3(1+w_0)}$, and the expansion rate becomes

$$E(z) = \left[\Omega_{\rm m}(1+z)^3 + \Omega_\Lambda(1+z)^{3(1+w_0)}\right]^{1/2} . \quad (21)$$

As our data currently have no constraining power on $w_0$, in this work we will only consider such a generalization as a robustness test (see Section 4.3). Our main results will be based on the flat–ΛCDM model with $w_0 = -1$.

### 2.4.2. *Parametrizations of Modified GW Propagation*

We also analyze our data in the context of cosmological modified–gravity models, which alter the behavior of cosmological perturbations. There is a large landscape of such models, introduced to explain dark energy (see Tsujikawa 2010; Clifton et al. 2012; Joyce et al. 2016; Ezquiaga & Zumalacárregui 2018; Ishak 2019, for reviews). Whilst this model space contains a wide variety of phenomenology, we focus here on a common (but not universal) feature, sometimes referred to as *GW friction* (Saltas et al. 2014; Pettorino & Amendola 2015; Nishizawa 2018; Amendola et al. 2018; Lagos et al. 2019). Under this effect, new terms in the GW propagation equation result in modifications to the GW amplitude received at the observer. This effect is indistinguishable from a change in the luminosity distance to the GW source. The result is that the luminosity distance $D_{\rm L}^{\rm GW}$ inferred for a GW source differs from the EM luminosity distance $D_{\rm L}$ given by Equation (19). Any measurement of the GW source luminosity distance obtained using EM observables would be unaffected, i.e., $D_{\rm L}^{\rm EM} = D_{\rm L}$. In models where the theory of gravity on cosmological scales is GR, the luminosity distance derived from GW events, $D_{\rm L}^{\rm GW}$, and that based on EM observations, $D_{\rm L}^{\rm EM}$, are instead identical and given by Equation (19).

Based on this, multiple studies (Belgacem et al. 2018b,a, 2019b,a; Mukherjee et al. 2021c; Finke et al. 2021a,b, 2022; Ezquiaga 2021; Finke et al. 2021a; Mancarella et al. 2022; Kalogera et al. 2021; Leyde et al. 2022; Liu et al. 2024; Branchesi et al. 2023; Chen et al. 2024a; Abac et al. 2025a) have considered the ratio $D_{\rm L}^{\rm GW}/D_{\rm L}^{\rm EM}$ as a convenient probe of departures from GR on cosmological scales. The ratio is always equal to one in GR, and in cosmological modified–gravity models can become a function of redshift. Rather than focusing on specific modified–gravity models, here we consider two commonly used parametrized forms for the GW–EM luminosity–distance ratio.

Two assumptions are relevant to both parametrizations. First, we assume the GW propagation speed, $c_T$, is luminal. Such a choice relies on the tight GW constraint on $c_T$ from GW170817 (Abbott et al. 2017d), which is made at redshift ∼0.01. Our data span up to redshift ∼0.9, so breaking this assumption would require a theory where the relative difference between $c_T$ and $c$ (the speed of light) grows from $\sim 10^{-15}$ by orders of magnitude in a redshift range between 0.01 to < 1. Furthermore, Ray et al. (2024) provide percent–level constraints on the GW propagation speed using dark BBHs candidates from GWTC-3.0. Given these results, significant deviations from luminal speed do not appear to be favored by current data; hence we do not consider non-luminal propagation in the present work. Non-luminal propagation at higher redshift could be incorporated in future analyses to explore potential deviations in the speed of GWs over a broader redshift range. Additionally, we do not consider frequency–dependent deviations in the speed of GWs. This remains the standard assumption in most cosmological tests of GR, and currently used waveforms based on GR suggest that any such deviations should be small (Abbott et al. 2019b, 2021d,e).

Second, we treat departures from GR impacting only the propagation phase of GW signals. In cosmological modified–gravity theories, changes to the strong–field gravitational regime are usually suppressed by screening mechanisms, in order to obey stringent tests of GR within the Solar System (see Joyce et al. 2015 for a review). As such, we do not consider here modifications to the generation of GWs, which would affect the waveform *at source*. Constraints on strong–field departures from GR are considered in Abac et al. (2025g,h,i); Abac et al. (2025h) also provides constraints on dispersive propagation effects.

$\Xi_0$–$n$ *Parametrization*—In this parametrization, $D_{\rm L}^{\rm GW}$ is described by (Belgacem et al. 2018a)

$$D_{\rm L}^{\rm GW} = D_{\rm L}^{\rm EM}\left(\Xi_0 + \frac{1-\Xi_0}{(1+z)^n}\right) , \quad (22)$$

where both parameters $\Xi_0$ and $n$ are positive. The primary parameter of interest is $\Xi_0$, which controls the overall amplitude of departures from GR. At low redshifts, $D_{\rm L}^{\rm GW}/D_{\rm L}^{\rm EM} \to 1$ (irrespective of $\Xi_0$), which models changes to $D_{\rm L}^{\rm GW}$ as an effect which accumulates with propagation distance. At high redshifts, $D_{\rm L}^{\rm GW}/D_{\rm L}^{\rm EM} \to \Xi_0$ as changes to $D_{\rm L}^{\rm GW}$ should saturate at redshifts where the fractional energy density of dark energy, $\Omega_\Lambda(z)$, is negligible. This holds under the assumption that deviations from GR are associated to the late–time emergence of dark energy. The power–law index $n$ controls the rate of transition between these two regimes.

The $\Xi_0$–$n$ parametrization is a direct phenomenological parametrization of the gravitational luminosity distance. The specific form of the parametrization is an assumption, and Equation (22) was calibrated to cover a large spectrum of known luminal modified–gravity theories (see Belgacem et al. 2019b, for a thorough discussion). These include $f(R)$ gravity (Hu & Sawicki 2007; Song et al. 2007; Starobinsky 2007), Jordan–Brans–Dicke (Brans & Dicke 1961), Galileon theories (Chow & Khoury 2009), nonlocal gravity (Maggiore 2014; Maggiore & Mancarella 2014). Notable exceptions, for which more complex parametrizations are needed to better capture the evolution of the distance ratio, are degenerate higher–order scalar–tensor theories (Langlois & Noui 2016), bigravity (Hassan & Rosen 2012), and extra–dimensional paradigms (Dvali et al. 2000); see Abbott et al. (2019c) and Corman et al. (2022) for constraints on the number of spacetime dimensions. The GR limit of the theory is $\Xi_0 \to 1$ (for any value of $n$). However, the parametrization is imperfectly behaved, since $n \to 0$ also recovers the GR behavior $D_{\rm L}^{\rm GW} = D_{\rm L}^{\rm EM}$.

*$\alpha_M$ Parametrization*—This parametrization is inspired by Horndeski gravity (Horndeski 1974; Deffayet et al. 2011; Kobayashi et al. 2011), which is the most general family of scalar–tensor gravity models with second–order equations of motion. In the widespread basis of Bellini & Sawicki (2014), adopted for describing linear cosmological perturbations of Horndeski theories around a FLRW solution, $\alpha_M(z)$ is the rate of change of the effective Planck mass, and hence the effective gravitational coupling strength (Bellini & Sawicki 2014; Gleyzes et al. 2015b). This results in the following expression for $D_{\rm L}^{\rm GW}$ (Lagos et al. 2019):

$$D_{\rm L}^{\rm GW} = D_{\rm L}^{\rm EM} \exp\left\{ \frac{1}{2} \int_0^z \frac{{\rm d}z'}{1+z'} \alpha_M(z') \right\}, \quad (23)$$

where in this work we will use the following ansatz for $\alpha_M(z)$

$$\alpha_M(z) = c_M \frac{\Omega_\Lambda(z)}{\Omega_\Lambda} = c_M \frac{1}{E^2(z)}, \quad (24)$$

where $\Omega_\Lambda = \Omega_\Lambda(z=0)$ and $c_M$ is a constant of proportionality. For the dimensionless expansion rate, $E(z)$, we use Equation (20) which assumes a flat–ΛCDM model with constant dark energy density, as in this work we are not considering changes to the cosmological expansion history. In principle, $\alpha_M(z)$ also enters the background evolution equations; however, any resulting change can be absorbed into other functions such as the effective dark energy equation of state (Bellini & Sawicki 2014). In addition, these background effects are highly subdominant compared to the impact on the distance ratio (Belgacem et al. 2018b). Therefore, it is legitimate to treat the background expansion as fixed, and we explicitly verify that even when allowing the background to vary (through the dark energy equation-of-state parameter, see section 4), this has no impact on our constraints. The GR limit of the model is obtained for $c_M = 0$.

The redshift–dependent form of $\alpha_M(z)$ is a choice, which would be fixed in a fully specified theory of gravity. In particular, Equation (24) is motivated by the association of the onset of $\alpha_M(z)$ to the late–time emergence of dark energy (Bellini & Sawicki 2014). The form of Equation (24) has been widely adopted for large scale structure (LSS) constraints (Bellini et al. 2016; Noller & Nicola 2019; Baker & Harrison 2021; Seraille et al. 2024; Ishak et al. 2024), but also criticized for not accurately representing a large number of modified–gravity models (Linder et al. 2016; Linder 2017; Denissenya & Linder 2018, see however Gleyzes 2017 for counter-arguments).

In a full treatment of Horndeski gravity, there are additional effects to the cosmological expansion rate and growth of LSS that impact EM observables; we do not consider these here, as our focus is on GW data analysis. Also, general Horndeski gravity can allow non-luminal GW propagation, but as noted above we do not consider this possibility. See Kobayashi (2019) and references therein for a review of Horndeski gravity and its phenomena. Finally, we conduct our analysis in the Jordan frame, where the effect of possible non-standard couplings between matter fields and the metric, that could impact the background expansion as well as scalar perturbations (Gleyzes et al. 2015a, 2016), are not present.

*Comparison among Parametrizations*—The $\Xi_0$–$n$ parametrization directly describes the redshift evolution of the distance ratio. In contrast, in the $\alpha_M$ parametrization the observable distance ratio is related to the integral of the function $\alpha_M(z)$, which encodes deviations from GR. As long as the dark energy density $\Omega_\Lambda(z)$ causes the integral in Equation (23) to saturate at large redshift, the resulting distance ratio exhibits the same qualitative behavior as described by Equation (22). The two parametrizations can be matched analytically at $z \to \infty$ and $z \sim 0$. Under the assumption of a flat–ΛCDM cosmology, and under our ansatz in Equation 24, the following relations hold:

$$\ln \Xi_0 = \frac{c_M}{6\Omega_\Lambda} \ln \frac{1}{\Omega_{\rm m}}, \quad n = -\frac{3}{\ln \Omega_{\rm m}}, \quad (25)$$

as discussed in detail in Belgacem et al. (2019b); Baker & Harrison (2021); Mancarella et al. (2022). The $\alpha_M$ parametrization features only a single free parameter, with its time evolution fully specified by the dark energy density, whereas the $\Xi_0$–$n$ parametrization allows additional flexibility via the redshift evolution index $n$, over which we marginalize. This effectively makes the $\alpha_M$ parametrization a special case of the $\Xi_0$–$n$ parametrization for a fixed $n$, albeit one with a direct link to theoretical models within the Horndeski class.

## 3. DATA

### 3.1. *GW Events*

The analyses presented are based on GWTC-4.0 (Abac et al. 2025b,c,d) and based on the detection of GW candidates produced by merging compact binaries between O1 and the end of O4a. To reduce the noise contamination of the datasets used in cosmological studies, we select a subset of GW events with the lowest FAR among all search pipelines, ensuring all events have FAR $< 0.25\,{\rm yr}^{-1}$. The GW candidates collected during the engineering run directly preceding the start of O4a are not included in the analysis, to remain consistent with the principles deployed in previous LVK cosmology analyses.

A total of 142 CBC GW candidates with FARs below this threshold have been detected by our search pipelines from O1 to O4a. Following the GWTC-4.0 classification of candidates into unambiguous BBHs and potential NS-binaries (Abac et al. 2025d), 137 out of 142 events are believed to originate from the coalescence of BBH candidates and 5 from binaries where at least one component mass could have been a NS. From the list of events that pass the sensitivity cut in O4a, we exclude GW231123_135430 (Abac et al. 2025j), as some of its inferred properties, such as the binary masses or its luminosity distance, appear to be more sensitive to the choice of waveform model than those of other events in our dataset, and in this work we prefer to use results from a single waveform model for each event, as discussed below.

This analysis shares 45 dark sirens with our previous cosmological analysis (Abbott et al. 2023a), which used 46 dark sirens. The event not used in the present work is GW200105_162426, which is excluded here due to its low probability of being of astrophysical origin (Abbott et al. 2023b). Thus, our analysis contains 96 additional dark sirens; 76 of these come from O4a, whilst 20 are additional events from O3 which were not used previously. This is due to the fact that Abbott et al. (2023a) selected candidates based on both a FAR and signal-to-noise ratio (SNR) threshold, resulting in fewer events analyzed. In this work we apply *only* the FAR threshold stated above. Added to the 141 dark sirens is the special case of the multi-messenger event GW170817. This is treated differently from the others, and will be used in the rest of the paper as a bright siren.

Compared to previous GW candidates (O1–O3), the O4a detections cover a similar parameter space in terms of luminosity distance and masses. Figure 2 shows the distribution of the 90% credible region (CR) of the sky-localization of CBC events observed in the same LVK observing runs, as well as that of the O4a events only. The sky localization of the GW events detected during the O4a observing run is, on average, relatively broad (see Figures 2 and 3). This is due to the fact that, during O4a, Virgo was not online resulting in two detector localizations. A full list of luminosity distances, binary component masses, and sky uncertainties of the GW candidates considered in our study can be found in Appendix E.

Different waveform models have been used to perform the parameter estimation (PE) for each GW candidate across the observing runs (Abac et al. 2025c). For our analysis, we use posterior samples produced with a single waveform approximant rather than a mixture of samples from different waveforms (Abac et al. 2025c,d). This choice mitigates potential difficulties in reweighting the PE samples if different waveforms use slightly different prior bounds, such as those on the luminosity distance, for a given candidate. In particular, for candidates from the O1, O2, and O3 runs, we use the posterior samples based on the IMRPhenomXPHM waveform model (Pratten et al. 2021), where for GW200115_042309 (Abbott et al. 2021f) we use the large-spin magnitude prior posterior samples, while for GW190425_081805 (Abbott et al. 2020a) and GW170817 (Abbott et al. 2021a) we use the large-spin magnitude prior posterior samples obtained with the IMRPhenomPv2_NRTidal (Dietrich et al. 2017, 2019) and a prior allowing for high-spin and low-spin magnitudes, respectively. For events from the O4a observing run, we use the posterior samples produced with the IMRPhenomXPHM_SpinTaylor model (Colleoni et al. 2025), except for GW230529_181500, for which we use posterior samples produced using the IMRPhenomXPHM waveform model and released in Abac et al. (2024). In this study we do not consider the impact of waveform systematics, as they are expected to be relevant only in population analyses of GW events with SNR above 100 (Kapil et al. 2024).

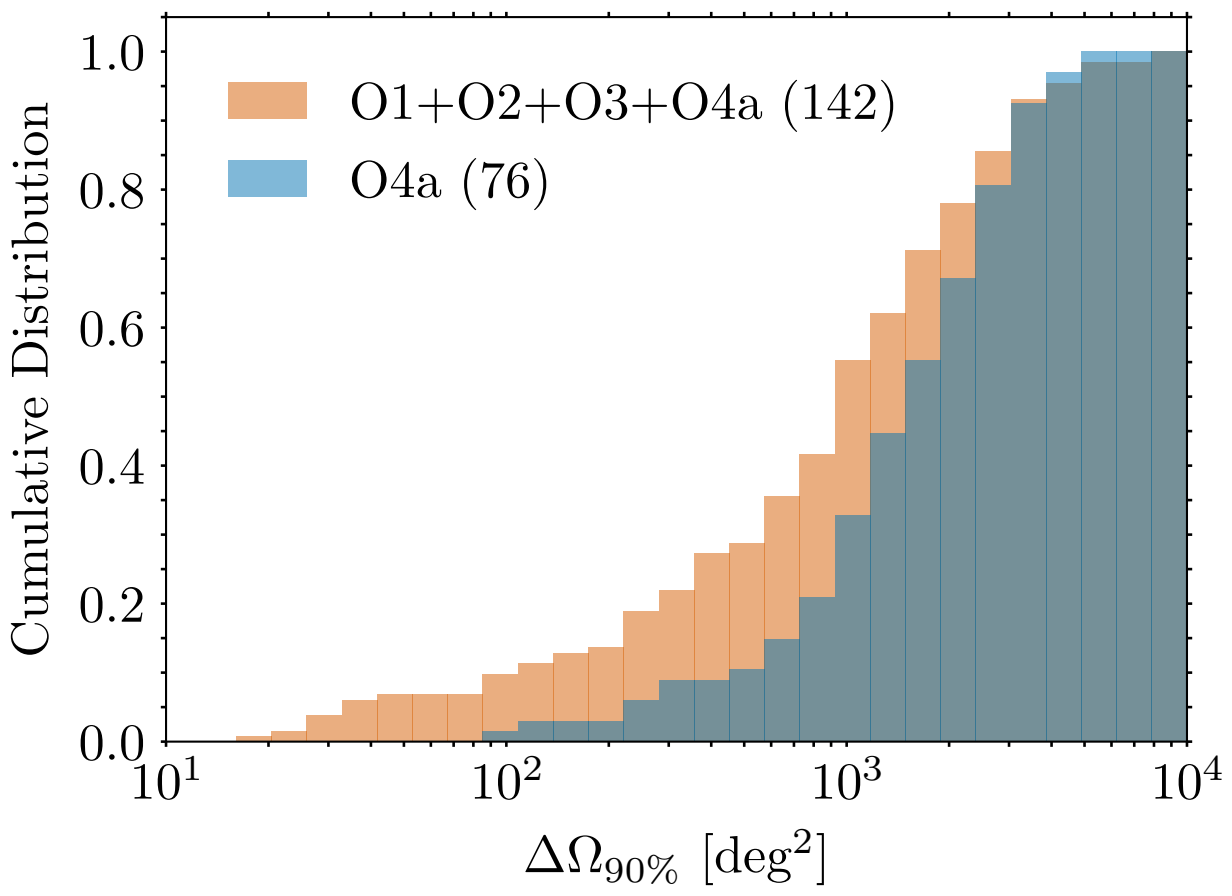

**Figure 2.** Cumulative distribution of the size of the 90% CR of the sky localization of CBC candidates observed during O1+O2+O3+O4a in orange (142 total events including GW170817) and O4a-only in blue (76 events). It can be seen that the typical sky localization of O4a events was larger than that of O1–O3 events.

During the final stages of this project, a normalization error was discovered in the noise-weighted inner product employed in the GW PE likelihood function (Abac et al. 2025c; Talbot et al. 2025). Although there is a version of the PE samples that accounts for the correct likelihood via a reweighting prescription (Abac et al. 2025c; Talbot et al. 2025), in this work we do not use these samples, but those released in the first digital version of the GWTC-4.0 catalog (LIGO Scientific Collaboration, Virgo Collaboration, and KAGRA Collaboration 2025). Furthermore, we discovered that incorrect priors were used when marginalizing over the uncertainty in the LIGO detector calibration for candidates detected during the first three observing runs (Abac et al. 2025c). As discussed in Abac et al. (2025c), we have checked that the impact on the most affected events is individually negligible, so that our previous results on the full population are also unaffected. We have checked that this error's impact on our cosmological analyses is negligible compared to other sources of systematic error.

Finally, we estimate the GW detection probability in Equation (2) by using a set of simulated GW signals (called injections) described in Essick et al. (2025); Abac et al. (2025c). More details on how the injections are used to compute Equation (2) can be found in Appendix A.

### 3.2. *Galaxy Catalog*

We use the GLADE+ galaxy catalog (Dálya et al. 2018, 2022) for our galaxy catalog method analysis. GLADE+ is an all-sky galaxy catalog containing around 22 million galaxies, which has been created from six different astronomical datasets: the Gravitational Wave Galaxy Catalog (GWGC, White et al. 2011), HyperLEDA (Makarov et al. 2014), the 2 Micron All-Sky Survey Extended Source Catalog (2MASS XSC, Skrutskie et al. 2006), the 2MASS Photometric Redshift Catalog (2MPZ, Bilicki et al. 2014), the WISExSCOS Photometric Redshift Catalog (WISExSCOSPZ, Bilicki et al. 2016) and the Sloan Digital Sky Survey quasar catalog from the 16th data release (SDSS-DR16Q, Lyke et al. 2020). The catalog provides nearly isotropic coverage of the whole sky, apart from the band of the Milky Way, towards which dust and stars reduce the visibility of galaxies. The redshifts in the catalog are corrected for the peculiar motions of the galaxies using a method proposed in Mukherjee et al. (2021a) which relies on the Bayesian Origin Reconstruction from Galaxies (BORG) formalism (Jasche & Wandelt 2013) up to a redshift of $z = 0.05$. The importance of peculiar velocity corrections diminishes above this range. GLADE+ also provides uncertainties on peculiar velocities. The median relative uncertainty of these estimates, compared to the galaxy redshifts, is 1.1%. Therefore, we do not expect peculiar velocity uncertainties to significantly affect our analysis.

GLADE+ reports galaxy magnitudes in 7 different bands, from which we chose to use the $K_s$ band (reported in the Vega system) for our main results, referred to as the *K-band* in this paper. This choice is motivated by our earlier studies (Abbott et al. 2023a) on how well the number density of galaxies in different bands follow the theoretical luminosity Schechter function (Schechter 1976). We found that the K-band absolute magnitude distribution of GLADE+ galaxies is well described by a Schechter function with parameters $M_{*,\mathrm{K}} = -23.39$ and $\alpha_\mathrm{K} = -1.09$ taken from Kochanek et al. (2001). GLADE+ contains K-band magnitudes for a subset of its entries, approximately 1.16 million sources. This subset that we used in our analysis mostly has photometric redshifts available with an absolute error of $\sigma_{\mathrm{z,obs}}{\sim}0.015$ (Bilicki et al. 2014). Spectroscopic redshifts are available for ~23% of this subsample. The top panel of Figure 3 shows the catalog's completeness fraction (see Appendix B) in the K-band as a function of redshift. The different curves are calculated for a given percentage of the sky coverage of the catalog, i.e., by excluding the ~5% of the sky where the catalog does not contain any galaxies with K–band magnitudes. For example, 20% of the coverage of GLADE+ has a completeness fraction lower than the blue curve and 80% of the coverage has a higher completeness fraction. The label also shows the apparent magnitude thresholds corresponding to these curves. The apparent magnitude threshold was obtained as the median magnitude of the galaxies in the pixel. This conservative approach excludes all galaxies fainter than the calculated threshold in the pixel from the analysis.

The bottom panel of Figure 3 presents the sky localizations of the ten best-localized GW events from O4a included in our analysis, in superposition with a sky map showing the directional dependence of the K-band apparent magnitude threshold for the GLADE+ galaxies. Outside of the Galactic plane, the apparent magnitude threshold is typically $m_\mathrm{thr} \sim 13.5$ for the K–band, while closer to the Galactic plane region the apparent magnitude threshold is significantly lower (i.e., brighter). Since Virgo did not observe during O4a, the localizations of events from this run are not as well constrained as they could be. Consequently, we can only expect a modest improvement in constraining power from the galaxy catalog relative to GWTC-3.0.

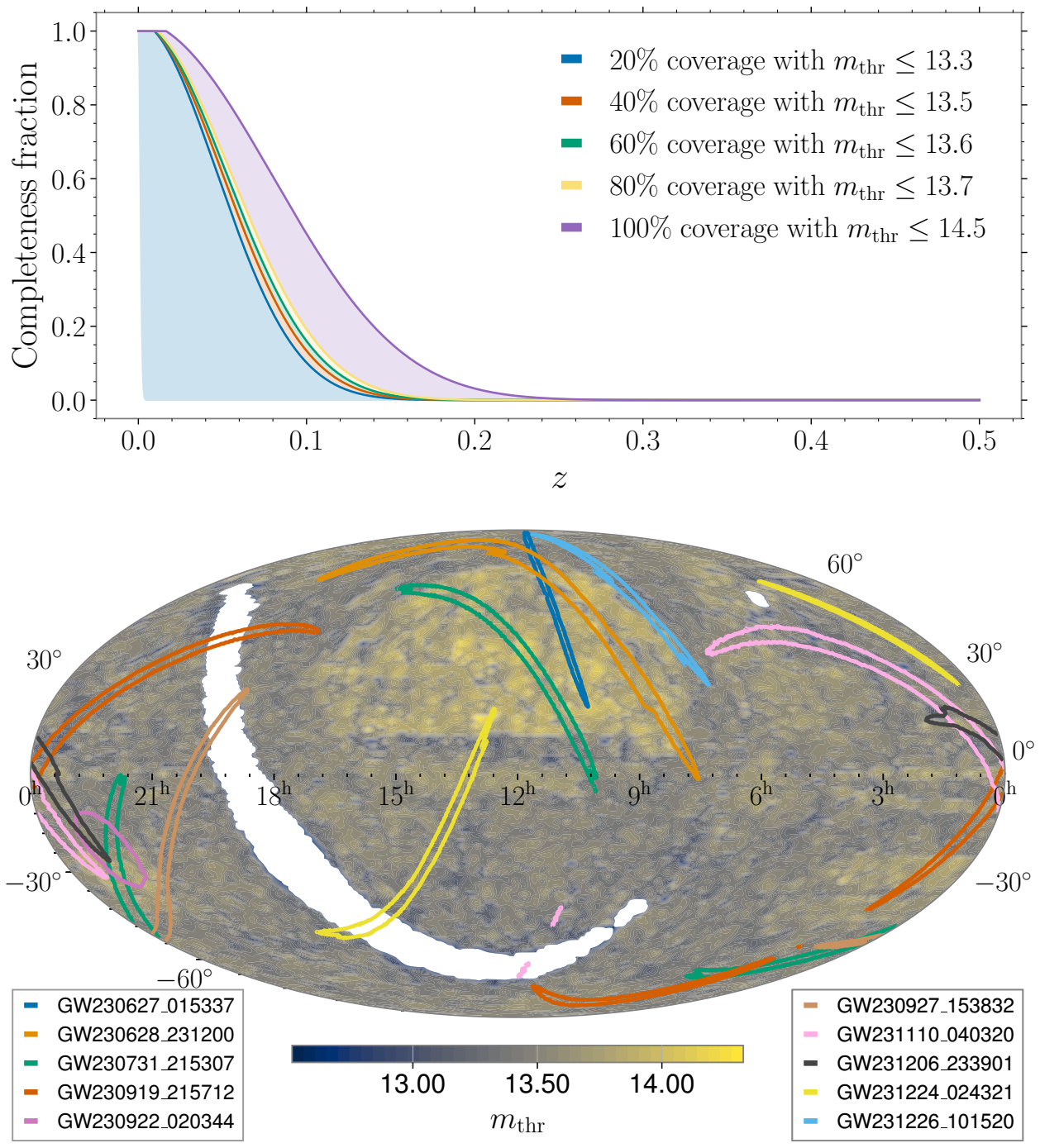

**Figure 3.** Top panel: Completeness fraction of GLADE+ in the K–band, indicating the probability that the catalog contains the host galaxy of a GW event, as a function of redshift for $H_0 = 67.9\ \mathrm{km\,s^{-1}\,Mpc^{-1}}$ and $\Omega_\mathrm{m} = 0.3065$. The different curves are calculated for a given percentage of sky coverage computed by dividing the sky in equal sized pixels of 3.35 deg$^2$, for which the apparent magnitude threshold is brighter than the corresponding $m_\mathrm{thr}$ value reported in the legend. The fraction of pixels with no galaxies is ~5%. Bottom panel: sky map showing the GLADE+ K–band apparent magnitude threshold, $m_\mathrm{thr}$, generated by dividing the sky into 3.35 deg$^2$ pixels. A mask is applied that removes all pixels (white region) with $m_\mathrm{thr} < 12.5$ in order to improve the figure readability. Also shown are the 90% CR sky localizations for the 10 best-localized O4a GW events included in our analysis.

For this analysis, we set bright and dim cutoffs at $M_\mathrm{min} = -27.0$ and $M_\mathrm{max} = -19.0$, respectively. These choices correspond to the limits we used in our previous analysis (Abbott et al. 2023a).

The dark siren analysis requires a pixelization of the galaxy catalog; we adopt the `healpix` pixelization algorithm (Górski et al. 2005; Zonca et al. 2019) with `nside` = 64 and verify that the pixel size remains below the localization scale of the best-constrained GW events, rendering finer resolution unnecessary.

## 4. RESULTS

In this Section, we present our cosmological results based on dark siren, which are derived from the joint inference of cosmological and population hyperparameters. These include parameters that describe the assumed mass distribution and merger rate models. For the $H_0$ results, we will also combine our dark siren constraints with those from the bright siren GW170817.

We sample the posterior in Equation (1) with the normalizing-flows-enhanced nested-sampling package `nessai` (Williams et al. 2021; Williams 2021).

Unless otherwise stated, we present combined results from `icarogw` and `gwcosmo` as posterior distributions built from an equal-weighted mixture of samples (50% from each pipeline). This approach ensures that our final constraints incorporate any residual (small) systematic uncertainty associated with differences in the numerical implementation of the likelihood.

Section 4.1 focuses on the measurement of $H_0$ in a flat-$\Lambda$CDM model, obtained by combining population information with galaxy catalog data from GLADE+ (Dálya et al. 2018, 2022). Section 4.2 presents constraints on modified GW propagation, and finally, Section 4.3 presents robustness checks for our results. When quoting results, we report the median value plus its 68.3% (90%) symmetric credible interval (CI). We use the relative decrease in average uncertainty, computed from the CI, as a metric to measure the improvement of our results.

### 4.1. *ΛCDM Cosmology*

Figure 4 presents the marginalized posterior distributions of the Hubble constant for different cases. In particular, the best estimate of $H_0$ comes from the combined posterior between the dark siren, luminosity-weighting analysis result with our fiducial mass model, FULLPOP-4.0, and the bright siren result of GW170817. This yields $H_0 = 76.6^{+13.0}_{-9.5}$ $(76.6^{+25.2}_{-14.0})$ $\mathrm{km\,s^{-1}\,Mpc^{-1}}$ (Figure 4, black curve).

From the dark siren measurement alone (Figure 4, blue curve), we obtain $H_0 = 81.6^{+21.5}_{-15.9}$ $(81.6^{+42.0}_{-26.8})$ $\mathrm{km\,s^{-1}\,Mpc^{-1}}$. This estimate of $H_0$, based solely on dark sirens, gives a posterior distribution of the Hubble constant which is still slightly broader than that obtained from the bright siren GW170817 (Figure 4, yellow curve), namely $H_0 = 78.4^{+25.7}_{-12.0}$ $(78.4^{+51.2}_{-16.6})$ $\mathrm{km\,s^{-1}\,Mpc^{-1}}$. We obtained our GW170817 $H_0$ posterior by using the same low-spin prior PE samples as in Abbott et al. (2021a), but with an enlarged $H_0$ prior and a different injection set to estimate the GW detection probability, in order to match those used with the $\Lambda$CDM spectral and dark siren analyses presented in this study. Our GW170817 $H_0$ estimate is consistent with those reported in Abbott et al. (2017a, 2021a, 2023a).

With the current set of dark sirens, most of the information on the Hubble constant still comes from the presence of mass features in the population. We assess this by comparing to the case where galaxy-catalog information is not included and constraints on $H_0$ are solely driven by our population assumption, which corresponds to the spectral siren result (Figure 4, orange curve), $H_0 = 76.4^{+23.0}_{-18.1}$ $(76.4^{+41.2}_{-28.6})$ $\mathrm{km\,s^{-1}\,Mpc^{-1}}$. The spectral siren analysis is further discussed in Appendix D.

From GLADE+, we find that the inclusion of K-band information improves the spectral siren constraints on the Hubble constant by approximately 8.6% (3.1%). The most informative dark sirens can be identified by computing the posterior probability on $H_0$ while fixing the population hyperparameters to reference values, and identifying events for which the information from the in-catalog term provides the largest improvement in constraints with respect to not using the catalog. The additional constraining power primarily arises from a few GW events that are nearby and well-localized, and for which the galaxy catalog is sufficiently complete, notably GW190814 (Abbott et al. 2020c), GW230627_015337, GW230814_230901 (Abac et al. 2025k) and GW230529_181500 (Abac et al. 2024). The overall limited gain of information from the galaxy catalog can be attributed to the low completeness fraction of the GLADE+ K-band data at the distances of most of the GWTC-4.0 events, see Figure 3 and Appendix E.

In Figure 5 we illustrate the impact of mass models and galaxy weighting on the marginalized posteriors of $H_0$. All curves use K-band information from the GLADE+ galaxy catalog, while the event GW170817 is excluded from the dark siren inferences, as it is treated solely as a bright siren in this paper (this choice is validated and discussed in detail in Section 5.2). The left panel presents results based on three different source mass models in the galaxy luminosity-weighting case: the PLP, the MLTP, and the FULLPOP-4.0 models. The right panel, in contrast, explores the difference between the no-weighting and luminosity-weighting cases, while keeping the source mass model fixed to our fiducial mass model (FULLPOP-4.0).

Based on the left panel of Figure 5, we find some differences in the measurements of $H_0$ due to assumptions about the shape of the mass spectrum. Though systematic differences are visible when comparing the PLP with the MLTP and FULLPOP-4.0 results, these are well within the statistical uncertainty. The posterior distributions are wide and overlap with each other. In particular, assuming a uniform prior $H_0 \in \mathrm{U}(10, 200)$ $\mathrm{km\,s^{-1}\,Mpc^{-1}}$ and considering the single-population BBH models, with the MLTP model we obtain $H_0 = 87.3^{+48.0}_{-28.2}$ $(87.3^{+85.6}_{-42.7})$ $\mathrm{km\,s^{-1}\,Mpc^{-1}}$, while with the PLP model we find $H_0 = 124.8^{+45.2}_{-39.4}$ $(124.8^{+63.2}_{-60.1})$ $\mathrm{km\,s^{-1}\,Mpc^{-1}}$.

The MLTP distribution shows better agreement with the FULLPOP-4.0 distribution. Its higher-mass peak occurs in a different location than that of the PLP model, and it is narrower. We find that the PLP distribution tends to drive the $H_0$ estimate toward higher values compared to the MLTP model. This is due to the fact that a single peak is unable to fit the complex low-mass structure of the BBH primary mass spectrum, as will be explained in more detail below when discussing the reconstructed mass spectrum. Moreover, the PLP results are constrained by the assumed upper $H_0$-prior

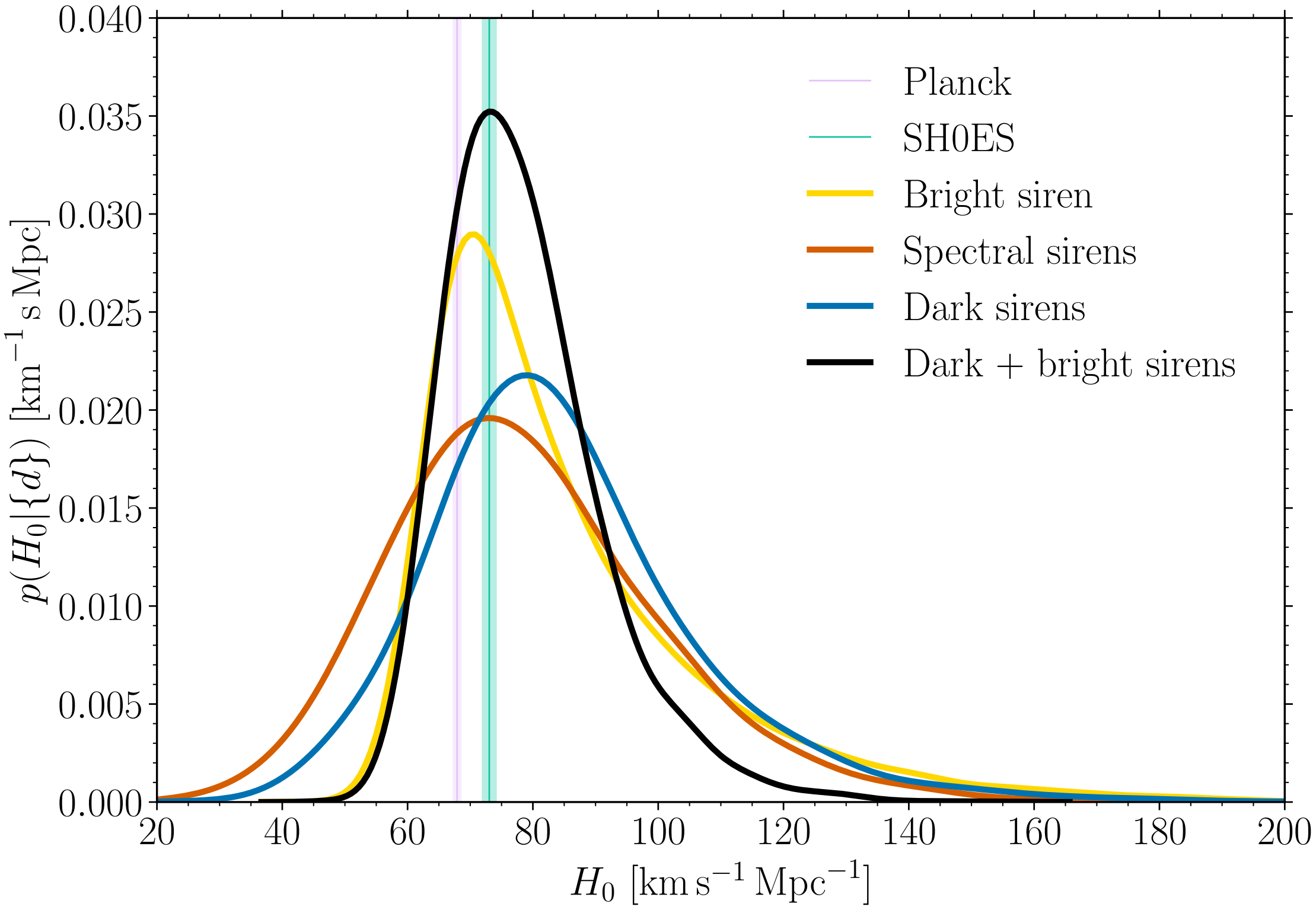

**Figure 4.** Hubble constant posterior for different cases. Yellow curve: posterior obtained from the bright siren GW170817 and its EM counterpart. Orange curve: posterior obtained with the spectral siren method and the FULLPOP-4.0 mass model. Blue curve: posterior obtained using all dark sirens with GLADE+ K-band in the luminosity-weighting case ($\epsilon = 1$) and the FULLPOP-4.0 mass model. Black curve: posterior after combining the dark and bright siren results. The pink and green shaded areas identify the 68% CI constraints on $H_0$ inferred from CMB anisotropies (Ade et al. 2016) and in the local Universe from SH0ES (Riess et al. 2022), respectively.

bound, which brings the PLP and MLTP distributions into closer agreement.

Within the single-population (BBH-candidates-only) framework, we find that the MLTP model is mildly preferred over the PLP model, which was favored in the previous GWTC-3.0 analysis (Abbott et al. 2023a,c). The Bayes factor between these two models is $\log_{10} \mathcal{B} = 0.30$, which does not directly allow us to discriminate between the two mass models. However, this conclusion strongly depends on the prior choice for the position of the two peaks in the MLTP model. Specifically, in the MLTP model, we let both peaks span the mass range $\mathrm{U}(5, 100)\ M_\odot$, differently from the PLP model where, following results from previous GWTC-3.0 analysis (Abbott et al. 2023a,c), the prior range for its single peak is restricted to $\mathrm{U}(20, 50)\ M_\odot$.

We verified that adopting narrower priors for the peaks of the MLTP model, which are compatible with those adopted in Abac et al. (2025f) for the BROKEN POWER LAW + 2 PEAKS model, namely $\mathrm{U}(7, 12)\ M_\odot$ and $\mathrm{U}(20, 50)\ M_\odot$, the PLP model is now strongly disfavored with a Bayes factor of $\log_{10} \mathcal{B} = 2.3$, while the posteriors are not constrained by the narrower priors. Finally, we are unable to perform a model-selection comparison between the single- and multi-population models, as they rely on different datasets.

Another key finding illustrated in the left panel of Figure 5 is the impact of incorporating the full population of CBCs, rather than restricting the analysis to BBHs candidates alone. Beside making the overall analysis more agnostic (by making no assumption about the nature of each GW candidate), the adoption of a multi-population mass model such as FULLPOP-4.0 significantly improves the dark siren constraints on $H_0$, despite the inclusion of just 5 additional candidates with at least a potential NS. We respectively find an improvement of $\sim 56\%\,(\sim 44\%)$ and $\sim 51\%\,(\sim 46\%)$ by using the multi-population mass model with respect to the PLP and MLTP models, although the latter two models provide substantially different medians from each other, with a relative difference of $\sim$47%. This is explained by the inclusion of further characteristic scales in the mass spectrum, related to the mass gap between BHs and NSs (Abac et al. 2025f). Importantly, the FULLPOP-4.0 model assumes identical redshift evolution for both BH and NS merger rates; we verified that allowing different evolutionary tracks does not lead to statistically significant changes in our results.

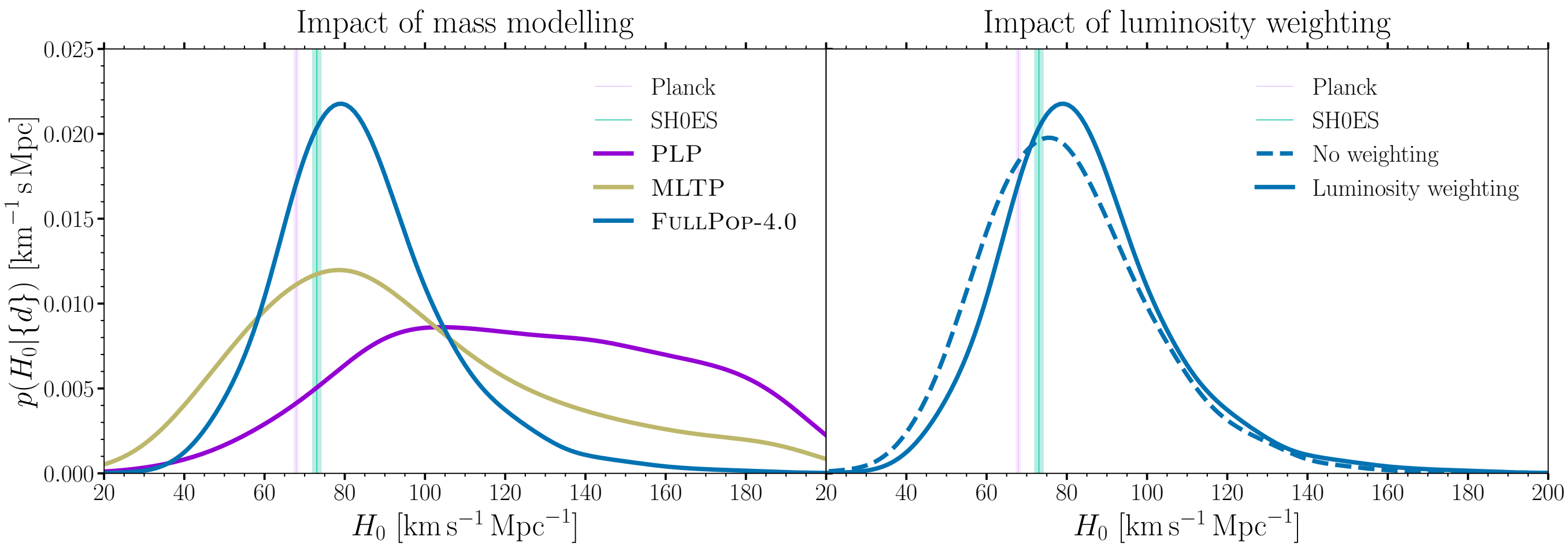

**Figure 5.** Left panel: Hubble constant posteriors with the dark siren method using GLADE+ K-band, assuming three different mass models in the luminosity-weighting case: PLP (magenta curve), MLTP (gold curve), and FullPop-4.0 (blue curve). See Section 2.3 and Appendix C for definitions of these models. Right panel: Hubble constant posteriors with the dark siren analysis in the no-weighting and luminosity-weighted schemes for host galaxies (blue dashed and solid curves, respectively). All analyses assume the FullPop-4.0 mass model. See Section 2.2 for details of the galaxy weighting scheme. In both panels the pink and green shaded areas identify the 68% CI constraints on $H_0$ inferred, respectively, from CMB anisotropies (Ade et al. 2016) and in the local Universe from SH0ES (Riess et al. 2022).

The right panel of Figure 5 shows the effect of the choice of different luminosity weights—either $\epsilon = 0$ or $\epsilon = 1$, see Equation (8). These choices balance computational cost and avoid likelihood inaccuracies that may arise with more extreme weightings. Although fixing these weights introduces a potential systematic uncertainty (Perna et al. 2024; Hanselman et al. 2025), the results for the no-weighting and luminosity-weighting cases are in good agreement. The ability to constrain luminosity weights would have astrophysical value, but we find no strong evidence, based on Bayes factors, to favor uniform weighting over luminosity-based weighting. We find $\log_{10}\mathcal{B} = -0.02$ in the luminosity-weighting case vs no-weighting case, indicating no significant evidence for CBCs to occur in more luminous galaxies in the present data. This outcome reflects the relatively limited impact of the galaxy catalog on the inference with the datasets used here. We expect that these differences will become significant with larger datasets and better-localized events, in which case marginalizing over the weighting power-law index may offer a more robust approach.

The left panel of Figure 6 shows the reconstructed primary mass spectrum using the PLP, MLTP, and FullPop-4.0 mass models in the dark siren scenario. The MLTP and FullPop-4.0 models identify two peaks around $8.7^{+0.4}_{-0.6}\,(8.7^{+0.7}_{-1.3})M_\odot$ and $26.2^{+2.6}_{-2.7}\,(26.2^{+4.2}_{-4.6})M_\odot$, where the error budgets are given by the uncertainties on each Gaussian peak (values from the FullPop-4.0 mass model). The PLP model, in contrast, can only identify a single peak at $27.5^{+4.2}_{-4.3}\,(27.5^{+6.1}_{-6.2})\,M_\odot$, which is compatible with the value found in (Abbott et al. 2023a), although a bit lower. As a consequence, the PLP model prefers lower masses to account for the missing first peak, which puts GW sources at higher redshifts, therefore leading to higher $H_0$ values, as shown in Figure 5. The overly simplistic structure of the PLP model is not able to capture the full complexity of the observed mass spectrum (Abac et al. 2025f), and provides fewer mass scales to inform the $H_0$ measurement. With the use of the FullPop-4.0 model, we also gain access to the NS mass range. In particular, we find support for a minimum mass value around $1.0^{+0.2}_{-0.3}\,(1.0^{+0.3}_{-0.5})M_\odot$, as well as the presence of two local maxima in the CBC mass spectrum at $2.4^{+0.4}_{-0.6}\,(2.4^{+0.5}_{-0.8})M_\odot$ and $7.2^{+1.1}_{-1.6}\,(7.2^{+1.5}_{-2.1})M_\odot$ (see Abac et al. 2025f for further discussions). Overall, the multi-population mass model reconstructs features in agreement with our favored single-population model, namely the MLTP.

The right panel of Figure 6 presents the reconstruction of the CBC merger rate, defined as $p(z|\{d\}) \propto (\mathrm{d}V_\mathrm{c}/\mathrm{d}z)\psi(z|\mathbf{\Lambda})/(1+z)$, (see Section 2 for definitions of these quantities) as derived in the same dark siren scenarios. While the uncertainties remain large at redshifts beyond $z = 0.5$, we find that the reconstructed redshift distributions are consistent across the three mass models, with the FullPop-4.0 and MLTP models predicting higher merger rate values than the PLP model. In absence of observations falling in the region around or above the expected peak, any conclusion about the shape of the redshift distribution at the corresponding redshifts is driven by the assumed parametric form of the merger rate and by the prior range of the associated parameters.

The above results for the merger rate evolution and mass spectrum reconstruction are based on the luminosity-weighted analysis. We verified that all key conclusions hold unchanged also in the no-weighing case, as well as in the spectral siren case—see Appendix D for details on the spectral siren results.

Finally, Figure 7 shows a reduced corner plot highlighting a subset of the population and cosmological hyperparame-ters inferred using our fiducial mass model in the dark siren analysis and luminosity-weighting case. We observe a strong correlation between $H_0$ and the locations of the two BH mass peaks, $\mu_{\rm g}^{\rm low}$ and $\mu_{\rm g}^{\rm high}$ (see Table 5), consistent with trends seen in our previous analysis (Abbott et al. 2023a). Changing $H_0$ shifts the inferred redshift of the sources, which in turn rescales their intrinsic masses, so the mass spectrum shifts alongside $H_0$ to match the observed signals. In contrast, the maximum mass parameter $m_{\rm max}$ does not seem to correlate significantly with $H_0$, although its posterior shows a long tail up to the upper prior boundary. The NS region of the mass spectrum exhibits very weak correlations with $H_0$, likely due to the lack of significant structure and the smaller number of events in that mass range. Overall, the Hubble constant appears to correlate only with certain mass scales, showing no significant correlation with merger-rate parameters such as the power-law index $\gamma$.

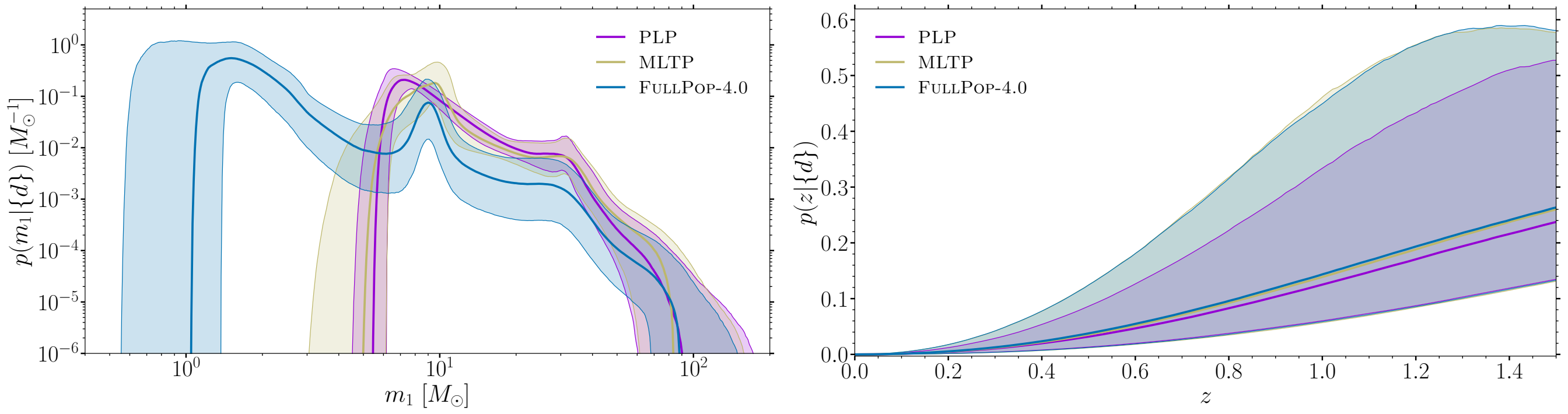

**Figure 6.** Left panel: Reconstructed source-frame primary-mass distribution (solid curve: median; shaded region: 90% CI). Right panel: reconstructed CBC merger rate as defined in the main text. Results in both panels are obtained from dark siren analyses using the PLP, MLTP, and the FullPop-4.0 mass models and the GLADE+ K-band in the luminosity-weighting case.

All constraints obtained in this section assume a fixed value of $\Omega_{\rm m} = 0.3065$ as well as a fixed dark energy equation-of-state parameter $w_0 = -1$. Inferring the values of these parameters independently with dark sirens is not possible at present using our methods, due to the computational cost of constructing redshift priors with varying $\Omega_{\rm m}$ and $w_0$. However, we can examine the impact of varying $\Omega_{\rm m}$ and $w_0$ with dedicated spectral siren analyses. We find that the posterior distributions of these parameters are consistent with the priors, due to the limited constraining power of our data at high redshift, while the uncertainties on other parameters of interest are only marginally affected (as discussed in Section 4.3). This confirms that allowing these parameters to vary does not influence our main results.

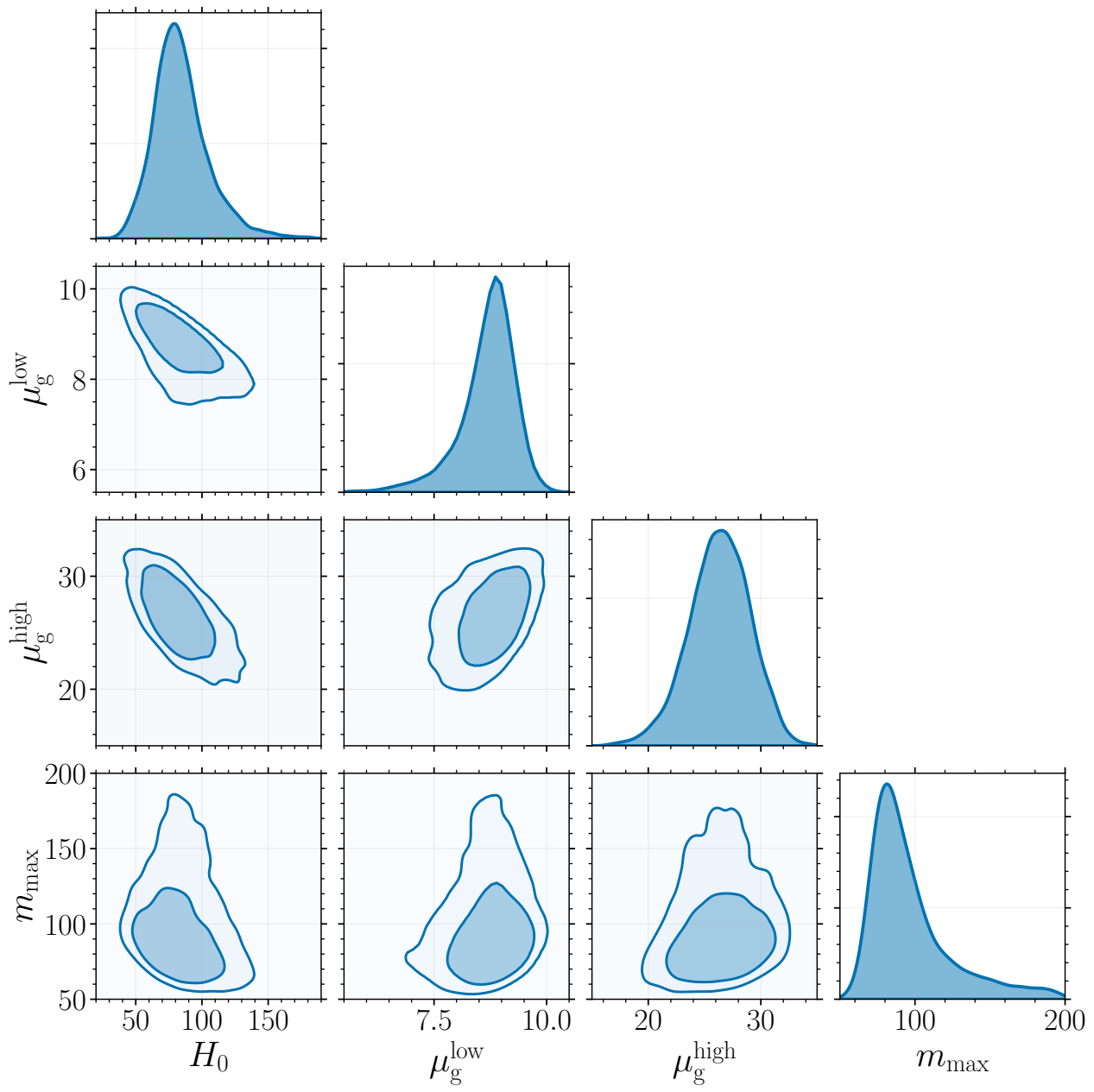

**Figure 7.** Corner plot showing $H_0$ and a subset of population parameters, obtained using a dark sirens analysis assuming the FullPop-4.0 mass model and with the GLADE+ K-band in the luminosity-weighing case. $\mu_{\rm g}^{\rm low}$ and $\mu_{\rm g}^{\rm high}$ are the central locations of the two peaks in the mass model, whilst $m_{\rm max}$ is the maximum allowed mass for either binary component. The solid contours indicate the 68.3% and 90% CR.

### 4.2. *Modified Gravity*

In this Section we present the results obtained by introducing parameterized deviations from GR that affect the luminosity distance ratio, $D_{\rm L}^{\rm GW}/D_{\rm L}^{\rm EM}$, as described in Section 2.4.2. The analysis is carried out using our fiducial mass model FullPop-4.0.

For each parametrization, we consider two different (flat) priors for the Hubble constant: a wide prior, $H_0 \in \mathrm{U}(10, 120)$ $\mathrm{km\,s^{-1}\,Mpc^{-1}}$, and a narrow prior, $H_0 \in \mathrm{U}(65, 77)$ $\mathrm{km\,s^{-1}\,Mpc^{-1}}$ (for the narrow prior analysis, we present results obtained with a single pipeline rather than a mixture of posterior samples). This choice is motivated by the following considerations. In general, $H_0$ and any parameter governing modified GW propagation are correlated to some extent, as both affect the luminosity distance–redshift relation. Consequently, the most agnostic approach to constraining deviations from GR involves marginalizing over $H_0$ using a broad enough prior—hence the adoption of the wider range. The broad prior adopted for $H_0$ in this Section is narrower than the one used for the ΛCDM case. This is because, for certain extreme combinations of $H_0$ and $\Xi_0$, a wider prior on $H_0$ would lead to assigning very high redshifts—beyond $z \gtrsim 10$—to the GW sources in the sample. Our redshift priors, by construction, do not cover these redshifts as we assume they are highly improbable. Furthermore, this would cause instability in our treatment of selection effects because at these very high redshifts the stability criterion for Monte Carlo (MC) integration could fail (see Appendix A). To avoid these issues, we restrict the $H_0$ prior accordingly. Conversely, it is also valuable to explore constraints on GR under the assumption of prior knowledge of other cosmological parameters, which motivates our second choice of a narrower prior which encompasses the region of the current Hubble tension at approximately $4\sigma$ (Aghanim et al. 2020; Riess et al. 2022; Di Valentino & Brout 2024).

Here we present the dark siren results with luminosity weighting, and focus on the comparison between broad and narrow $H_0$ priors in the figures. We also performed a spectral siren analysis and find consistent results with the dark sirens method. We find that modified gravity results are not significantly improved by the inclusion of galaxy-catalog information, as the non-GR effects primarily emerge at redshifts where the catalog is significantly incomplete (see discussion in Section 2.4.2).

First, we discuss results for the $\Xi_0$–$n$ parametrization, see Equation (22). The uniform priors used in this analysis are $\Xi_0 \in \mathrm{U}(0.435, 10)$ and $n \in \mathrm{U}(0.1, 10)$. The left panel of Figure 8 shows the 2D corner plot for $\Xi_0$ and the low-redshift power-law slope of the merger rate, $\gamma$. We find that $\gamma$ shows a strong correlation with the parameters describing deviations from GR (in this case $\Xi_0$), in agreement with previous findings (Mancarella et al. 2022; Leyde et al. 2022; Chen et al. 2024a). This correlation occurs because $\Xi_0$ and $n$ modify the relationship between $D_\mathrm{L}^\mathrm{GW}$ and $z$, therefore affecting the observability of GW sources as a function of redshift. A similar change could be reproduced by adjusting the merger rate of CBCs as a function of redshift, which is what $\gamma$ controls, leading to degeneracy.

Adopting a wide $H_0$-prior, we find $\Xi_0 = 1.2^{+0.8}_{-0.4}\,(1.2^{+2.4}_{-0.5})$, while with the narrow $H_0$-prior we obtain $\Xi_0 = 1.0^{+0.4}_{-0.2}\,(1.0^{+1.1}_{-0.4})$. This result is consistent with GR, recovered in the limit $\Xi_0 = 1$. The parameter $n$ is poorly constrained, and we do not show it in Figure 8.

The right panel of Figure 8 shows the results for the $\alpha_M$ parametrization (see Equations 23 and 24), now displaying the parameters $c_M$ and $\gamma$. The uniform prior used for this analysis is $c_M \in \mathrm{U}(-10, 50)$. We find $c_M = 0.3^{+1.6}_{-1.4}\,(0.3^{+2.9}_{-2.2})$ for the dark siren analysis with a wide $H_0$-prior, and $c_M = -0.3^{+1.4}_{-1.0}\,(-0.3^{+2.5}_{-1.5})$ with a narrow $H_0$-prior. This is also consistent with GR, recovered in the limit $c_M = 0$. The degeneracy with the merger rate parameter $\gamma$ is visibly pronounced, for the same reasons discussed for $\Xi_0$.

In full Horndeski gravity, the function $\alpha_M$ can also be constrained through its effects on the CMB. However, in this work we have fixed $\Omega_\mathrm{m}$ to a value inferred in a flat-$\Lambda$CDM analysis of Planck data (Ade et al. 2016). This may introduce a small bias in constraints on $c_M$, which we do not expect to be significant given the order-of-magnitude of the constraints obtained here. A fully correct approach would be to jointly analyze the Planck data alongside our GW events, which is beyond the scope of the present work; a related discussion is found in Lagos et al. (2019).

As expected, we find a correlation between modified gravity parameters and the Hubble constant. In particular, both $\Xi_0$ and $c_M$ are positively correlated with $H_0$, with Pearson correlation coefficients $0.31$ and $0.58$ respectively. This explains the narrower error-bars when restricting $H_0$ with a narrower prior.

Finally, Figure 9 presents the reconstructed relation between redshift and GW luminosity distance $D_\mathrm{L}^\mathrm{GW}$ for both the $\Xi_0$–$n$ and $\alpha_M$ parametrizations, obtained with the dark siren analysis. Figure 9 shows no deviation from the GR prediction (in which the distance ratio is always one), considering both large and narrow priors on the local expansion rate of the Universe. The slight asymmetry of the contours around $D_\mathrm{L}^\mathrm{GW}/D_\mathrm{L}^\mathrm{EM} = 1$, particularly visible in the wide $H_0$-prior case, is inherited from the asymmetry of the marginalized posteriors on $\Xi_0$ and $c_M$ visible in Figure 8.

In order to check consistency among the parametrizations, we can map the constraint on $c_M$ (under our fiducial value of $\Omega_\mathrm{m}$) into a corresponding constraint on $\Xi_0$ using Equation (25). This map implies that a flat prior on $c_M$ results in a prior which is not flat in $\Xi_0$. Therefore, for a fair comparison, we reweight the samples by the Jacobian implied by Equation (25). Averaging one hundred realizations to reduce the effect of random fluctuations, we obtain $\Xi_0 = 1.3^{+1.0}_{-0.5}\,(1.3^{+2.1}_{-0.7})$ with a wide $H_0$-prior, and $\Xi_0 = 1.0^{+0.6}_{-0.3}\,(1.0^{+1.5}_{-0.4})$ with a narrow $H_0$-prior. These values are consistent with the bounds obtained directly from the $\Xi_0$ analysis, although with slightly larger uncertainties, in particular at the high tail of the posterior. This can be attributed to the fact that, as explained in Section 2.4.2, the time evolution of the distance ratio in the two parametrizations is not fully equivalent: the $\alpha_M$ parametrization adopts a fixed time evolution for the distance ratio, while in the $\Xi_0$–$n$ parametrization the time evolution is encoded in the parameter $n$, over which we marginalize. The fixed time evolution results in a more marked correlation between $c_M$ and the parameter $\gamma$ describing the low-redshift evolution of the merger rate, which leads to a broader marginal posterior on $\Xi_0$. We verify that this is the case with an analysis where we vary $\Xi_0$ while keeping $n$ fixed to the value predicted by Equation (25), which is $n \approx 2.54$. In this case, we recover a consistent correlation between $\gamma$ and $\Xi_0$ across the two parametrizations. The reconstructed distance ratio (Figure 9) also shows consistency among the two parametrizations, while also displaying explicitly the slight difference of the contours as functions of redshift due to the different time evolution.

### 4.3. *Systematics Tests*

Finally, we summarize checks conducted to ensure robustness of our results. Figure 10 shows a summary of the constraints on $H_0$ varying several assumptions discussed in Sections 4.1 and 4.2, with additional numerical stability checks that we discuss below. In particular, we display the effects of luminosity weighting, varying mass models, varying other parameters of the cosmic expansion history, and varying choices related to the accuracy of the likelihood evaluation. For these tests, we used the dark siren approach with the FULLPOP-4.0 mass model and a luminosity-weighting scheme. The posteriors shown in this plot do *not* include constraints from GW170817, and are obtained with the `icarogw` pipeline only.

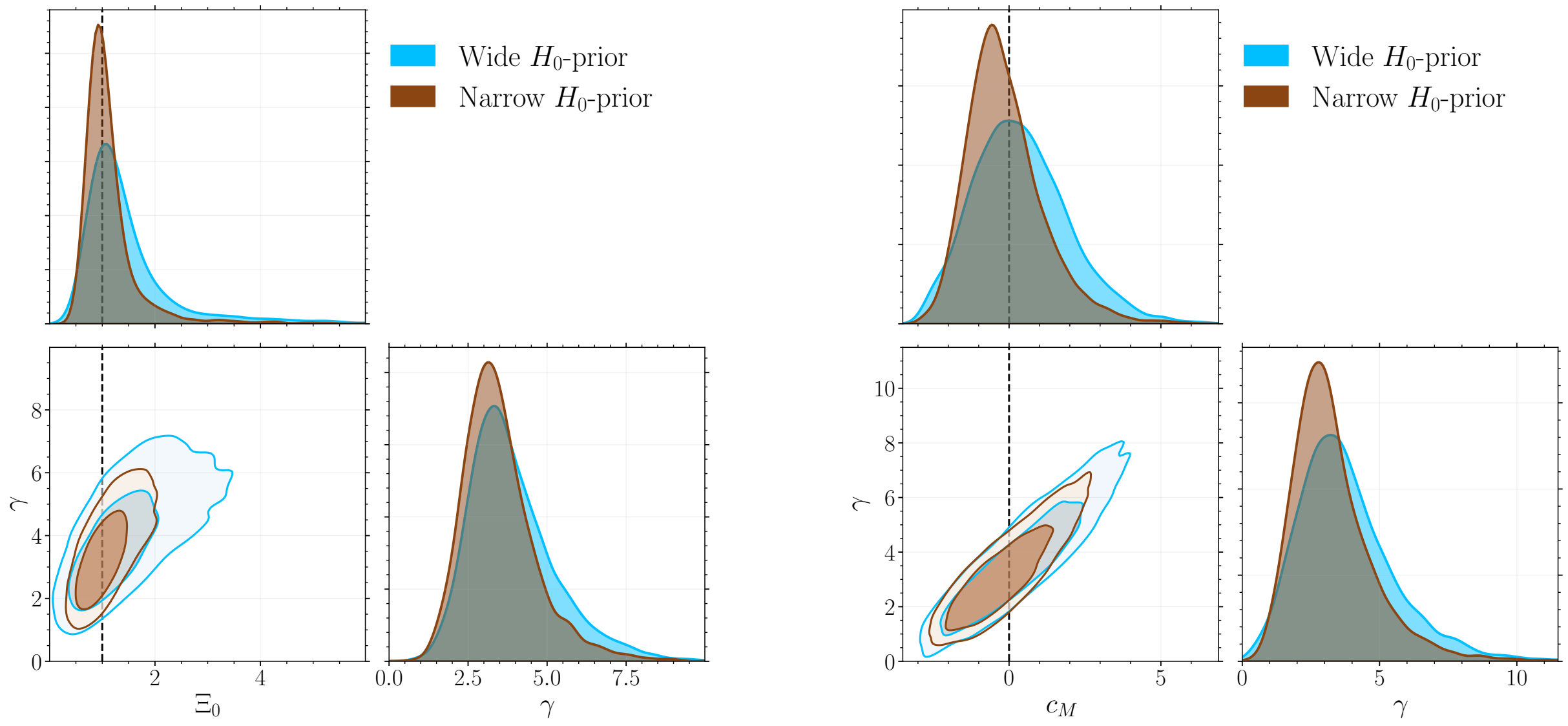

**Figure 8.** Corner plots of the modified gravity parameters $\Xi_0$ (left) and $c_M$ (right), and the merger rate parameter $\gamma$, obtained with the dark siren method and assuming the FULLPOP-4.0 mass model. Vertical dashed lines in the abscissa indicate the GR limit of the respective modified gravity parameters. The contours indicate the 68.3% and 90% CR.

Most of these cases have already been discussed in previous sections, so here we focus on the numerical stability tests. Accurate likelihood evaluation relies on line-of-sight redshift integrals. In particular, one of our two pipelines employs MC integration with a threshold on the effective number of PE samples $n_{\rm eff,PE}$ in the MC integral (see Appendix A for details). We check the effect of changing this threshold from 10 (which is our baseline choice) to 50, or eliminating the threshold altogether (Figure 10, "$n_{\rm eff,PE} > 50$" and "No threshold on $n_{\rm eff,PE}$" labels, respectively). Raising the threshold corresponds to a more stringent condition on the precision of the MC integration. However, adopting a cut that is too high may lead to an artificial shift of the posterior towards a low variance region. This motivates the need for a check of the stability under this choice. We also consider the possibility of thresholding on the total likelihood variance instead (Figure 10, "Log-likelihood variance $< 1$", see Appendix A for details).

Although these thresholds effectively modify the likelihood by introducing a data-dependent condition, we find their effect negligible for the models considered in this work. Nonetheless, this conclusion holds specifically for the set of models considered here, and should not be taken as a general statement. In particular, this strategy could become problematic when dealing with highly-peaked integrands, such as those resulting from extreme luminosity-weighting schemes, where redshift priors are dominated by spikes from bright galaxies.

MC integration is also employed to compute the selection effect term (see Appendix A). We follow the criterion (Farr et al. 2019) requiring the number of effective MC samples $n_{\rm eff,inj}$ to exceed four times the number of observed events (e.g., at least 564 for a sample of 141 events). We also test more stringent thresholds, including increasing the requirement to 2000 or removing it entirely (Figure 10, "$n_{\rm eff,inj} > 2000$" and "No threshold on $n_{\rm eff,inj}$" labels, respectively), finding no evidence of systematic bias resulting from these changes.

As mentioned in Section 2.1, our population models implicitly assume that the CBC spin distribution is isotropic with uniform distribution in the spin magnitudes (Abac et al. 2025d). However, we verified that including spin distributions for the BBH population using the DEFAULT model (Abbott et al. 2023c; Abac et al. 2025f) has no significant impact on the current cosmological constraints (see Section 5). For the spin-informed tests, we adopted the MLTP mass model, as this model better fits the BBH mass spectrum of the GW candidates used in our analysis (Abac et al. 2025f).

## 5. DISCUSSION AND PERSPECTIVES

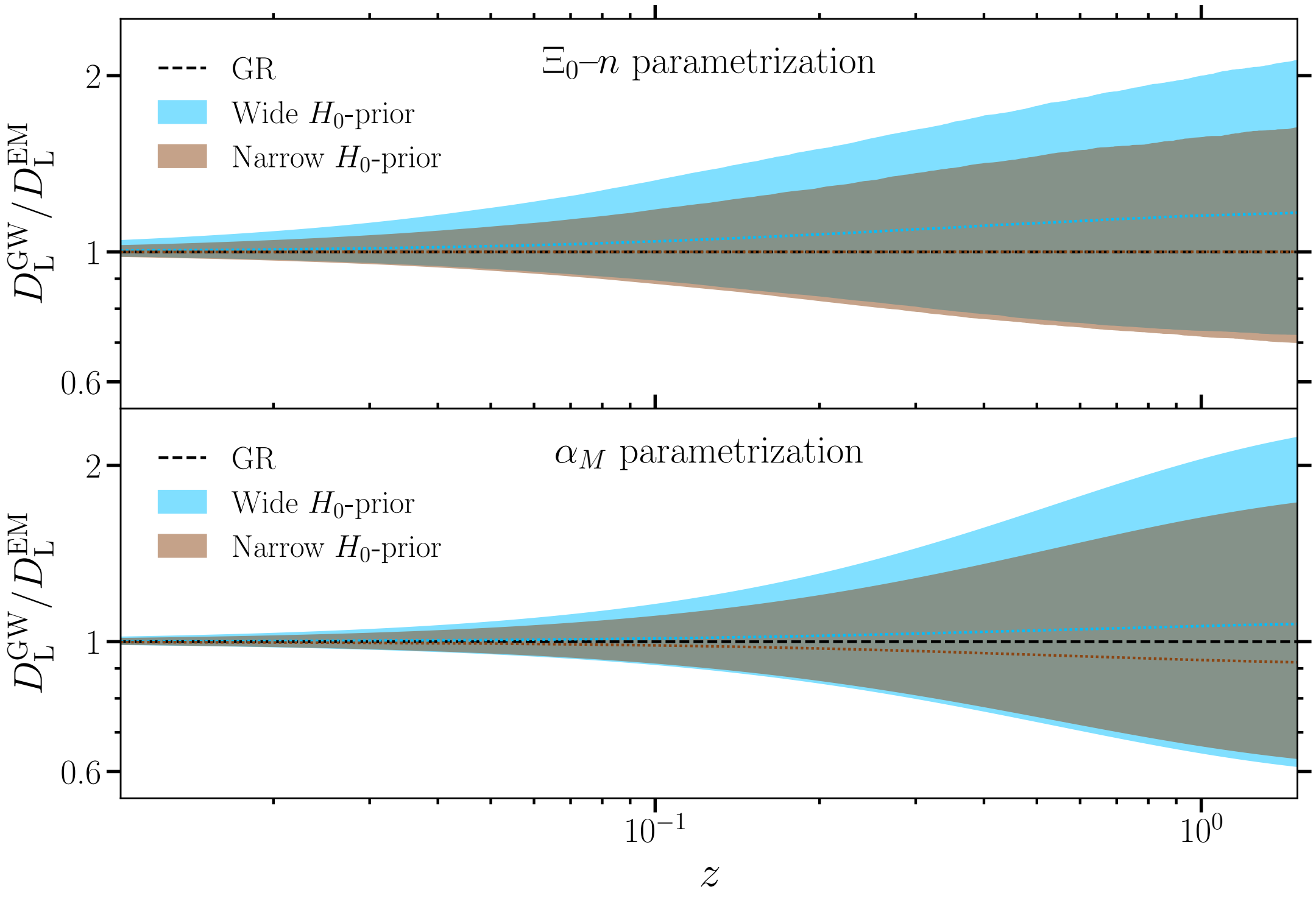

**Figure 9.** Reconstructed ratio $D_L^{GW}/D_L^{EM}$ as a function of cosmological redshift $z$, for the two modified gravity parametrizations considered, $\Xi_0$–$n$ and $\alpha_M$. In both cases the contours show the 90% CI with median (dotted curve) reconstructed from the wide-$H_0$ prior (orange) and narrow-$H_0$ prior (blue) analyses with the FULLPOP-4.0 mass model. The black dashed curve represents the GR limit. Note that the reconstructed distance ratio is asymmetric at higher redshifts.

In this Section, we compare our results to the literature, and discuss possible improvements and future developments which constitute negligible systematics at present.

### 5.1. *Comparison with Existing Results*

We begin by discussing our constraints on the Hubble constant. Figure 11 summarizes our findings alongside previous measurements from the LVK.

Our baseline result is obtained through a dark siren analysis that improves upon previous LVK measurements (Abbott et al. 2023a) by implementing a more advanced methodology, as detailed in Section 2. In particular, we perform a full marginalization over the CBC mass distribution and merger rate parameters, even with the inclusion of redshift information coming from a galaxy catalog. In contrast, prior LVK analyses with galaxy catalogs relied on fixed population parameters. This makes a direct comparison unfair, as fixing population parameters leads to overly optimistic constraints on $H_0$. In contrast, our approach achieves comparable precision while providing a more statistically robust treatment by fully accounting for population uncertainties. This explains the narrower error bar associated to the measurement of Abbott et al. (2023a) in Figure 11 as compared to the result of this work, despite the former being obtained with a smaller sample of GW data.

The spectral siren analysis, on the other hand, is directly comparable to the GWTC-3.0 results obtained with the same method. In this case, without the use of GW170817 as a bright siren, our new measurement yields a $\sim 60\%$ improvement with respect to Abbott et al. (2023a), driven by the increased number of events and by the adoption of the more comprehensive FULLPOP-4.0 population mass model.

Finally, all our results remain statistically consistent with the values reported by the Planck (Ade et al. 2016; Aghanim et al. 2020) and SH0ES (Riess et al. 2021) collaborations at the 90% CI.

We now turn to results on modified GW propagation. Previous constraints on modified gravity parameters with the GWTC-3.0 catalog were obtained by Mancarella et al. (2022); Leyde et al. (2022); Mastrogiovanni et al. (2023); Chen et al. (2024a), while Ezquiaga (2021) previously had bound $c_M$ with GWTC-2.0. In particular, assuming a value of $H_0$ compatible with Ade et al. (2016), Mancarella et al. (2022) and Mastrogiovanni et al. (2023) found $\Xi_0 = 1.3^{+0.9}_{-0.5}$ and $\Xi_0 = 1.44^{+1.17}_{-0.93}$ (68% CI), respectively, while Mastrogiovanni et al. (2023) also measured $c_M = 1.0^{+2.6}_{-3.4}$ (68% CI) and Ezquiaga (2021) found $c_M = -3.2^{+3.4}_{-2.0}$ (68% CI). In Leyde et al. (2022), different constraints are reported depending on the mass model and selection cut applied to the data. Here, we refer to the result with a PLP model and 35 BBH with SNR > 12. Adopting a prior on $H_0$ restricted to the tension region, Leyde et al. (2022) found $\Xi_0 = 1.4^{+1.8}_{-0.8}$ and $c_M = 0.4^{+3.2}_{-3.0}$ (90% CI). Finally, including also three

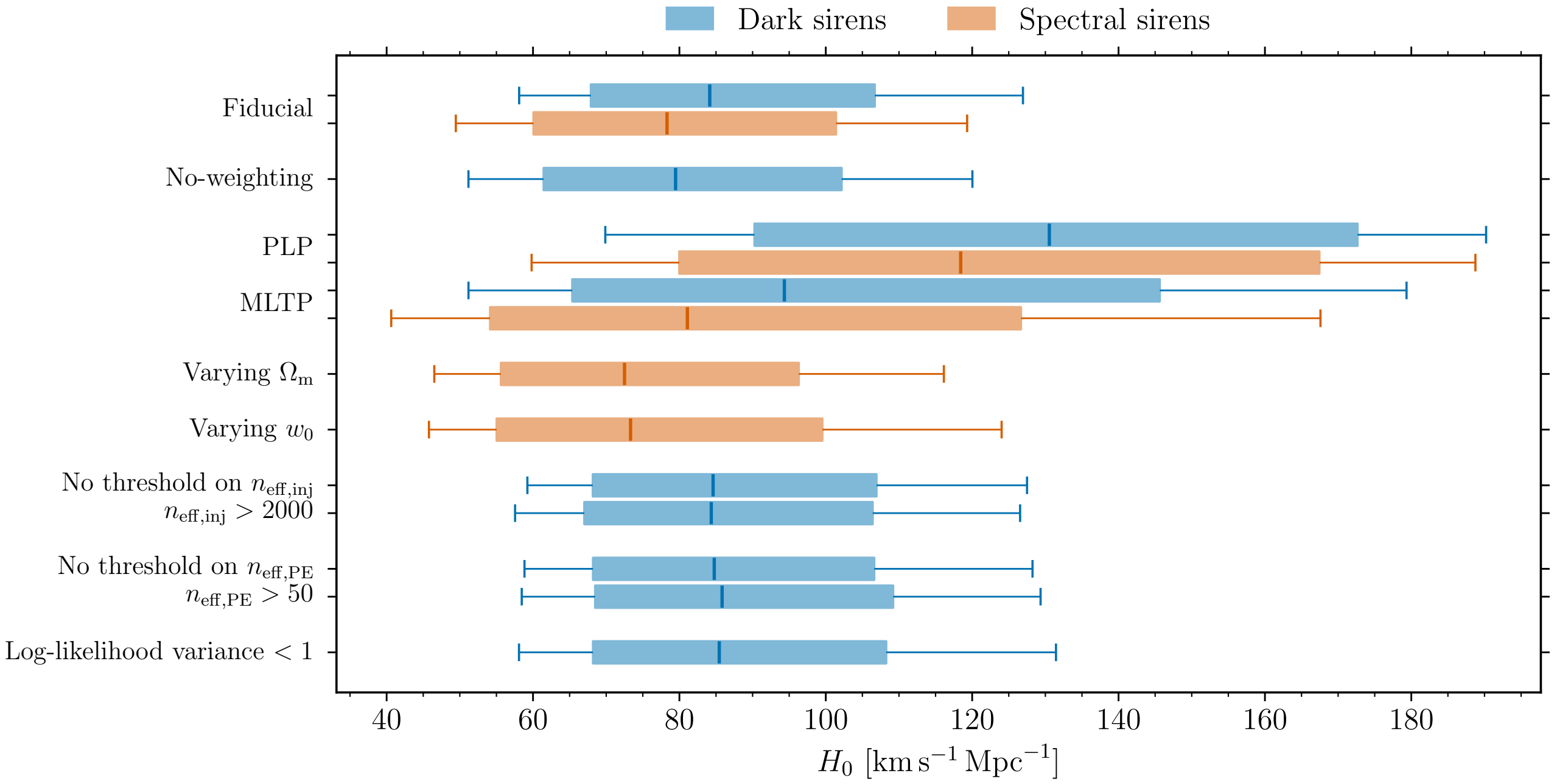

**Figure 10.** Robustness checks against various systematics discussed in Section 4, compared to the fiducial results ($\Lambda$CDM, FULLPOP-4.0, and luminosity-weighting case for the dark sirens case). The box-plots show the median value as a vertical segment. The colored boxes stretches to the 68.3% CI, while the whiskers extend to encompass the 90% CI. The labels indicate variations with respect to the fiducial results. The posteriors shown in this plot do *not* include bounds from GW170817. In the analyses with varying $\Omega_{\rm m}$ and $w_0$ the priors used are $\mathrm{U}(0,1)$ and $\mathrm{U}(-3,0)$, respectively. All displayed checks were generated exclusively using `icarogw`, and not by merging posterior samples from both pipelines, as done for the main results.

NSBH mergers from GWTC-3.0, Chen et al. (2024a) found $c_M = 1.5^{+2.2}_{-2.1}$ and $\Xi_0 = 1.29^{+0.93}_{-0.94}$ (68% CI) with a wide $H_0$ prior. With respect to the best among those results, our result with a wide $H_0$-prior gives a $\sim$36% improvement for $\Xi_0$, and a $\sim$30% improvement for $c_M$.

Our bound with a narrow $H_0$-prior gives instead a $\sim$42% improvement for $\Xi_0$ and a $\sim$35% improvement for $c_M$. These improvements are due to the additional events from O4a and to the use of the FULLPOP-4.0 population model.

Assuming $\alpha_M$ is sourced by a scalar degree of freedom within the effective field theory (EFT) framework, and in the class of Horndeski-type theories, our constraints from GW observations can be compared to those from LSS and the CMB. When analyzing LSS and CMB data, it is essential to ensure that the scalar sector remains free from ghost and gradient instabilities. These theoretical consistency requirements further restrict the allowed parameter space. Recent LSS analyses that assume luminal tensor propagation (Noller & Nicola 2019; Baker & Harrison 2021; Seraille et al. 2024; Ishak et al. 2024) impose such constraints. In a GW-only analysis, we assume that stability can be enforced by appropriate choices of additional EFT operators—particularly the braiding parameter $\alpha_B$—which influence only the scalar sector. For theories with $\alpha_B = 0$, regions with $\alpha_M < 0$ are typically ruled out by stability arguments.

The latest available LSS bounds correspond to the clustering measurements from DESI 2024 (Ishak et al. 2024). This work finds the bound $c_M < 1.14$ (95% CI), assuming vanishing braiding and a $\Lambda$CDM background. Relaxing the braiding assumption and marginalizing over it yields a constraint of $c_M = 1.05 \pm 0.96$ at 68% CI. More stringent constraints can be obtained by combining different LSS observables. In particular, the integrated Sachs–Wolfe (ISW) effect from galaxy–CMB cross-correlations has been shown to provide significant improvements (Renk et al. 2017; Seraille et al. 2024). Combining LSS and CMB observables to ISW, Seraille et al. (2024) find $c_M = 0.54^{+0.90}_{-0.60}$ at 95% CI after marginalization over the braiding parameter. Although consistent with these bounds, our best result is weaker by approximately $\sim$25% and $\sim$60% relative to the latter two, respectively. Despite this, our results are based on an entirely independent dataset with different systematics.

### 5.2. *Perspectives*

*Considerations on the Spectral Siren Analysis*—Spectral siren information is driven by the shape of the mass spectrum of compact objects. Our mass models are based on specific parametric forms. We choose a set of hyperparametric prior bounds from previous studies Abbott et al. (2023a,c). Extending the prior range of the population parameters to significantly wider priors may significantly change the reconstructed mass spectra and therefore effectively represent dif-ferent mass models, even though the analytical formulation of the mass model is the same Gennari et al. (2025). We do not consider this possibility here. As discussed in Section 4.1, we observe some variation in the results obtained with our simplest mass model, the PLP model, compared to those obtained with the MLTP and FULLPOP-4.0 models, which are able to better describe the observed primary mass distribution (Abac et al. 2025f), as shown in Figure 5, Figure 6, and Figure 10. Compared to different galaxy-weighting schemes, this represents the dominant systematic in our analysis.

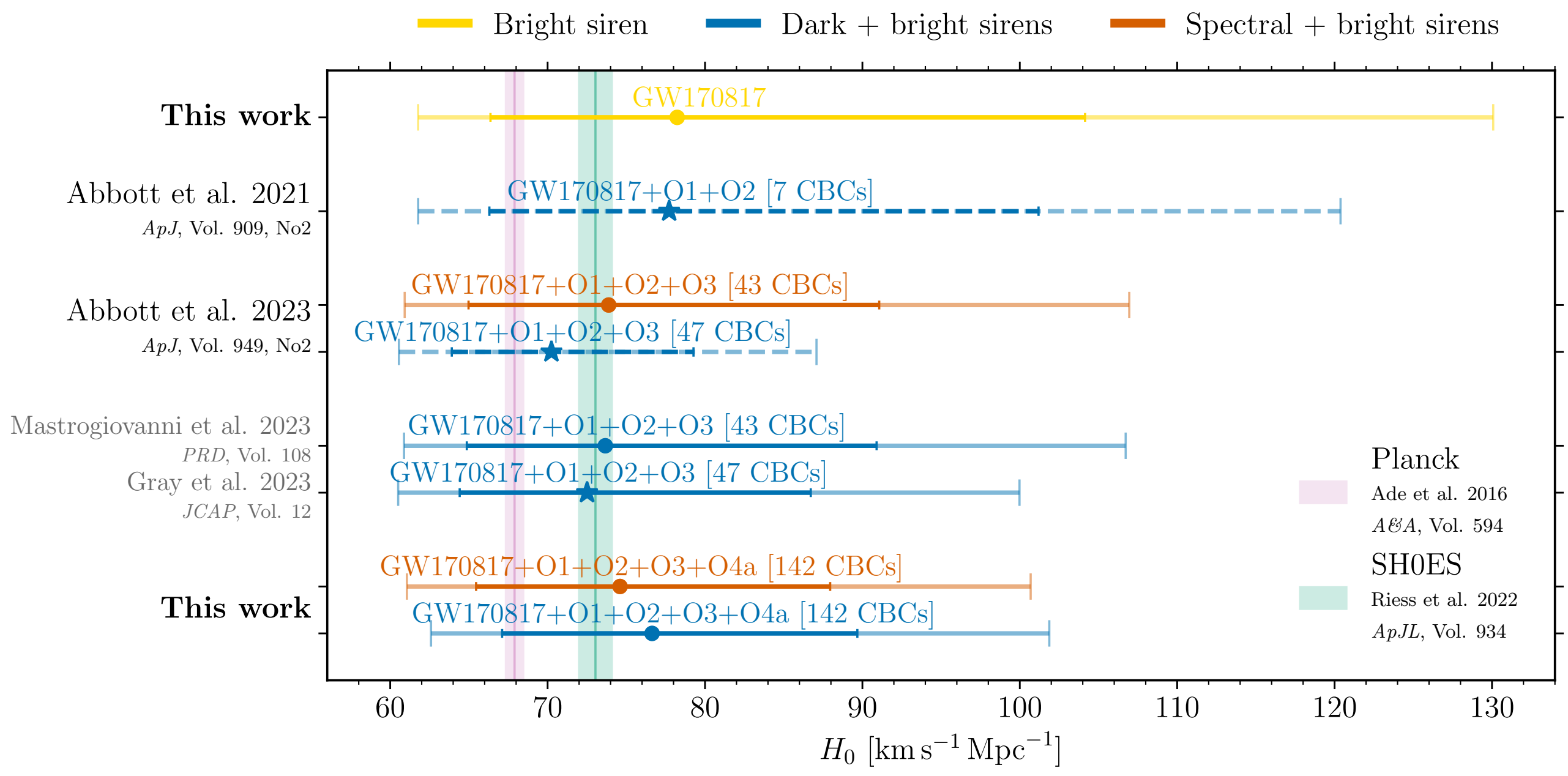

**Figure 11.** Summary of $H_0$ measurements from GW detections, combining bright with dark or spectral siren analyses conducted by LVK pipelines from O1 up to O4a. Non-LVK works are shown are labeled in light gray. In yellow, we report the bright siren result that has been recalculated for this work. Note that previous papers combined results use the bright siren samples from Abbott et al. (2021a). We report the dark siren results in blue and spectral siren results in orange, including the bright siren in both cases. The darker-shaded line covers the symmetric 68.3% CI, and extends till 90% CI. Studies that assumed a fixed population model are marked with a dashed line style, and a star as a marker for the median value. The study from Gray et al. (2023) is marked with a star and a solid line, as it assumed a fixed BNS population model, but not-fixed models for the BBH and NSBH populations. The total number of CBC events used in the analysis is indicated in square brackets on top of each result. For details on the analysis settings, see the respective publications. The pink and green vertical bands indicate the Planck (Ade et al. 2016) and SH0ES (Riess et al. 2022) median and $1\sigma$ values, respectively. The error bars obtained in this work are based on our fiducial mass model FULLPOP-4.0.

Related to this, we did not consider extra correlations between population features, such as mass–redshift and mass–spin interplay. However, their possible existence is being increasingly investigated. For example, Li et al. (2024b); Pierra et al. (2024a) report correlations between spin magnitude and mass, which could influence cosmological inferences as current constraints are closely linked to the mass distribution, while Abac et al. (2025f) finds support for the evolution of the spin distribution with redshift (Biscoveanu et al. 2022). Tong et al. (2025) studies the impact of spin information in spectral siren cosmology, showing that its inclusion can mitigate systematics related to mismodeling of the mass spectrum. Additionally Li et al. (2024a) find that modelling two populations with different spin and mass distributions yields an improvement in Hubble constant bounds from GWTC-3.0 data.

Similarly, while current data do not robustly support evolution of the mass distribution with redshift (Heinzel et al. 2025; Lalleman et al. 2025; Gennari et al. 2025; Abac et al. 2025f), considering this effect may become important as GW detector sensitivity improves. Such evolution could introduce biases if not properly modeled (Pierra et al. 2024b; Agarwal et al. 2025). Nevertheless, because cosmological effects imprint a coherent and predictable modulation on the mass spectrum observed across different redshifts, it is expected that appropriate modeling should allow disentanglement of these from astrophysical evolution (Ezquiaga & Holz 2022; Chen et al. 2024b). In future studies, it would be valuable to incorporate comprehensive correlation modeling or adopt data-driven approaches (Farah et al. 2025), which offer increased flexibility and robustness by reconstructing features directly from the observations, without strong parametric assumptions.

*Combination with Bright Sirens*—When combining dark siren events with bright sirens such as GW170817, particularly within a sample that includes both BNS mergers with and without EM counterparts, the correct approach would be to model the joint GW and EM detection probabilities and perform a unified hierarchical inference. At present, while our pipelines fully account for GW selection effects, they do not yet model the EM detection probability (potential systematics related to EM selection effects, and related mitigation strategies, can be found in Chen 2020; Chen et al. 2024c; Mancarella et al. 2024; Müller et al. 2024; Salvarese & Chen 2024). Consequently, we exclude GW170817 from the dark siren inference and instead combine its posterior with that of the dark sirens a posteriori. We verify that this choice does not introduce any bias by checking that the exclusion of GW170817 does not affect the inferred BNS mass spectrum. We therefore conclude that the a posteriori combination used here is robust and does not impact the final cosmological constraints. Nevertheless, this effect will need to be included when more bright siren events occur.

*Considerations on the Analysis with Galaxy catalog* —Given the catalog's incompleteness, assumptions must be made about the distribution of missing galaxies. We model the expected number density with a redshift-independent Schechter function (see Section 3.2) and assume that missing galaxies are uniformly distributed in comoving volume and isotropically in sky position. While the latter is the most conservative choice, viable alternative assumptions include having them trace the distribution of cataloged galaxies (Finke et al. 2021a) or follow prior knowledge of large-scale structure (Dalang & Baker 2024; Leyde et al. 2024; Dalang et al. 2024; Leyde et al. 2025). Future work could examine the sensitivity of results to these choices and their potential to improve constraints. Possible systematics related to neglecting a putative evolution of the Schechter function might be also considered when using deeper catalogs. Furthermore, in this work, we model the uncertainty on galaxy redshift using a Gaussian distribution. However, this assumption likely represents an oversimplification, as photometric redshift error distributions can be more complex and even vary on a galaxy-by-galaxy basis. More comprehensive approaches, such as the use of full photo-$z$ PDFs, have been explored in the literature (e.g., Palmese et al. 2020). Redshift uncertainties can propagate into derived quantities that depend on redshift, such as K-corrections and absolute magnitudes (or luminosities). Turski et al. (2023) investigated two common error models (Gaussian and modified Lorentzian) and found that, under current levels of uncertainty, the choice of redshift error model does not significantly affect constraints on the Hubble constant. Nonetheless, this conclusion may not hold as future catalogs become more complete and systematic uncertainties are reduced, potentially making the choice of redshift uncertainty model more consequential.

*Considerations on Modified Gravity* —The considerations in the previous paragraphs apply also to modified-gravity analyses. The possible evolution of mass features with redshift could be potentially more impactful in this case, due to the redshift dependence of modified GW propagation.

A specific point to address in this case is the parametrization choice. While the parametrizations adopted here are widespread and cover most known theories, they are not fully universal. To avoid limitations imposed by specific parametrization choices, it would be valuable to consider model-independent approaches to constrain the GW-to-EM distance ratio directly from data.

## 6. CONCLUSIONS

We have presented cosmological constraints obtained from the GWTC-4.0 catalog of GW events detected by the LVK detectors. Our headline results are updated bounds on the Hubble constant: when 141 events with FAR $< 0.25\,\mathrm{yr}^{-1}$ are analyzed as dark sirens with the GLADE+ galaxy catalog, and combined with the bright siren GW170817, we obtain a bound of $H_0 = 76.6^{+13.0}_{-9.5}\,(76.6^{+25.2}_{-14.0})\,\mathrm{km\,s^{-1}\,Mpc^{-1}}$ (using the FullPop-4.0 mass model and applying luminosity weighting to the galaxy catalog). A summary of the different $H_0$ values obtained using different data sets and model assumptions can be seen in Table 1.

The $H_0$ bounds obtained from applying a spectral sirens analysis to GWTC-4.0 are improved relative to those with GWTC-3.0 by $\sim 60\%$. When spectral siren posteriors are combined with those from the bright siren GW170817 the change between GWTC-4.0 and GWTC-3.0 is somewhat reduced, as can be seen by comparing the two orange bounds in Figure 11. This is consistent with GW170817 still being an important component of our constraints.

The comparison of some dark sirens bounds between GWTC-3.0 and GWTC-4.0 is not straightforward given the major upgrades in methodology that we have presented in this work, such as marginalization over mass distribution and merger rate parameters. The bounds shown in Figure 11 that use a fixed or partially-fixed CBC mass distribution (indicated by a star) should be considered as artificially tight for this reason. Comparing the results of this work to that of Mastrogiovanni et al. (2023), which did vary merger rate and BBH mass distribution parameters, one can see that the improvement in the dark sirens results is essentially driven by the improvement in the spectral sirens component (comparing this work to the orange line from Abbott et al. 2023a.)

We have considered here a range of parameterized models for the mass distribution of compact objects. The PLP model is now mildly disfavored relative to the MLTP model. However, the tightest constraints on $H_0$ are obtained using the FullPop-4.0 model which enables the NS and BH distributions to be jointly analyzed. In comparison, the choice of luminosity weight applied to the host-galaxy probabilities has negligible importance. Increasing our certainty about types and locations of features in the CBC mass distribution is a major route to tightening GW bounds on the Hubble constant, via the spectral siren method.

In addition, we have presented bounds on parameterized deviations from GR affecting the GW luminosity distance. A summary of these constraints can be seen in Table 2. Using two commonly-used parametrizations, we obtain the dark siren bounds of $\Xi_0 = 1.2^{+0.8}_{-0.4}\,(1.2^{+2.4}_{-0.5})$ and $c_M = 0.3^{+1.6}_{-1.4}\,(0.3^{+2.9}_{-2.2})$, where the GR limit is recovered in the cases $\Xi_0 = 1$ and $c_M = 0$, respectively. Hence, our re-sults show good consistency with GR on cosmological distance scales. The improvement in constraints on these parameters is quite substantial (∼35–42% when using a narrow $H_0$ prior) relative to previous GW analyses. This is because these constraints in particular utilize higher-redshift GW events and are scarcely impacted by galaxy catalog limitations; so they benefit strongly from the ∼3-fold increase in GW events in GWTC-4.0. Whilst not on an equal footing with EM constraints (for parameters where these are available), this work demonstrates the potential of LVK events to act as an independent probe of cosmological modified gravity. In future, these tests could be honed on selected modified gravity parameters that are inaccessible through galaxy surveys.

| ΛCDM – Dark sirens | | | |
|---|---|---|---|
| Population model | GW candidates | $H_0$ (Dark sirens) [$\mathrm{km\,s^{-1}\,Mpc^{-1}}$] | $H_0$ (Dark + bright sirens) [$\mathrm{km\,s^{-1}\,Mpc^{-1}}$] |
| POWER LAW + PEAK | 137 (138) | $124.8^{+45.2}_{-39.4}$ ($124.8^{+63.2}_{-60.1}$) | $85.9^{+24.4}_{-15.9}$ ($85.9^{+48.8}_{-22.1}$) |
| MULTI PEAK | 137 (138) | $87.3^{+48.0}_{-28.2}$ ($87.3^{+85.6}_{-42.7}$) | $77.0^{+17.6}_{-10.9}$ ($77.0^{+35.3}_{-15.7}$) |
| FULLPOP-4.0 | 141 (142) | $81.6^{+21.5}_{-15.9}$ ($81.6^{+42.0}_{-26.8}$) | $76.6^{+13.0}_{-9.5}$ ($76.6^{+25.2}_{-14.0}$) |

**Table 1.** Values of the Hubble constant measured in this study using different data sets and analysis methods, adopting a uniform prior $H_0 \in \mathrm{U}(10, 200)\,\mathrm{km\,s^{-1}\,Mpc^{-1}}$. Columns are: population mass model assumed in the analysis (first column), number of GW candidates analyzed, including GW170817 in parentheses (second column), $H_0$ dark siren measurement reported as a median with 68.3% and 90% symmetric CI, before (third column) and after (fourth column) combination with the bright siren (GW170817) measurement.

The advances in methodology presented have been possible due to upgrades in computational efficiency of our software pipelines. We anticipate that continuing improvements will open up more flexible parameterized models (e.g., evolving mass-distribution models) in near-future analyses, and will ultimately allow non-parametric analyses (such as binned approaches, splines or Gaussian processes). This will allow us to lift the assumptions of parameterized forms for the mass distribution and luminosity distance ratio in modified-gravity tests.

A limiting factor affecting our present results is the completeness and redshift depth of the galaxy catalog used in our analysis. For most events analyzed here, the bulk of the luminosity-distance posterior distribution lies beyond the redshift range of the K-band GLADE+ catalog for the preferred values of $H_0$. This means that for many events our results are largely uninformed by the distribution of potential galaxy hosts; instead, features in the mass distribution of CBCs dominate the constraints. However, the absence of Virgo during O4a means that O4a candidates have luminosity distance errors that are, on average, larger than those of GWTC-3.0. Hence they are somewhat less informative than O3 events, even when used as spectral sirens.

Fortunately, both of these limiting factors have near-term solutions. The additional contribution of the Virgo detector in the remainder of O4, when combined with the two LIGO detectors, is expected to result in better-localized events, on average. Source localization depends upon the number of GW observatories that can detect a source (Schutz 2011; Abbott et al. 2020b; Abac et al. 2025b). Good sky localization requires data from at least three observatories (Wen & Chen 2010; Singer et al. 2014; Pankow et al. 2020), while volume localization also depends upon the signal-to-noise ratio (Cutler & Flanagan 1994; Del Pozzo et al. 2018), which also improves with more observatories. Precise localization aids both follow-up to search for EM counterparts and cross-referencing with galaxy catalogs (Nissanke et al. 2013b; Gehrels et al. 2016; Singer et al. 2016; Chen & Holz 2016; Pankow et al. 2020). Consequently, prospects for GW measurements of $H_0$ are significantly enhanced when there is a network of at least three comparable-sensitivity GW observatories online (Chen et al. 2018; Kiendrebeogo et al. 2023; Emma et al. 2024; Soni et al. 2024).

Meanwhile, a deeper successor to the GLADE+ galaxy catalog, UpGLADE, is in preparation for future release. The next few years will also see further data releases from Stage IV galaxy surveys such as the Dark Energy Spectroscopic Instrument (Aghamousa et al. 2016), Euclid (Laureijs et al. 2011; Mellier et al. 2025), and the start of observations by the Vera Rubin Observatory (Ivezić et al. 2019). Using data from these surveys is expected to strengthen the informativeness of the galaxy catalog component of the dark sirens method. Forecasts using simulations of the 100 highest signal-to-noise ratio events in O4 and O5 with a complete galaxy catalog are presented in Borghi et al. (2024); these yield bounds on $H_0$ better than 10% in O5 for both photometric and spectroscopic galaxy catalogs. Having this kind of galaxy data in hand will accelerate the progress towards competitive GW bounds on the Hubble constant presented in Figure 11.

It remains possible that future runs of the LVK detectors will yield a further bright siren detection(s), although these are rare. Such an event would likely give GW measurements of $H_0$ a rapid boost in constraining power. However, with or without such events, the methods and analyses of this paper demonstrate that dark and spectral sirens can provide steady progress towards the goals of GW cosmology.

*Data Availability:* All strain data analyzed as part of GWTC-4.0 are publicly available through Gravitational Wave Open Science Center (GWOSC). The details of this data release and information about the digital version of the GWTC are described in detail in Abac et al. (2025l). The data products generated by the methods described within this work are available from Zenodo (LIGO Scientific Collaboration et al. 2025).

| Modified gravity – Dark sirens | |
|---|---|
| Parametrization $\Xi_0$–$n$ | $\Xi_0$ |
| Wide $H_0$-prior | $1.2^{+0.8}_{-0.4}$ $(1.2^{+2.4}_{-0.5})$ |
| Narrow $H_0$-prior | $1.0^{+0.4}_{-0.2}$ $(1.0^{+1.1}_{-0.4})$ |
| Parametrization $\alpha_M$ | $c_M$ |
| Wide $H_0$-prior | $0.3^{+1.6}_{-1.4}$ $(0.3^{+2.9}_{-2.2})$ |
| Narrow $H_0$-prior | $-0.3^{+1.4}_{-1.0}$ $(-0.3^{+2.5}_{-1.5})$ |

**Table 2.** Values of the modified-gravity parameters $\Xi_0$ and $c_M$ constrained assuming two different parametrizations of modified GW propagation. Both analyses are carried out assuming our fiducial population model FULLPOP-4.0 (141 GW candidates) with the dark siren method. We explore wide and narrow priors for $H_0$, i.e., $H_0 \in \mathrm{U}(10, 120)\,\mathrm{km\,s^{-1}\,Mpc^{-1}}$ and $H_0 \in \mathrm{U}(65, 77)\,\mathrm{km\,s^{-1}\,Mpc^{-1}}$, respectively (see the main text for more details). We adopt uniform priors for $\Xi_0 \in \mathrm{U}(0.435, 10)$ and $n \in \mathrm{U}(0.1, 10)$ (not reported in this table), and a uniform prior for $c_M \in \mathrm{U}(-10, 50)$. Columns are: $H_0$ prior chosen for the analysis (first column), modified gravity parameter ($\Xi_0$ or $c_M$) measurement reported as a median with 68.3% (second column, first value) and 90% (second column, second value) symmetric CI. Note that, in contrast to Table 1, the bright siren GW170817 is not used as it is uninformative in this analysis.

## ACKNOWLEDGEMENTS

This material is based upon work supported by NSF's LIGO Laboratory, which is a major facility fully funded by the National Science Foundation. The authors also gratefully acknowledge the support of the Science and Technology Facilities Council (STFC) of the United Kingdom, the Max-Planck-Society (MPS), and the State of Niedersachsen/Germany for support of the construction of Advanced LIGO and construction and operation of the GEO 600 detector. Additional support for Advanced LIGO was provided by the Australian Research Council. The authors gratefully acknowledge the Italian Istituto Nazionale di Fisica Nucleare (INFN), the French Centre National de la Recherche Scientifique (CNRS) and the Netherlands Organization for Scientific Research (NWO) for the construction and operation of the Virgo detector and the creation and support of the EGO consortium. The authors also gratefully acknowledge research support from these agencies as well as by the Council of Scientific and Industrial Research of India, the Department of Science and Technology, India, the Science & Engineering Research Board (SERB), India, the Ministry of Human Resource Development, India, the Spanish Agencia Estatal de Investigación (AEI), the Spanish Ministerio de Ciencia, Innovación y Universidades, the European Union NextGenerationEU/PRTR (PRTR-C17.I1), the ICSC - CentroNazionale di Ricerca in High Performance Computing, Big Data and Quantum Computing, funded by the European Union NextGenerationEU, the Comunitat Autonòma de les Illes Balears through the Conselleria d'Educació i Universitats, the Conselleria d'Innovació, Universitats, Ciència i Societat Digital de la Generalitat Valenciana and the CERCA Programme Generalitat de Catalunya, Spain, the Polish National Agency for Academic Exchange, the National Science Centre of Poland and the European Union - European Regional Development Fund; the Foundation for Polish Science (FNP), the Polish Ministry of Science and Higher Education, the Swiss National Science Foundation (SNSF), the Russian Science Foundation, the European Commission, the European Social Funds (ESF), the European Regional Development Funds (ERDF), the Royal Society, the Scottish Funding Council, the Scottish Universities Physics Alliance, the Hungarian Scientific Research Fund (OTKA), the French Lyon Institute of Origins (LIO), the Belgian Fonds de la Recherche Scientifique (FRS-FNRS), Actions de Recherche Concertées (ARC) and Fonds Wetenschappelijk Onderzoek - Vlaanderen (FWO), Belgium, the Paris Île-de-France Region, the National Research, Development and Innovation Office of Hungary (NKFIH), the National Research Foundation of Korea, the Natural Sciences and Engineering Research Council of Canada (NSERC), the Canadian Foundation for Innovation (CFI), the Brazilian Ministry of Science, Technology, and Innovations, the International Center for Theoretical Physics South American Institute for Fundamental Research (ICTP-SAIFR), the Research Grants Council of Hong Kong, the National Natural Science Foundation of China (NSFC), the Israel Science Foundation (ISF), the US-Israel Binational Science Fund (BSF), the Leverhulme Trust, the Research Corporation, the National Science and Technology Council (NSTC), Taiwan, the United States Department of Energy, and the Kavli Foundation. The authors gratefully acknowledge the support of the NSF, STFC, INFN and CNRS for provision of computational resources.

This work was supported by MEXT, the JSPS Leading-edge Research Infrastructure Program, JSPS Grant-in-Aid for Specially Promoted Research 26000005, JSPS Grant-in-Aid for Scientific Research on Innovative Areas 2402: 24103006, 24103005, and 2905: JP17H06358, JP17H06361 and JP17H06364, JSPS Core-to-Core Program A. Advanced Research Networks, JSPS Grants-in-Aid for Scientific Research (S) 17H06133 and 20H05639, JSPS Grant-in-Aid for Transformative Research Areas (A) 20A203: JP20H05854, the joint research program of the Institute for Cosmic Ray Research, University of Tokyo, the National Research Foundation (NRF), the Computing Infrastructure Project of the Global Science experimental Data hub Center (GSDC) at KISTI, the Korea Astronomy and Space Science Institute (KASI), the Ministry of Science and ICT (MSIT) in Korea, Academia Sinica (AS), the AS Grid Center (ASGC) and the National Science and Technology Council (NSTC) in Taiwan under grants including the Science Vanguard Research Program, the Advanced Technology Center (ATC) of NAOJ, and the Mechanical Engineering Center of KEK.

Additional acknowledgements for support of individual authors may be found in the following document: `https://dcc.ligo.org/LIGO-M2300033/public`.
 We request that citations to this article use 'A. G. Abac *et al.* (LIGO-Virgo-KAGRA Collaboration), ...' or similar phrasing, depending on journal convention.

*Software:* Calibration of the LIGO strain data was performed with GstLAL-based calibration software pipeline (Viets et al. 2018). Data-quality products and event-validation results were computed using the DMT (John Zweizig 2006), DQR (LIGO Scientific Collaboration and Virgo Collaboration 2018), DQSEGDB (Fisher et al. 2020), GWDETCHAR (Urban et al. 2021), HVETO (Smith et al. 2011), IDQ (Essick et al. 2020), OMICRON (Robinet et al. 2020) and PYTHONVIRGOTOOLS (Virgo Collaboration 2021) software packages and contributing software tools. Analyses in this catalog relied upon the LALSUITE software library (LIGO Scientific Collaboration et al. 2018; Wette 2020). The detection of the signals and subsequent significance evaluations in this catalog were performed with the GSTLAL-based inspiral software pipeline (Messick et al. 2017; Sachdev et al. 2019; Hanna et al. 2020; Cannon et al. 2020), with the MBTA pipeline (Adams et al. 2016; Aubin et al. 2021), and with the PYCBC (Usman et al. 2016; Nitz et al. 2017; Davies et al. 2020) and the CWB (Klimenko & Mitselmakher 2004; Klimenko et al. 2011, 2016) packages. Estimates of the noise spectra and glitch models were obtained using BAYESWAVE (Cornish & Littenberg 2015; Littenberg et al. 2016; Cornish et al. 2021). Source-parameter estimation was performed with the BILBY library (Ashton et al. 2019; Romero-Shaw et al. 2020) using the DYNESTY nested sampling package (Speagle 2020). PESUMMARY was used to postprocess and collate parameter-estimation results (Hoy & Raymond 2021). The various stages of the parameter-estimation analysis were managed with the ASIMOV library (Williams et al. 2023). Plots were prepared with MATPLOTLIB (Hunter 2007), SEABORN (Waskom 2021) and GWPY (Macleod et al. 2021). NUMPY (Harris et al. 2020) and SCIPY (Virtanen et al. 2020) were used in the preparation of the manuscript. We made use of the software packages `gwcosmo`, see https://git.ligo.org/lscsoft/gwcosmo/-/releases/v3.0.0 and `icarogw`, see https://github.com/simone-mastrogiovanni/icarogw/releases/tag/v2.0.3.

# APPENDIX

## A. DETAILS ON THE LIKELIHOOD EVALUATION

In this appendix, we provide more details on the likelihood evaluation. Assuming spins are neglected, the integrals in Equation (1) span five dimensions, encompassing the component masses, sky position, and luminosity distance. A common strategy in GW population studies is to evaluate these integrals via MC integration. The posterior distributions for individual events $p\big(\boldsymbol{\theta}_i^{\rm det}|d_i\big)$ are provided as discrete sets of samples, which can be repurposed to compute MC sums.

In `icarogw` the evaluation of the integrals at the numerator of the posterior in Equation (1) uses MC integration. Consider a normalized probability distribution $p(x)$ with $\int {\rm d}x\, p(x) = 1$, and the expectation value of a function $f(x)$, i.e.,

$$\langle f \rangle = \int {\rm d}x\, f(x)\, p(x)\,. \tag{A1}$$

This can be approximated with the following MC estimator:

$$\langle \hat{f} \rangle = \frac{1}{N_{\rm draw}} \sum_{k=1}^{N_{\rm draw}} f(x_k)\,, \tag{A2}$$

where the points $x_k$ are drawn from $p(x)$, for a total of $N_{\rm draw}$ draws. In the case of the posterior distribution in Equation (1), for each observed event labeled by $i$, we are given $N_{{\rm s},i}$ samples $\boldsymbol{\theta}_{k,i} \sim p(\boldsymbol{\theta}_i^{\rm det}|d_i)$ from the corresponding posterior. The estimator of each integral in the product sign (denoted here as $\hat{\mathcal{L}}_i$) is therefore:

$$\hat{\mathcal{L}}_i = \frac{1}{N_{{\rm s},i}} \sum_{k=1}^{N_{{\rm s},i}} \frac{1}{\pi_{\rm PE}(\boldsymbol{\theta}_{i,k}^{\rm det})} \left[ \left| \frac{{\rm d}\boldsymbol{\theta}_i^{\rm det}(\boldsymbol{\theta}_{i,k}, \boldsymbol{\Lambda}_c)}{{\rm d}\boldsymbol{\theta}_i} \right|^{-1} p_{\rm pop}(\boldsymbol{\theta}_{i,k}|\boldsymbol{\Lambda}) \right]_{\boldsymbol{\theta}_{i,k}=\boldsymbol{\theta}_i(\boldsymbol{\theta}_{i,k}^{\rm det}, \boldsymbol{\Lambda}_c)}\,. \tag{A3}$$

The advantage of MC integration lies in its ability to handle high-dimensional integrals efficiently, provided sufficient convergence is achieved. Specifically, MC integration introduces sampling variance that must be carefully managed (Farr 2019; Essick & Farr 2022; Talbot & Golomb 2023). The variance associated to the estimator in Equation (A2) can be written as

$$\mathrm{var}(\langle f \rangle) = \frac{1}{N_{\mathrm{draw}}} \left[ \langle f^2 \rangle - \langle f \rangle^2 \right] . \tag{A4}$$

To ensure accuracy, one typically requires the variance to be small enough. A common diagnostic is the effective sample size (Farr 2019),

$$n_{\mathrm{eff}} \equiv N_{\mathrm{draw}} \frac{\langle f \rangle^2}{\langle f^2 \rangle} . \tag{A5}$$

We adopt the default choice $n_{\mathrm{eff,PE}} > 10$ for the `icarogw` analyses presented in this work. Here, the suffix PE denotes the threshold for the estimator in Equation (A3).

In contrast, `gwcosmo` employs a one-dimensional kernel density estimation (KDE) method. This approach first re-weights the posterior samples $\boldsymbol{\theta}_{i,k}$ based on a given population model. For each sample the reweighting is calculated as

$$w_{i,k} = \frac{p_{\mathrm{pop}}(\boldsymbol{\theta}_{i,k}|\boldsymbol{\Lambda})}{\pi_{\mathrm{PE}}(\boldsymbol{\theta}_{i,k}|\boldsymbol{\Lambda})}. \tag{A6}$$

This is followed by the construction of a redshift kernel within each sky pixel $\Omega$. For each re-weighted sample a redshift $z_{i,k}$ is calculated from its luminosity distance $D^{\mathrm{GW}}_{\mathrm{L},i,k}$ given $\boldsymbol{\Lambda}_c$. The KDE is then used to construct the redshift probability distribution of the event $i$ in the pixel $\Omega$,

$$p(z|d_i, \Omega, \boldsymbol{\Lambda}) \approx \sum_{k \in \Omega_j} w'_{i,k} K(z - z_{i,k}, h), \tag{A7}$$

with $K(\cdot, h)$ being a kernel with bandwidth $h$ and $w'_{i,k}$ the normalized weights. Including the catalog information explicitly, the likelihood for a single event becomes

$$\mathcal{L}_i(\boldsymbol{\Lambda}) \propto \sum_j p(\Omega_j|d_i, \boldsymbol{\Lambda}) \int \mathrm{d}z \, p(z|\Omega_j, \boldsymbol{\Lambda}) \, \frac{\psi(z|\boldsymbol{\Lambda})}{1+z} \, p(z|d_i, \Omega_j, \boldsymbol{\Lambda}) \, . \tag{A8}$$

In this equation $p(z|d_i, \Omega_j, \boldsymbol{\Lambda})$ is the population-weighted KDE in the pixel $\Omega_j$, $p(\Omega_j|d_i, \boldsymbol{\Lambda})$ is the per-pixel probability given event $d_i$, and $p(z|\Omega_j, \boldsymbol{\Lambda})$ is the prior from the catalog information. Here, the sum over pixels $j$ effectively discretizes the integral over solid angle, so that each pixel $\Omega_j$ covers a finite $\Delta\Omega_j$.

While this approach effectively incorporates galaxy catalog information while avoiding the issues of numerical stability found with the MC integration method, it is susceptible to systematic uncertainties if re-weighted sample sizes are too small.

The selection function $\xi(\lambda)$ is estimated by both `icarogw` and `gwcosmo` through MC reweighting of simulated GW injections campaigns (Tiwari 2018; Farr 2019). A number of $N_{\mathrm{draw}}$ injections are generated with parameters $\boldsymbol{\theta}^{\mathrm{det}}_k$ drawn from a reference distribution $p_{\mathrm{draw}}(\boldsymbol{\theta}^{\mathrm{det}}_k)$. Then, the same detection threshold used for building the observed GW catalog is applied. As a result, the detection probability is set to $P(\mathrm{det}|\boldsymbol{\theta}^{\mathrm{det}}_k) = 1$ for the $N_{\mathrm{det}}$ injections that pass the threshold, and $P(\mathrm{det}|\boldsymbol{\theta}^{\mathrm{det}}_k) = 0$ for the rest. The MC estimator of the integral in Equation (2) is then

$$\hat{\xi}(\lambda) = \frac{1}{N_{\mathrm{draw}}} \sum_{k=1}^{N_{\mathrm{det}}} \frac{1}{p_{\mathrm{draw}}(\boldsymbol{\theta}^{\mathrm{det}}_k)} \left[ \left| \frac{\mathrm{d}\boldsymbol{\theta}^{\mathrm{det}}_i(\boldsymbol{\theta}_k, \boldsymbol{\Lambda}_c)}{\mathrm{d}\boldsymbol{\theta}_i} \right|^{-1} p_{\mathrm{pop}}(\boldsymbol{\theta}_k|\boldsymbol{\Lambda}) \right]_{\boldsymbol{\theta}_k = \boldsymbol{\theta}_k(\boldsymbol{\theta}^{\mathrm{det}}_k, \boldsymbol{\Lambda}_c)} . \tag{A9}$$

More details about the set of injections used in this study can be found in Essick et al. (2025); Abac et al. (2025c).

For the selection function $\xi(\lambda)$, we follow the condition $n_{\mathrm{eff,inj}} > 4N_{\mathrm{obs}}$ to ensure sufficient coverage of the parameter space by the injections (Farr 2019).

When computing the per-event likelihoods entering the posterior in Equation (1), `icarogw` additionally imposes $n_{\mathrm{eff,PE}} > 10$ for each term; as an alternative diagnostic, Talbot & Golomb (2023) suggests using the total variance of the population log-likelihood:

$$\mathrm{var}(\ln \mathcal{L}) = \sum_{i=1}^{N_{\mathrm{obs}}} \frac{\mathrm{var}(\mathcal{L}_i)}{\hat{\mathcal{L}}_i^2} + N_{\mathrm{obs}}^2 \frac{\mathrm{var}(\xi)}{\hat{\xi}^2} . \tag{A10}$$

A threshold $\mathrm{var}(\ln \mathcal{L}) < 1$ is found to be sufficient for reliable inference. We also assess the impact of this condition when using `icarogw`.

In practice, these criteria act as regularization tools: they restrict the sampler from exploring regions where MC estimates are unreliable. These conditions are inherently data-dependent and effectively modify the likelihood surface. Nevertheless, they stem from numerical limitations, not the likelihood's theoretical form.

## B. LUMINOSITY FUNCTION AND COMPLETENESS ESTIMATES

### B.1. *Schechter Luminosity Function*

The out-of-catalog term in the redshift prior term requires us to estimate the incompleteness of our galaxy catalog due to the flux limits of imaging or spectroscopic surveys they are obtained from. The luminosity function of galaxies which quantifies the number density of galaxies in the Universe is used to quantify this incompleteness. We use a parameterized Schechter function to describe the luminosity function such that

$$\Phi(L,\lambda)\mathrm{d}L = \phi_* \left[\frac{L}{L_*}\right]^{\alpha} \exp\left[-\frac{L}{L_*}\right] \frac{\mathrm{d}L}{L_*}\,, \tag{B11}$$

where the parameters $\lambda$ consist of: the normalization $\phi_*$, representing the galaxy number density at the characteristic luminosity, the characteristic luminosity $L_*$, where the function transitions from a power-law to an exponential cutoff, and the faint-end slope $\alpha$ that determines the abundance of low-luminosity galaxies.

The luminosity function can be further expressed in terms of magnitudes by using

$$\mathrm{Sch}(M,\lambda)\mathrm{d}M = \Phi(L,\lambda)\mathrm{d}L\,, \tag{B12}$$

and the relation between absolute magnitude and luminosity. To express the luminosity function in terms of absolute magnitude $M$, we use the standard relation between luminosity and magnitude:

$$\frac{L}{L_*} = 10^{0.4(M_*-M)}\,, \tag{B13}$$

where $M_*$ is the characteristic magnitude corresponding to $L_*$. Substituting this into Equation (B11), we obtain the Schechter function in terms of magnitude:

$$\mathrm{Sch}(M;\lambda)\,\mathrm{d}M = 0.4\ln(10)\,\phi_*\,10^{0.4(\alpha+1)(M_*-M)}\,\exp\left[-10^{0.4(M_*-M)}\right]\,\mathrm{d}M\,, \tag{B14}$$

where $M_*$ is the characteristic magnitude corresponding to $L_*$. In summary, $\lambda = \{\alpha, \phi_*, M_*\}$. Finally, we assume the Schechter function to be defined in the interval $M_{\rm min} \le M \le M_{\rm max}$, and being zero outside.

In practice, the Schechter parameters are often provided assuming $H_0 = 100\,\mathrm{km\,s^{-1}\,Mpc^{-1}}$. To convert them to a cosmology with arbitrary $H_0$, the following scaling relations apply:

$$M_*(h) = M_*(h=1) + 5\log_{10} h\,, \tag{B15}$$

$$\phi_*(h) = \phi_*(h=1)\,h^3\,, \tag{B16}$$

where we have defined $h = H_0/100\,\mathrm{km^{-1}\,s\,Mpc}$. We also assume the luminosity function of galaxies to be non-evolving, i.e., independent of the redshift.

For the K-band luminosity function, we use $M_*^K - 5\log h = -23.55$, $\alpha = -1.09$ and $\phi_* = 1.16\times10^{-2} h^3\,\mathrm{Mpc^{-3}}$ based on the results from the 2MASS galaxy survey (Kochanek et al. 2001). The parameter $\phi_*$ can be reabsorbed in a normalization factor, therefore its value does not impact our results (Mastrogiovanni et al. 2023; Gray et al. 2023).

Finally, we discuss the calculation of the out-of-catalog contribution in Equation (12). One has

$$\begin{aligned}\frac{\mathrm{d}N^{\rm eff}_{\rm gal,out}(z,\Omega)}{\mathrm{d}z\mathrm{d}\Omega} &= \frac{\mathrm{d}V_c(z,\Omega)}{\mathrm{d}z\mathrm{d}\Omega}\int_{M_{\rm thr}(z,m_{\rm thr}(\Omega))}^{M_{\rm max}} \mathrm{d}M\,10^{-0.4\epsilon(M-M_*)}\mathrm{Sch(M,}\lambda) = \\ &= \frac{\mathrm{d}V_c(z,\Omega)}{\mathrm{d}z\mathrm{d}\Omega}\,\phi_*\int_{L_{\rm thr}(z,m_{\rm thr}(\Omega))/L*}^{L_{\rm max}/L*}\mathrm{d}x\,x^{\alpha+\epsilon}e^{-x}\,,\end{aligned} \tag{B17}$$

where from the first to the second line we changed variables to $x \equiv L/L_*$ and used Equation (B11).

For $\alpha > -1$, one can compute the integral in the second line of Equation (B17) as the difference of incomplete gamma functions, obtaining:

$$\frac{\mathrm{d}N^{\rm eff}_{\rm gal,out}(z,\Omega)}{\mathrm{d}z\mathrm{d}\Omega} = \frac{\mathrm{d}V_c(z,\Omega)}{\mathrm{d}z\mathrm{d}\Omega}\phi_*\left[\Gamma_{\rm inc}(\alpha+\epsilon+1,x_{\rm thr}) - \Gamma_{\rm inc}(\alpha+\epsilon+1,x_{\rm max})\right] \qquad \Big(\alpha+\epsilon > -1\Big)\,, \tag{B18}$$

where $x_{\rm thr} = 10^{0.4[M_*-M_{\rm thr}(z,m_{\rm thr}(\Omega))]}$, $x_{\rm max} = 10^{0.4(M_*-M_{\rm max})}$.

However, for values $\alpha+\epsilon < -1$ (as is the case here), the incomplete Gamma functions in square brackets on the right-hand side of Equation (B18) are divergent, while their difference and the integral in the second line of Equation (B17) remain finite. In this case, we compute the integral directly via numerical integration.

### B.2. *Over-density of Galaxies and Incompleteness of the Galaxy Catalog*

For any given GW event, the over-density of galaxies towards the line-of-sight to an event can be defined as

$$\mathcal{O}(z,\Omega;\lambda) = \left[\frac{\mathrm{d}N_{\rm gal}(z,\Omega)}{\mathrm{d}z\mathrm{d}\Omega} + \frac{\mathrm{d}V_{\rm c}(z,\Omega)}{\mathrm{d}z\mathrm{d}\Omega}\int_{M_{\rm thr}}^{M_{\rm max}(z,\Omega)} \mathrm{d}M\,\mathrm{Sch}(M;\lambda)\right]\left[\frac{\mathrm{d}V_{\rm c}(z,\Omega)}{\mathrm{d}z\mathrm{d}\Omega}\int_{-\infty}^{M_{\rm max}(z,\Omega)} \mathrm{d}M\,\mathrm{Sch}(M;\lambda)\right]^{-1} . \quad \text{(B19)}$$

The first term in the numerator is evaluated based on the galaxies present in the galaxy catalog, while the integrals are performed based on the assumed luminosity function parameters. If this ratio is greater (less) than unity, it indicates an over(under)-density of galaxies along the line of sight toward the GW source. We compute the mean of this quantity for all pixels at a given redshift, and then list the minimum and maximum of these values over the range of redshifts encompassing the 90% credible localization intervals in Table 8.

In Table 8, we list the minimum and maximum values of this quantity within the redshift range that covers the 90% credible localization for each event.

The incompleteness of the galaxy catalog in a given direction for the case of luminosity weighting is defined as:

$$\mathcal{I}(z,\Omega;\lambda) = \left[\frac{\mathrm{d}V_{\rm c}(z,\Omega)}{\mathrm{d}z\mathrm{d}\Omega}\int_{-\infty}^{M_{\rm thr}(z,\Omega)} \mathrm{d}M\,10^{-0.4(M-M_*)}\mathrm{Sch}(M;\lambda)\right]\left[\frac{\mathrm{d}V_{\rm c}(z,\Omega)}{\mathrm{d}z\mathrm{d}\Omega}\int_{-\infty}^{M_{\rm max}(z,\Omega)} \mathrm{d}M\,10^{-0.4(M-M_*)}\mathrm{Sch}(M;\lambda)\right]^{-1} . \quad \text{(B20)}$$

At every redshift, we compute the median of this quantity over all pixels corresponding to the 90% localization. We report values corresponding to the minimum and maximum redshift that encompasses the 90% credible intervals as a range in Table 8.

## C. MASS AND MERGER RATE MODELS

In this appendix, we describe the population models that we have considered in this paper, both in terms of mass and merger rate of CBCs. All the adopted population models are composed of various simple mathematical functions which we describe below.

The truncated power law $\mathcal{P}(x|x_{\rm min},x_{\rm max},\alpha)$ is described by slope $\alpha$, and lower and upper bounds $x_{\rm min},x_{\rm max}$ where the distribution shows hard cutoffs,

$$\mathcal{P}(x|x_{\rm min},x_{\rm max},\alpha) \propto \begin{cases} x^\alpha & (x_{\rm min}\le x\le x_{\rm max}) \\ 0 & \text{otherwise} \end{cases} . \quad \text{(C21)}$$

The truncated Gaussian distribution with mean $\mu$ and standard deviation $\sigma$ with support at $[a,b]$ is defined as

$$\mathcal{G}(x|\mu,\sigma,a,b) = \frac{G(a,b)}{\sigma\sqrt{2\pi}}\exp\left[-\frac{(x-\mu)^2}{2\sigma^2}\right] , \quad \text{(C22)}$$

with the normalization $G(a,b)$ implicitly determined through

$$\int_a^b \mathcal{G}(x|\mu,\sigma,a,b)\mathrm{d}x = 1 . \quad \text{(C23)}$$

In the PLP and MLTP population models, we apply a smoothing function at low masses ($m=m_{\rm min}$), also called high-pass filter, so that

$$p(m_1,m_2|\mathbf{\Lambda}) \propto p(m_1|\mathbf{\Lambda})S_{\rm h}(m_1|m_{\rm min},\delta_{\rm m})p(m_2|m_1,\mathbf{\Lambda})S_{\rm h}(m_2|m_{\rm min},\delta_{\rm m}) . \quad \text{(C24)}$$

Here, $S_{\rm h}(m|m_{\rm min},\delta_{\rm m})$ is a sigmoid-like smoothing function that rises from 0 to 1 over the interval $(m_{\rm min}, m_{\rm min}+\delta_{\rm m})$ given by

$$S_{\rm h}(m|m_{\rm min},\delta_{\rm m}) = \begin{cases} 0 & (m<m_{\rm min}) \\ \left[f(m-m_{\rm min},\delta_{\rm m})+1\right]^{-1} & (m_{\rm min}\le m<m_{\rm min}+\delta_{\rm m}) , \\ 1 & (m\geqslant m_{\rm min}+\delta_{\rm m}) \end{cases} \quad \text{(C25)}$$

where $\delta_{\rm m}$ is a smoothing scale parameter and

$$f(m',\delta_{\rm m}) = \exp\left(\frac{\delta_{\rm m}}{m'} + \frac{\delta_{\rm m}}{m'-\delta_{\rm m}}\right) . \quad \text{(C26)}$$

The notation in Equation (C24) has been slightly misused because, due to the smoothing functions, the marginal distribution $p(m_{1,2}|\mathbf{\Lambda})$ is no longer obtained by marginalization of Equation (C24) over $m_{2,1}$.

POWER LAW + PEAK

| Parameter | Description | Prior |
|---|---|---|
| $\alpha$ | Spectral index of primary mass power law | $\mathrm{U}(1.5, 12)$ |
| $\beta$ | Spectral index of secondary mass power law | $\mathrm{U}(-4, 12)$ |
| $m_{\min}$ | Minimum primary mass $[M_\odot]$ | $\mathrm{U}(2, 10)$ |
| $m_{\max}$ | Maximum primary mass $[M_\odot]$ | $\mathrm{U}(50, 200)$ |
| $\delta_{\mathrm{m}}$ | Smoothing parameter $[M_\odot]$ | $\mathrm{U}(10^{-3}, 10)$ |
| $\mu_{\mathrm{g}}$ | Location of the peak $[M_\odot]$ | $\mathrm{U}(20, 50)$ |
| $\sigma_{\mathrm{g}}$ | Width of the peak $[M_\odot]$ | $\mathrm{U}(0.4, 10)$ |
| $\lambda_{\mathrm{g}}$ | Fraction of events in the peak | $\mathrm{U}(0, 1)$ |

**Table 3.** Summary of the hyperparameters priors used for the PLP model. U stands for uniform prior.

The PLP model (Talbot & Thrane 2018) describes the primary mass distribution as a combination of two components: a truncated power law with slope $-\alpha$, defined between a minimum mass $m_{\min}$ and a maximum mass $m_{\max}$, and a truncated Gaussian distribution with mean $\mu_{\mathrm{g}}$ and standard deviation $\sigma_{\mathrm{g}}$ defined in the range $[m_{\min}, m_{\max}]$, with the parameter $\lambda_{\mathrm{g}}$ denoting the fraction of events belonging to the Gaussian component; the secondary mass is modeled with a separate power law, defined between $m_{\min}$ and $m_1$ and characterized by the slope $\beta$,

$$\begin{aligned} p(m_1|\mathbf{\Lambda}) = & \; (1-\lambda_{\mathrm{g}})\,\mathcal{P}\big(m_1 \mid m_{\min}, m_{\max}, -\alpha\big) \\ & + \lambda_{\mathrm{g}}\,\mathcal{G}\big(m_1 \mid \mu_{\mathrm{g}}, \sigma_{\mathrm{g}}, m_{\min}, m_{\max}\big)\,, \end{aligned} \tag{C27}$$

$$p(m_2|m_1, \mathbf{\Lambda}) = \mathcal{P}(m_2|m_{\min}, m_1, \beta)\,. \tag{C28}$$

We report the parameter priors of the PLP model in Table 3.

The MLTP model (Abbott et al. 2021c) is the direct extension of the PLP model. The primary mass distribution is based on Equation (C24) and consists of one power law combined with two Gaussian peaks,

$$\begin{aligned} p(m_1|\mathbf{\Lambda}) = & \; (1-\lambda_{\mathrm{g}})\mathcal{P}(m_1|m_{\min}, m_{\max}, -\alpha) \\ & + \lambda_{\mathrm{g}}\lambda_{\mathrm{g}}^{\mathrm{low}}\mathcal{G}(m_1|\mu_{\mathrm{g}}^{\mathrm{low}}, \sigma_{\mathrm{g}}^{\mathrm{low}}, m_{\min}, m_{\max}) \\ & + \lambda_{\mathrm{g}}(1-\lambda_{\mathrm{g}}^{\mathrm{low}})\mathcal{G}(m_1|\mu_{\mathrm{g}}^{\mathrm{high}}, \sigma_{\mathrm{g}}^{\mathrm{high}}, m_{\min}, m_{\max})\,, \end{aligned} \tag{C29}$$

where the two means of the Gaussian components are given by $\mu_{\mathrm{g}}^{\mathrm{low}}$ and $\mu_{\mathrm{g}}^{\mathrm{high}}$, and their respective standard deviations by $\sigma_{\mathrm{g}}^{\mathrm{low}}$ and $\sigma_{\mathrm{g}}^{\mathrm{high}}$. Once again, the respective fraction of events in the first and second Gaussian peak are given by $\lambda_{\mathrm{g}}$ and $\lambda_{\mathrm{g}}^{\mathrm{low}}$. The secondary mass distribution is still modeled as in Equation (C28). We report the parameter priors of the MLTP model in Table 4.

The FULLPOP-4.0 model (Abac et al. 2025f) spans the full mass distribution of CBCs and therefore includes BNSs, NSBHs, and BBHs. It consists of a broken power-law continuum, Gaussian peaks, and smoothing at the edges of the distribution. It additionally includes notch filters to allow for both lower and upper mass gaps (Ozel et al. 2010; Farr et al. 2011; Fryer et al. 2012; Belczynski et al. 2012). The depth of these mass gaps is a free parameter: the data can determine whether the rate goes to zero within the gap or if the gap is partially or totally filled. This model is an extension of the POWERLAW–DIP–BREAK model described in Fishbach et al. (2020); Farah et al. (2022), and is the same as the FULLPOP-4.0 model described in Abac et al. (2025f). The primary and secondary mass distributions of FULLPOP-4.0 are described by the following equation,

$$\begin{aligned} p(m|\mathbf{\Lambda}) = & \Big[(1-\lambda_{\mathrm{g}})\mathcal{B}(m|m_{\min}, m_{\max}, \alpha_1, \alpha_2, b) + \lambda_{\mathrm{g}}\lambda_{\mathrm{g}}^{\mathrm{low}}\mathcal{G}(m|\mu_{\mathrm{g}}^{\mathrm{low}}, \sigma_{\mathrm{g}}^{\mathrm{low}}, m_{\min}, m_{\max}) \\ & + \lambda_{\mathrm{g}}(1-\lambda_{\mathrm{g}}^{\mathrm{low}})\mathcal{G}(m|\mu_{\mathrm{g}}^{\mathrm{high}}, \sigma_{\mathrm{g}}^{\mathrm{high}}, m_{\min}, m_{\max})\Big]\,. \end{aligned} \tag{C30}$$

In Equation (C30), the distributions $\mathcal{G}$ are the same Gaussian components as for the MLTP mass model, hence the hyperparameters governing the fractions of events in the peaks or the position of the mass features are named similarly. The function $\mathcal{B}$ is a broken power law constructed from two truncated power-law distributions that are joined at the point $b$ such that:

$$b = \frac{m_{\mathrm{break}} - m_{\min}}{m_{\max} - m_{\min}}, \tag{C31}$$

with

$$m_{\rm break} = 0.5(m_{\rm d}^{\rm low} + m_{\rm d}^{\rm high} + \delta_{\rm d}^{\rm min} - \delta_{\rm d}^{\rm max}), \tag{C32}$$

namely, the center of the mass gap between the NS and BH regions. The probability density distribution of the broken power law $\mathcal{B}$ for the primary and secondary masses is hence written as

$$\mathcal{B}(m|m_{\rm min}, m_{\rm max}, \alpha_1, \alpha_2, b) = \frac{1}{N_{\mathcal{B}}}\left[\mathcal{P}(m|m_{\rm min}, b, -\alpha_1) + \frac{\mathcal{P}(b|m_{\rm min}, b, -\alpha_1)}{\mathcal{P}(b|b, m_{\rm max}, -\alpha_2)}\mathcal{P}(m|b, m_{\rm max}, -\alpha_2)\right], \tag{C33}$$

where $N_{\mathcal{B}}$ is the normalization factor. Finally, the primary and secondary mass distributions are combined using low-, high-, and notch filters to construct the FULLPOP-4.0 mass model. The total distribution is then given by the product of $p(m|\mathbf{\Lambda})$ with a high-pass filter $S_{\rm h}$ at $m_{\rm min}$ governed by $\delta_{\rm m}^{\rm min}$, a low-pass filter $S_{\rm l}$ at $m_{\rm max}$ governed by $\delta_{\rm m}^{\rm max}$, and a notch filter $S_{\rm n}$ between $m_{\rm d}^{\rm low}$ and $m_{\rm d}^{\rm high}$, governed by $\delta_{\rm d}^{\rm min}$ and $\delta_{\rm d}^{\rm max}$. The low-pass filter is constructed similarly to the high-pass filter,

$$S_{\rm l}\left(m|m_{\rm max}, \delta_{\rm m}\right) = \begin{cases} 0 & (m > m_{\rm max}) \\ \left[f\left(m_{\rm max} - m, \delta_{\rm m}\right) + 1\right]^{-1} & \left(m_{\rm max} - \delta_{\rm m} \leq m < m_{\rm max}\right), \\ 1 & (m \leq m_{\rm max} - \delta_{\rm m}) \end{cases} \tag{C34}$$

where the function $f(.)$ is the same as the one defined above in Equation (C26). The notch filter is defined as a combination of the low- and high-pass filters,

$$S_{\rm n}(m|m_{\rm min}, \delta_{\rm m}^{\rm min}, m_{\rm max}, \delta_{\rm m}^{\rm max}) = 1 - A\, S_{\rm l}(m|m_{\rm max}, \delta_{\rm m}^{\rm max}) S_{\rm h}(m|m_{\rm min}, \delta_{\rm m}^{\rm min}), \tag{C35}$$

where $A \in [0, 1]$ is a parameter governing the deepness of the dip. Following the above definitions and using Equation (C30), we can define

$$p_{\rm S}(m_1|\mathbf{\Lambda}) \propto p(m_1|\mathbf{\Lambda}) S_{\rm h}(m_1|m_{\rm min}, \delta_{\rm m}^{\rm min}) S_{\rm l}(m_1|m_{\rm max}, \delta_{\rm m}^{\rm max}) S_{\rm n}(m_1|m_{\rm d}^{\rm low}, \delta_{\rm d}^{\rm min}, m_{\rm d}^{\rm high}, \delta_{\rm d}^{\rm max}), \tag{C36}$$

and similarly for the secondary mass. The normalized joint probability density is then given by the product of both $p_{\rm S}(m_1|\mathbf{\Lambda})$ and $p_{\rm S}(m_2|\mathbf{\Lambda})$ with the pairing function $f(m_1, m_2|\mathbf{\Lambda})$ defined by

$$f(m_1, m_2|\beta_1, \beta_2, m_{\rm break}) = \begin{cases} \left(\dfrac{m_2}{m_1}\right)^{\beta_1} & (m_2 < m_{\rm break}) \\ \left(\dfrac{m_2}{m_1}\right)^{\beta_2} & (m_2 \geq m_{\rm break}), \end{cases} \tag{C37}$$

so that

$$p(m_1, m_2|\mathbf{\Lambda}) = \frac{1}{N_S} p_{\rm S}(m_1|\mathbf{\Lambda}) p_{\rm S}(m_2|\mathbf{\Lambda}) f(m_1, m_2|\mathbf{\Lambda}), \tag{C38}$$

where $N_S$ is a normalization constant to be computed numerically. Note that due to the pairing formalism the marginal mass distributions are different from the marginalized joint distribution (Abac et al. 2025f). The full set of parameter priors, descriptions and notations, are shown in Table 5.

Finally, we describe the merger rate evolution as a function of the redshift, modeled with a Madau–Dickinson parametrization (Madau & Dickinson 2014), which is characterized by parameters $\{\gamma, \kappa, z_{\rm p}\} \in \mathbf{\Lambda}$, where $\gamma$ and $\kappa$ are the power-law slopes respectively before and after the redshift turning point between the two power-law regimes, $z_{\rm p}$. Explicitly,

$$\psi\left(z|\gamma, \kappa, z_{\rm p}\right) = \left[1 + \left(1 + z_{\rm p}\right)^{-\gamma-\kappa}\right] \frac{(1+z)^{\gamma}}{1 + \left[(1+z)/\left(1+z_{\rm p}\right)\right]^{\gamma+\kappa}}. \tag{C39}$$

The parameter priors are shown in Table 6.

## D. SPECTRAL SIREN RESULTS

In this Appendix we report details on results using the spectral sirens method. Figure 12 displays the marginalized posteriors for the Hubble constant estimated with each of the three mass models considered. As for the galaxy catalog results (see Figure 4), we show the marginalized posterior for $H_0$ from the spectral siren analysis, with different mass models, as well as the posterior for the FULLPOP-4.0 model combined with the bright siren GW170817 (blue curve). The analyses using the PLP, MLTP, and FULLPOP-4.0 mass models yield $H_0 = 112.7^{+51.0}_{-35.9}(112.7^{+74.0}_{-55.1})\,{\rm km\,s^{-1}\,Mpc^{-1}}$, $H_0 = 77.1^{+40.8}_{-26.3}(77.1^{+83.5}_{-39.1})\,{\rm km\,s^{-1}\,Mpc^{-1}}$, and $H_0 = 76.4^{+23.0}_{-18.1}(76.4^{+41.2}_{-28.6})\,{\rm km\,s^{-1}\,Mpc^{-1}}$, respectively. We observe that, as for the dark siren analysis, the best precision

MULTI PEAK

| Parameter | Description | Prior |
|---|---|---|
| $\alpha$ | Spectral index of primary-mass power law | $\mathrm{U}(1.5, 12)$ |
| $\beta$ | Spectral index of secondary-mass power law | $\mathrm{U}(-4, 12)$ |
| $m_{\rm min}$ | Minimum primary mass $[M_{\odot}]$ | $\mathrm{U}(2, 10)$ |
| $m_{\rm max}$ | Maximum primary mass $[M_{\odot}]$ | $\mathrm{U}(50, 200)$ |
| $\delta_{\rm m}$ | Smoothing parameter $[M_{\odot}]$ | $\mathrm{U}(10^{-3}, 10)$ |
| $\mu_{\rm g}^{\rm low}$ | Location of the first peak $[M_{\odot}]$ | $\mathrm{U}(5, 100)$ |
| $\sigma_{\rm g}^{\rm low}$ | Width of the first peak $[M_{\odot}]$ | $\mathrm{U}(0.4, 5)$ |
| $\mu_{\rm g}^{\rm high}$ | Location of the second peak $[M_{\odot}]$ | $\mathrm{U}(5, 100)$ |
| $\sigma_{\rm g}^{\rm high}$ | Width of the second peak $[M_{\odot}]$ | $\mathrm{U}(0.4, 10)$ |
| $\lambda_{\rm g}$ | Fraction of sources in the peaks | $\mathrm{U}(0, 1)$ |
| $\lambda_{\rm g}^{\rm low}$ | Fraction of sources in the first peak | $\mathrm{U}(0, 1)$ |

**Table 4.** Summary of the hyperparameters priors used for the MLTP model. U stands for uniform prior.

FULLPOP-4.0

| Parameter | Description | Prior |
|---|---|---|
| $\alpha_1$ | Spectral index of the power law before $b$ | $\mathrm{U}(-4, 12)$ |
| $\alpha_2$ | Spectral index of the power law after $b$ | $\mathrm{U}(-4, 12)$ |
| $\beta_1$ | Spectral index of the pairing function before $m_{\rm break}$ | $\mathrm{U}(-4, 12)$ |
| $\beta_2$ | Spectral index of the pairing function after $m_{\rm break}$ | $\mathrm{U}(-4, 12)$ |
| $m_{\rm min}$ | Minimum primary and secondary mass $[M_{\odot}]$ | $\mathrm{U}(0.4, 1.4)$ |
| $m_{\rm max}$ | Maximum primary and secondary mass $[M_{\odot}]$ | $\mathrm{U}(50, 200)$ |
| $\delta_{\rm m}^{\rm min}$ | 1st smoothing parameter of the low mass $[M_{\odot}]$ | $\mathrm{LU}(10^{-2}, 1)$ |
| $\delta_{\rm m}^{\rm max}$ | 2nd smoothing parameter of the low mass $[M_{\odot}]$ | $\mathrm{LU}(10^{-3}, 1)$ |
| $\mu_{\rm g}^{\rm low}$ | Location of the first peak $[M_{\odot}]$ | $\mathrm{U}(5, 150)$ |
| $\sigma_{\rm g}^{\rm low}$ | Width of the first peak $[M_{\odot}]$ | $\mathrm{U}(0.4, 5)$ |
| $\mu_{\rm g}^{\rm high}$ | Location of the second peak $[M_{\odot}]$ | $\mathrm{U}(5, 150)$ |
| $\sigma_{\rm g}^{\rm high}$ | Width of the second peak $[M_{\odot}]$ | $\mathrm{U}(0.4, 10)$ |
| $\lambda_{\rm g}$ | Fraction of sources in peaks | $\mathrm{U}(0, 1)$ |
| $\lambda_{\rm g}^{\rm low}$ | Fraction of sources in the first peak | $\mathrm{U}(0, 1)$ |
| $m_{\rm d}^{\rm low}$ | Left side of the dip $[M_{\odot}]$ | $\mathrm{U}(1.5, 3)$ |
| $m_{\rm d}^{\rm high}$ | Right side of the dip $[M_{\odot}]$ | $\mathrm{U}(5, 9)$ |
| $\delta_{\rm d}^{\rm min}$ | Smoothing of the left side of the dip $[M_{\odot}]$ | $\mathrm{LU}(0.01, 2)$ |
| $\delta_{\rm d}^{\rm max}$ | Smoothing of the right side of the dip $[M_{\odot}]$ | $\mathrm{LU}(0.01, 2)$ |
| A | Amplitude of the dip | $\mathrm{U}(0, 1)$ |

**Table 5.** Summary of the hyperparameters priors used for the FULLPOP-4.0 mass model. U (LU) stands for uniform (log-uniform) prior.

is also achieved using the FULLPOP-4.0 population mass model, which benefits from a larger number of GW events and more mass features. Our most precise estimate is obtained by combining the FULLPOP-4.0 model with GW170817, which leads to a value of $H_0 = 74.6^{+13.4}_{-9.1} (74.6^{+26.1}_{-13.5})\,\mathrm{km\,s^{-1}\,Mpc^{-1}}$, similar to the dark sirens results.

Figure 13 shows the reconstructed primary mass spectrum from the spectral analysis using the PLP, MLTP, and FULLPOP-4.0 mass models. As for the dark siren analysis, the MLTP and FULLPOP-4.0 models identify two peaks at $8.9^{+0.4}_{-0.6}(8.9^{+0.7}_{-1.1}) M_{\odot}$ and $26.8^{+2.6}_{-2.7}(26.8^{+4.4}_{-4.4}) M_{\odot}$, while the PLP model only identifies the latter at $28.6^{+3.9}_{-4.9}(28.6^{+6.0}_{-7.0}) M_{\odot}$. For the NS region, the results are again consistent with the galaxy catalog analysis, supporting the presence of a shallow dip between $2.3^{+0.4}_{-0.5}(2.3^{+0.6}_{-0.7})M_\odot$ and $7.2^{+1.1}_{-1.6}(7.2^{+1.5}_{-2.0})M_\odot$.

| Merger rate model | | |
|---|---|---|
| **Parameter** | **Description** | **Prior** |
| $\gamma$ | Slope of the power law before the point $z_p$ | $U(0, 12)$ |
| $\kappa$ | Slope of the power law after the point $z_p$ | $U(0, 6)$ |
| $z_p$ | Redshift turning point between the power laws | $U(0, 4)$ |

**Table 6.** Summary of the hyperpriors used in the merger rate evolution model. U stands for uniform prior.

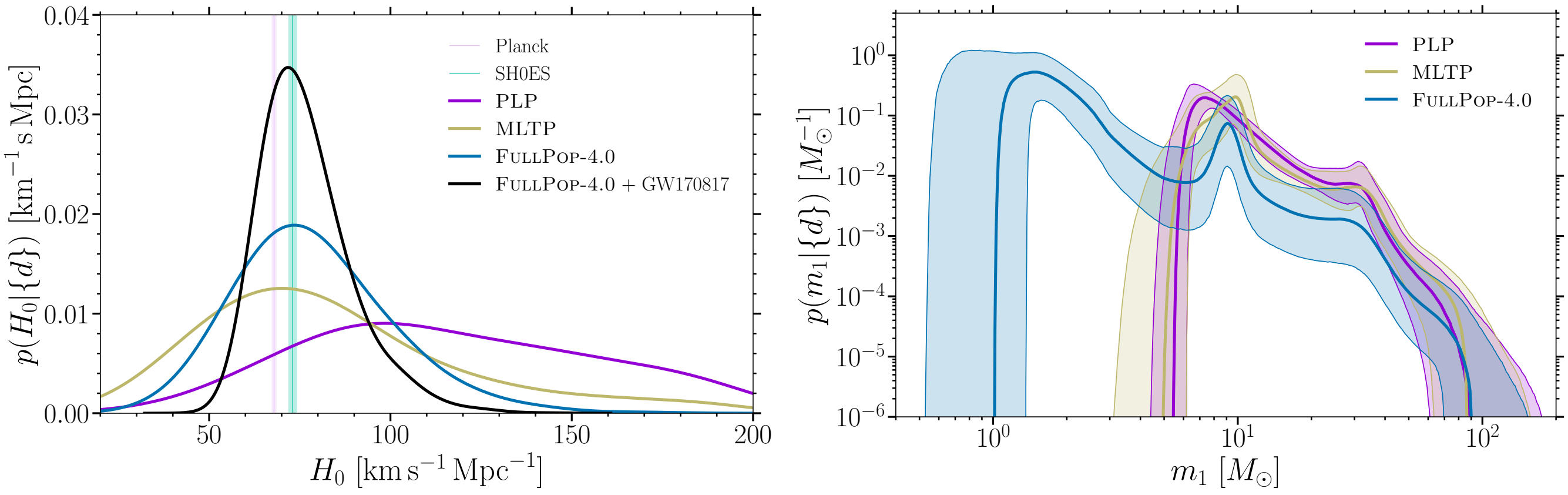

**Figure 12.** Left panel: Hubble constant posteriors with the spectral sirens method assuming different population mass models, namely the PLP (magenta curve), MLTP (gold curve) and FullPop-4.0 (blue curve). The black curve corresponds to the combined posterior between the FullPop-4.0 result and the bright siren posterior measured with GW170817. The pink and green shaded areas identify the 68% CI constraints on $H_0$ inferred from CMB anisotropies (Ade et al. 2016) and in the local Universe from SH0ES (Riess et al. 2022) respectively. Right panel: reconstructed source-frame primary mass distribution with the spectral siren method assuming the PLP, MLTP, and the FullPop-4.0 mass models (solid curve: median; shaded region: 90% CI).

Figure 13 presents the reduced corner plot showing the most interesting population and cosmological parameters derived from the spectral siren analysis with the FullPop-4.0 mass model, as in Figure 7. In addition, in Figure 13 we display results obtained with our two pipelines separately, to show explicitly their consistency. These results are consistent with those obtained from the dark siren analysis.

Finally, in addition to constraints on the Hubble constant, with the spectral siren approach in principle we are able to infer the present-day matter density of the Universe, $\Omega_m$ and the dark energy equation-of-state parameter $w_0$. To facilitate comparison with the results of Section 4.1, the main results of this section keep $\Omega_m$ fixed. See Section 4.1 and Figure 10 for a discussion of the impact of varying $\Omega_m$ and $w_0$.

A summary of the different $H_0$ values obtained using different data sets and model assumptions can be seen in Table 7.

## E. EVENT LIST

In this Appendix we provide a list of the events used in our analyses with their main properties relevant for our analysis. For the details on the PE and waveform models used, see Sec. 3.1. For each of the 142 events used in our analyses, Table 8 reports the following properties:

- SNR: we give the value of the search pipeline which has reported the lowest FAR.
- FAR: in units of inverse years, we report the lowest FAR among the pipelines.
- $m_1^{\rm det}$, $m_2^{\rm det}$, $D_L$, and $z$: detector-frame masses of the primary and secondary components, the luminosity distance to the source and the corresponding redshift, calculated from the distance samples assuming Planck-15 (Ade et al. 2016) cosmology. We give the median of the samples and the 90% CI, cutting away 5% of samples at the edges of the posterior distribution

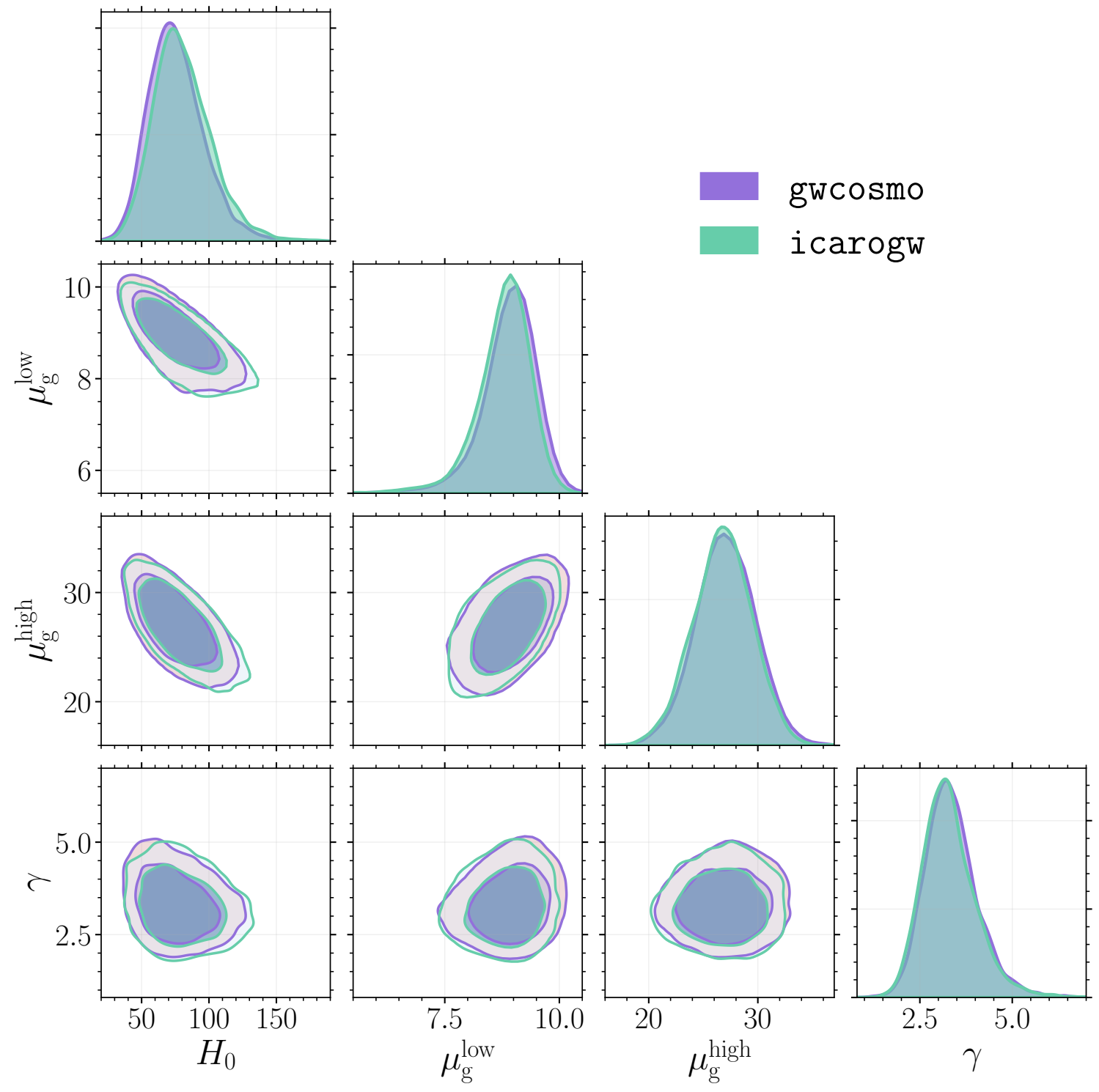

**Figure 13.** Spectral siren reduced corner plot of the Hubble constant and a subset of the FULLPOP-4.0 model mass parameters obtained with `gwcosmo` and `icarogw`. The contours indicate the 68.3% and 90% CR.

| ΛCDM – Spectral sirens | | | |
|---|---|---|---|
| **Population model** | **GW sources** | $H_0$ (Spectral sirens) [$\mathrm{km\,s^{-1}\,Mpc^{-1}}$] | $H_0$ (Spectral + bright sirens) [$\mathrm{km\,s^{-1}\,Mpc^{-1}}$] |
| POWER LAW + PEAK | 137 (138) | $112.7^{+51.0}_{-35.9}$ ($112.7^{+74.0}_{-55.1}$) | $74.8^{+15.5}_{-9.9}$ ($74.8^{+31.6}_{-14.4}$) |
| MULTI PEAK | 137 (138) | $77.1^{+40.8}_{-26.3}$ ($77.1^{+83.5}_{-39.1}$) | $74.8^{+15.5}_{-9.9}$ ($74.8^{+31.6}_{-14.4}$) |
| FULLPOP-4.0 | 141 (142) | $76.4^{+23.0}_{-18.1}$ ($76.4^{+41.2}_{-28.6}$) | $74.6^{+13.4}_{-9.1}$ ($74.6^{+26.1}_{-13.5}$) |

**Table 7.** Values of the Hubble constant measured using different data sets and analysis methods, adopting a uniform prior $H_0 \in \mathcal{U}(10, 200)\,\mathrm{km\,s^{-1}\,Mpc^{-1}}$. Columns are: population mass model assumed in the analysis (first column), number of GW sources analyzed (second column), $H_0$ measurement reported as a median with 68.3% (third column) and 90% (fourth column) symmetric CI. The values in parentheses are those obtained after combining the dark and bright (GW170817) measurements.

- Sky localization $\Delta\Omega$: the localization area of the event calculated from the skymap as a fraction of pixels containing the 90% of the probability
- Localization volume $\Delta V$: localization volume of the event at 90% CI, calculated as the fraction corresponding to the 90% of the sky area (see above) of the spherical shell, at the 90% CI of the event's redshift distribution
- $N_{\mathrm{gal}}$, over(under)-density and incompleteness: the number of galaxies inside the 90% localization volume, the over(under)-density fraction (see Equation (B19)), and the catalog (K-band of the GLADE+ catalog) incompleteness percentage (see Equation (B20)).

**Table 8**. List of the 142 CBC events selected with FAR $< 0.25\,\mathrm{yr}^{-1}$. Columns with a '*' have been computed assuming the reference cosmology in Ade et al. (2016), i.e. $H_0 = 67.9\,\mathrm{km\,s^{-1}\,Mpc^{-1}}$ and $\Omega_\mathrm{m} = 0.3065$. This table reports all of the GW events considered in this work and summarizes some of their properties reported with their median values and 90% symmetric CI (Abbott et al. 2019a, 2021b, 2024, 2023b; Abac et al. 2025d). First, second and third columns: GW event label, detected SNR (the one corresponding to the lowest FAR among the different pipelines, see Abac et al. 2025d) and FAR. Fourth, fifth, sixth columns: estimated primary and secondary detector-frame masses, luminosity distance. Seventh, eighth and ninth columns: redshift, sky localization area, and 3D localization comoving volume. The tenth column lists the number of galaxies in GLADE+ inside the localization volume for each event with K-band observations, while the eleventh and twelfth columns report the under(over)-density of galaxies and the incompleteness fraction of the catalog for each event. The under(over)-density is computed as the fraction between the number of galaxies (in the K-band) corrected for incompleteness, and the effective number expected from the Schechter function. We report the minimum and maximum values, respectively. The lower and upper bounds on the incompleteness probabilities, instead, are derived from the incompleteness fractions at the boundaries of the 90% localization volume. The last three columns, number of galaxies, under(over)-density, and incompleteness fraction, do not apply to the GW170817 event. We report these values using a distance prior proportional to $D_\mathrm{L}^2$ and a uniform in detector-frame masses prior. For events released with GWTC-3.0, we use the first data release associated with Abbott et al. (2023b).

| Name | SNR | FAR | $m_1^\mathrm{det}$ | $m_2^\mathrm{det}$ | $D_\mathrm{L}$ | $z^*$ | $\Delta\Omega$ | $\Delta V^*$ | $N_\mathrm{gal}^*$ | Under(over)-density* | Incompleteness* (%) |
|---|---|---|---|---|---|---|---|---|---|---|---|
| – | – | $[\mathrm{yr}^{-1}]$ | $[M_\odot]$ | $[M_\odot]$ | [Mpc] | – | [deg$^2$] | [Gpc$^3$] | – | – | – |
| GW150914 | 24.4 | $1.1\times10^{-39}$ | $38^{+5}_{-3}$ | $32^{+3}_{-5}$ | $463^{+132}_{-142}$ | $0.10^{+0.03}_{-0.03}$ | 159 | $2.0\times10^{-3}$ | $1.4\times10^{3}$ | 0.98–1.15 | 55–90 |
| GW151012 | 10.0 | $7.9\times10^{-3}$ | $30^{+19}_{-7}$ | $16^{+6}_{-6}$ | $1056^{+621}_{-493}$ | $0.21^{+0.10}_{-0.09}$ | $1.5\times10^{3}$ | 0.29 | $6.7\times10^{3}$ | 0.98–1.02 | 88–100 |
| GW151226 | 13.1 | $2.0\times10^{-15}$ | $16^{+12}_{-4}$ | $8^{+2.6}_{-3.0}$ | $471^{+158}_{-196}$ | $0.10^{+0.03}_{-0.04}$ | $1.0\times10^{3}$ | 0.02 | $1.2\times10^{4}$ | ∼1–1.19 | 46–93 |
| GW170104 | 13.0 | $2.5\times10^{-9}$ | $35^{+8}_{-5}$ | $25^{+5}_{-6}$ | $1126^{+385}_{-466}$ | $0.22^{+0.07}_{-0.09}$ | 938 | 0.14 | $3.1\times10^{3}$ | 0.98–1.01 | 93–100 |
| GW170608 | 14.9 | $4.9\times10^{-16}$ | $12^{+5}_{-1.7}$ | $8^{+1.3}_{-2.3}$ | $334^{+123}_{-117}$ | $0.07^{+0.03}_{-0.02}$ | 392 | $2.5\times10^{-3}$ | $1.8\times10^{3}$ | 0.54–2.12 | 29–74 |
| GW170729 | 10.8 | 0.18 | $77^{+16}_{-14}$ | $46^{+17}_{-18}$ | $2874^{+1630}_{-1474}$ | $0.49^{+0.22}_{-0.23}$ | $1.1\times10^{3}$ | 1.9 | 41 | ∼1–∼1 | 100–100 |
| GW170809 | 12.4 | $4.6\times10^{-14}$ | $41^{+10}_{-6}$ | $29^{+6}_{-7}$ | $1117^{+298}_{-352}$ | $0.22^{+0.05}_{-0.06}$ | 269 | 0.03 | 487 | 0.98–∼1 | 96–100 |
| GW170814 | 15.9 | $1.2\times10^{-19}$ | $34^{+6}_{-3}$ | $28^{+3}_{-5}$ | $600^{+165}_{-226}$ | $0.12^{+0.03}_{-0.04}$ | 85 | $2.2\times10^{-3}$ | 744 | 0.95–1.05 | 63–97 |
| GW170817 | 33.0 | $7.9\times10^{-51}$ | $1.6^{+0.3}_{-0.2}$ | $1.2^{+0.2}_{-0.2}$ | $42^{+6}_{-13}$ | $0.01^{+0.00}_{-0.00}$ | - | - | -— | -— | -— |
| GW170818 | 11.3 | $4.2\times10^{-5}$ | $43^{+8}_{-5}$ | $34^{+5}_{-7}$ | $1178^{+408}_{-422}$ | $0.23^{+0.07}_{-0.08}$ | 29 | $4.5\times10^{-3}$ | 1 | 0.97–∼1 | 97–100 |
| GW170823 | 11.5 | $4.6\times10^{-12}$ | $52^{+12}_{-8}$ | $39^{+8}_{-11}$ | $2115^{+836}_{-910}$ | $0.38^{+0.12}_{-0.15}$ | $1.6\times10^{3}$ | 1.1 | 260 | ∼1–∼1 | 100–100 |
| GW190408_181802 | 14.7 | $2.1\times10^{-15}$ | $32^{+7}_{-4}$ | $24^{+4}_{-5}$ | $1575^{+439}_{-595}$ | $0.30^{+0.07}_{-0.10}$ | 277 | 0.07 | 234 | ∼1–1.01 | 100–100 |
| GW190412 | 19.0 | $1.9\times10^{-27}$ | $34^{+6}_{-5}$ | $10^{+1.3}_{-1.1}$ | $700^{+167}_{-187}$ | $0.14^{+0.03}_{-0.04}$ | 31 | $9.5\times10^{-4}$ | 91 | 0.89–0.98 | 78–97 |
| GW190413_134308 | 8.9 | 0.18 | $83^{+19}_{-15}$ | $54^{+17}_{-23}$ | $4607^{+2384}_{-2202}$ | $0.73^{+0.29}_{-0.30}$ | 579 | 2.1 | 0 | ∼1–∼1 | 100–100 |
| GW190421_213856 | 10.5 | $2.8\times10^{-3}$ | $61^{+13}_{-9}$ | $47^{+9}_{-14}$ | $2958^{+1450}_{-1359}$ | $0.51^{+0.20}_{-0.21}$ | $1.1\times10^{3}$ | 1.7 | 133 | ∼1–∼1 | 100–100 |
| GW190425 | 12.9 | 0.03 | $2.2^{+0.5}_{-0.4}$ | $1.4^{+0.3}_{-0.2}$ | $143^{+75}_{-61}$ | $0.03^{+0.02}_{-0.01}$ | $9.1\times10^{3}$ | $7.9\times10^{-3}$ | $8.6\times10^{3}$ | 0.62–2.77 | 4–31 |
| GW190503_185404 | 12.0 | $2.3\times10^{-6}$ | $54^{+12}_{-10}$ | $36^{+10}_{-13}$ | $1596^{+658}_{-640}$ | $0.30^{+0.10}_{-0.11}$ | 108 | 0.04 | 89 | ∼1–∼1 | 100–100 |
| GW190512_180714 | 12.2 | $7.7\times10^{-12}$ | $30^{+8}_{-7}$ | $16^{+5}_{-3}$ | $1534^{+470}_{-621}$ | $0.29^{+0.07}_{-0.11}$ | 264 | 0.07 | 322 | ∼1–1.02 | 99–100 |
| GW190513_205428 | 12.3 | $1.3\times10^{-5}$ | $51^{+14}_{-14}$ | $25^{+11}_{-7}$ | $2283^{+934}_{-772}$ | $0.41^{+0.14}_{-0.12}$ | 433 | 0.33 | 1 | ∼1–∼1 | 100–100 |
| GW190517_055101 | 10.3 | $3.5\times10^{-4}$ | $53^{+14}_{-10}$ | $32^{+10}_{-12}$ | $1920^{+1846}_{-976}$ | $0.35^{+0.27}_{-0.16}$ | 434 | 0.53 | 393 | ∼1–1.02 | 100–100 |
| GW190519_153544 | 12.4 | $2.2\times10^{-6}$ | $94^{+16}_{-12}$ | $59^{+18}_{-19}$ | $3057^{+2027}_{-1343}$ | $0.52^{+0.27}_{-0.20}$ | 579 | 1.2 | 17 | ∼1–∼1 | 100–100 |
| GW190521 | 13.6 | $1.3\times10^{-3}$ | $154^{+29}_{-19}$ | $102^{+38}_{-42}$ | $4527^{+2265}_{-2635}$ | $0.72^{+0.28}_{-0.37}$ | 899 | 3.3 | 0 | ∼1–∼1 | 100–100 |
| GW190521_074359 | 24.4 | $5.0\times10^{-33}$ | $52^{+6}_{-5}$ | $39^{+6}_{-7}$ | $961^{+498}_{-420}$ | $0.19^{+0.09}_{-0.08}$ | 457 | 0.06 | $2.5\times10^{3}$ | ∼1–1.07 | 87–100 |
| GW190527_092055 | 8.7 | 0.23 | $56^{+59}_{-14}$ | $35^{+35}_{-17}$ | $3221^{+4197}_{-1686}$ | $0.54^{+0.53}_{-0.25}$ | $3.7\times10^{3}$ | 17 | 169 | ∼1–∼1 | 100–100 |
| GW190602_175927 | 12.3 | $1.1\times10^{-7}$ | $107^{+25}_{-18}$ | $71^{+22}_{-31}$ | $3274^{+2026}_{-1438}$ | $0.55^{+0.27}_{-0.21}$ | 790 | 1.8 | 0 | ∼1–∼1 | 100–100 |
| GW190620_030421 | 10.9 | 0.01 | $88^{+25}_{-18}$ | $52^{+19}_{-23}$ | $3418^{+1670}_{-1556}$ | $0.57^{+0.22}_{-0.23}$ | $5.8\times10^{3}$ | 12 | 13 | ∼1–∼1 | 100–100 |
| GW190630_185205 | 15.2 | $1.4\times10^{-10}$ | $41^{+9}_{-6}$ | $29^{+5}_{-7}$ | $908^{+550}_{-380}$ | $0.18^{+0.10}_{-0.07}$ | $1.1\times10^{3}$ | 0.15 | $9.3\times10^{3}$ | 0.99–1.03 | 83–100 |
| GW190701_203306 | 11.7 | $5.7\times10^{-3}$ | $75^{+15}_{-11}$ | $56^{+12}_{-18}$ | $2217^{+807}_{-762}$ | $0.40^{+0.12}_{-0.12}$ | 43 | 0.03 | 6 | ∼1–∼1 | 100–100 |

**Table 8** *continued*

**Table 8** *(continued)*

| Name | SNR | FAR | $m_1^{\rm det}$ | $m_2^{\rm det}$ | $D_{\rm L}$ | $z^*$ | $\Delta\Omega$ | $\Delta V^*$ | $N_{\rm gal}^*$ | Under(over)-density* | Incompleteness* (%) |
|---|---|---|---|---|---|---|---|---|---|---|---|
| – | – | $[{\rm yr}^{-1}]$ | $[M_\odot]$ | $[M_\odot]$ | [Mpc] | – | $[{\rm deg}^2]$ | $[{\rm Gpc}^3]$ | – | – | – |
| GW190706_222641 | 12.5 | $5.0 \times 10^{-5}$ | $119^{+23}_{-20}$ | $65^{+26}_{-27}$ | $4420^{+2678}_{-2328}$ | $0.70^{+0.33}_{-0.33}$ | $2.4 \times 10^3$ | 9.5 | 0 | ∼1–∼1 | 100–100 |
| GW190707_093326 | 13.2 | $2.7 \times 10^{-15}$ | $13^{+3}_{-1.8}$ | $10^{+1.5}_{-1.8}$ | $842^{+320}_{-366}$ | $0.17^{+0.06}_{-0.07}$ | 926 | 0.07 | $4.6 \times 10^3$ | ∼1–1.08 | 83–100 |
| GW190708_232457 | 13.1 | $3.1 \times 10^{-4}$ | $21^{+7}_{-2.9}$ | $15^{+2.3}_{-4}$ | $963^{+304}_{-386}$ | $0.19^{+0.05}_{-0.07}$ | $1.2 \times 10^4$ | 1.1 | $3.9 \times 10^4$ | 0.97–∼1 | 87–100 |
| GW190720_000836 | 11.5 | $4.4 \times 10^{-8}$ | $16^{+8}_{-4}$ | $9^{+2.6}_{-2.5}$ | $775^{+571}_{-243}$ | $0.16^{+0.10}_{-0.05}$ | 109 | 0.01 | 348 | 0.95–1.01 | 84–100 |
| GW190727_060333 | 12.1 | $2.7 \times 10^{-10}$ | $59^{+13}_{-8}$ | $46^{+8}_{-13}$ | $3235^{+1210}_{-1184}$ | $0.55^{+0.16}_{-0.17}$ | 130 | 0.19 | 0 | ∼1–∼1 | 100–100 |
| GW190728_064510 | 13.4 | $5.4 \times 10^{-16}$ | $15^{+11}_{-3}$ | $9^{+2.2}_{-3}$ | $907^{+247}_{-399}$ | $0.18^{+0.04}_{-0.07}$ | 340 | 0.03 | $1.4 \times 10^3$ | 0.98–∼1 | 84–100 |
| GW190803_022701 | 9.1 | 0.07 | $58^{+13}_{-9}$ | $44^{+9}_{-13}$ | $3573^{+1676}_{-1520}$ | $0.59^{+0.22}_{-0.22}$ | $1.0 \times 10^3$ | 2.2 | 0 | ∼1–∼1 | 100–100 |
| GW190814 | 22.2 | $5.4 \times 10^{-12}$ | $24^{+1.6}_{-1.4}$ | $2.7^{+0.1}_{-0.1}$ | $233^{+43}_{-46}$ | $0.05^{+0.01}_{-0.01}$ | 22 | $2.6 \times 10^{-5}$ | 30 | 0.58–0.69 | 23–42 |
| GW190828_063405 | 16.3 | $5.0 \times 10^{-27}$ | $43^{+7}_{-5}$ | $35^{+5}_{-7}$ | $2191^{+599}_{-925}$ | $0.39^{+0.09}_{-0.15}$ | 315 | 0.18 | 37 | ∼1–1.02 | 100–100 |
| GW190828_065509 | 11.1 | $3.5 \times 10^{-5}$ | $30^{+8}_{-8}$ | $13^{+5}_{-2.7}$ | $1646^{+674}_{-660}$ | $0.31^{+0.10}_{-0.11}$ | 593 | 0.23 | 227 | ∼1–1.01 | 100–100 |
| GW190910_112807 | 13.4 | $2.9 \times 10^{-3}$ | $57^{+9}_{-7}$ | $45^{+7}_{-10}$ | $1878^{+986}_{-873}$ | $0.34^{+0.15}_{-0.15}$ | $8.0 \times 10^3$ | 5.2 | $3.8 \times 10^3$ | ∼1–∼1 | 100–100 |
| GW190915_235702 | 13.0 | $7.8 \times 10^{-6}$ | $43^{+9}_{-6}$ | $33^{+5}_{-8}$ | $1886^{+705}_{-681}$ | $0.35^{+0.11}_{-0.11}$ | 452 | 0.22 | 128 | ∼1–∼1 | 100–100 |
| GW190924_021846 | 13.0 | $5.0 \times 10^{-10}$ | $10^{+7}_{-2.7}$ | $5^{+1.6}_{-1.8}$ | $556^{+205}_{-225}$ | $0.12^{+0.04}_{-0.04}$ | 358 | $9.3 \times 10^{-3}$ | $2.3 \times 10^3$ | 0.67–0.97 | 51–95 |
| GW190925_232845 | 9.9 | $7.2 \times 10^{-3}$ | $25^{+9}_{-3}$ | $18^{+2.8}_{-5}$ | $937^{+377}_{-337}$ | $0.19^{+0.07}_{-0.06}$ | $1.1 \times 10^3$ | 0.12 | $5.9 \times 10^3$ | ∼1–1.02 | 89–100 |
| GW190929_012149 | 10.1 | 0.16 | $101^{+25}_{-19}$ | $45^{+26}_{-19}$ | $3769^{+3147}_{-1692}$ | $0.62^{+0.40}_{-0.24}$ | $1.7 \times 10^3$ | 6.3 | 0 | ∼1–∼1 | 100–100 |
| GW190930_133541 | 10.0 | 0.01 | $14^{+13}_{-2.8}$ | $9^{+2.1}_{-4}$ | $790^{+318}_{-325}$ | $0.16^{+0.06}_{-0.06}$ | $1.6 \times 10^3$ | 0.11 | $5.8 \times 10^3$ | 0.99–1.03 | 79–100 |
| GW191105_143521 | 9.8 | 0.01 | $13^{+5}_{-2.1}$ | $9^{+1.6}_{-2.3}$ | $1203^{+404}_{-474}$ | $0.23^{+0.07}_{-0.09}$ | 683 | 0.11 | $1.0 \times 10^3$ | 0.97–∼1 | 95–100 |
| GW191109_010717 | 15.2 | $1.8 \times 10^{-4}$ | $80^{+10}_{-8}$ | $59^{+17}_{-17}$ | $1318^{+1378}_{-666}$ | $0.25^{+0.21}_{-0.12}$ | $1.5 \times 10^3$ | 0.90 | $4.8 \times 10^3$ | 0.99–1.01 | 94–100 |
| GW191127_050227 | 10.3 | 0.25 | $96^{+58}_{-35}$ | $44^{+30}_{-28}$ | $4595^{+3459}_{-2627}$ | $0.73^{+0.42}_{-0.37}$ | 982 | 4.9 | 0 | ∼1–∼1 | 100–100 |
| GW191129_134029 | 13.3 | $1.3 \times 10^{-23}$ | $13^{+5}_{-2.6}$ | $8^{+1.9}_{-1.7}$ | $803^{+237}_{-320}$ | $0.16^{+0.04}_{-0.06}$ | 843 | 0.05 | $8.1 \times 10^3$ | 0.98–1.01 | 79–100 |
| GW191204_171526 | 15.6 | $3.5 \times 10^{-25}$ | $14^{+4}_{-2.3}$ | $9^{+1.7}_{-1.8}$ | $644^{+180}_{-219}$ | $0.13^{+0.03}_{-0.04}$ | 260 | $7.8 \times 10^{-3}$ | $1.6 \times 10^3$ | 0.92–∼1 | 73–98 |
| GW191215_223052 | 10.9 | $1.3 \times 10^{-6}$ | $33^{+9}_{-5}$ | $25^{+4}_{-5}$ | $2102^{+866}_{-935}$ | $0.38^{+0.13}_{-0.15}$ | 571 | 0.39 | 173 | ∼1–1.01 | 100–100 |
| GW191216_213338 | 18.6 | $3.1 \times 10^{-11}$ | $14^{+7}_{-3.0}$ | $8^{+2.0}_{-2.2}$ | $343^{+109}_{-129}$ | $0.07^{+0.02}_{-0.03}$ | 224 | $1.4 \times 10^{-3}$ | $1.3 \times 10^3$ | 0.65–1.44 | 31–76 |
| GW191222_033537 | 12.0 | $2.8 \times 10^{-13}$ | $68^{+15}_{-10}$ | $52^{+11}_{-15}$ | $3382^{+1623}_{-1768}$ | $0.57^{+0.21}_{-0.26}$ | $1.9 \times 10^3$ | 3.9 | 92 | ∼1–∼1 | 100–100 |
| GW191230_180458 | 10.4 | 0.05 | $83^{+18}_{-12}$ | $63^{+13}_{-20}$ | $4841^{+2194}_{-2099}$ | $0.76^{+0.27}_{-0.28}$ | $1.1 \times 10^3$ | 3.9 | 0 | ∼1–∼1 | 100–100 |
| GW200112_155838 | 17.6 | $8.0 \times 10^{-6}$ | $45^{+8}_{-6}$ | $34^{+6}_{-7}$ | $1299^{+427}_{-458}$ | $0.25^{+0.07}_{-0.08}$ | $3.3 \times 10^3$ | 0.62 | $3.4 \times 10^3$ | ∼1–∼1 | 98–100 |
| GW200115_042309 | 11.5 | $3.0 \times 10^{-10}$ | $6^{+2.6}_{-2.5}$ | $1.6^{+0.8}_{-0.4}$ | $292^{+133}_{-92}$ | $0.06^{+0.03}_{-0.02}$ | 432 | $2.3 \times 10^{-3}$ | $2.2 \times 10^3$ | 0.82–1.30 | 28–73 |
| GW200128_022011 | 9.9 | $4.3 \times 10^{-3}$ | $65^{+14}_{-9}$ | $50^{+10}_{-13}$ | $3871^{+2131}_{-1972}$ | $0.63^{+0.27}_{-0.28}$ | $2.2 \times 10^3$ | 6.2 | 1 | ∼1–∼1 | 100–100 |
| GW200129_065458 | 26.5 | $2.9 \times 10^{-33}$ | $44^{+10}_{-6}$ | $31^{+6}_{-9}$ | $973^{+219}_{-341}$ | $0.19^{+0.04}_{-0.06}$ | 29 | $2.2 \times 10^{-3}$ | 11 | 0.96–∼1 | 94–100 |
| GW200202_154313 | 11.3 | $5.2 \times 10^{-8}$ | $11^{+4}_{-1.6}$ | $8^{+1.3}_{-2.0}$ | $422^{+146}_{-160}$ | $0.09^{+0.03}_{-0.03}$ | 156 | $1.8 \times 10^{-3}$ | 321 | 0.36–0.88 | 34–81 |
| GW200208_130117 | 10.8 | $3.1 \times 10^{-4}$ | $53^{+12}_{-8}$ | $39^{+9}_{-11}$ | $2371^{+1029}_{-912}$ | $0.42^{+0.15}_{-0.14}$ | 29 | 0.03 | 0 | ∼1–∼1 | 100–100 |
| GW200209_085452 | 10.0 | 0.05 | $56^{+14}_{-10}$ | $43^{+11}_{-13}$ | $3935^{+1972}_{-1875}$ | $0.64^{+0.25}_{-0.27}$ | 834 | 2.3 | 0 | ∼1–∼1 | 100–100 |
| GW200219_094415 | 10.7 | $9.9 \times 10^{-4}$ | $59^{+13}_{-9}$ | $44^{+9}_{-13}$ | $3830^{+1677}_{-1735}$ | $0.63^{+0.22}_{-0.25}$ | 667 | 1.6 | 0 | ∼1–∼1 | 100–100 |
| GW200224_222234 | 18.9 | $2.4 \times 10^{-32}$ | $53^{+9}_{-6}$ | $42^{+6}_{-10}$ | $1772^{+473}_{-658}$ | $0.33^{+0.07}_{-0.11}$ | 43 | 0.01 | 29 | ∼1–∼1 | 100–100 |
| GW200225_060421 | 12.3 | $1.1 \times 10^{-5}$ | $24^{+5}_{-3}$ | $17^{+3}_{-5}$ | $1157^{+505}_{-492}$ | $0.23^{+0.08}_{-0.09}$ | 494 | 0.09 | $1.6 \times 10^3$ | ∼1–1.01 | 94–100 |
| GW200302_015811 | 10.6 | 0.11 | $49^{+9}_{-10}$ | $25^{+12}_{-7}$ | $1582^{+1028}_{-733}$ | $0.30^{+0.16}_{-0.13}$ | $5.7 \times 10^3$ | 3.1 | $5.2 \times 10^3$ | ∼1–1.01 | 99–100 |
| GW200311_115853 | 17.7 | $4.2 \times 10^{-34}$ | $42^{+9}_{-5}$ | $33^{+5}_{-8}$ | $1203^{+277}_{-388}$ | $0.23^{+0.05}_{-0.07}$ | 35 | $4.2 \times 10^{-3}$ | 1 | 0.97–∼1 | 97–100 |

**Table 8** *continued*

**Table 8** *(continued)*

| Name | SNR | FAR | $m_1^{\rm det}$ | $m_2^{\rm det}$ | $D_{\rm L}$ | $z^*$ | $\Delta\Omega$ | $\Delta V^*$ | $N^*_{\rm gal}$ | Under(over)-density* | Incompleteness* (%) |
|---|---|---|---|---|---|---|---|---|---|---|---|
| – | – | [yr$^{-1}$] | [$M_\odot$] | [$M_\odot$] | [Mpc] | – | [deg$^2$] | [Gpc$^3$] | – | – | – |
| GW200316_215756 | 10.1 | $8.9 \times 10^{-6}$ | $17^{+14}_{-4}$ | $9^{+2.8}_{-3}$ | $1134^{+434}_{-429}$ | $0.22^{+0.07}_{-0.08}$ | 203 | 0.03 | 297 | ∼1–1.05 | 97–100 |
| GW230529_181500 | 11.4 | $2.2 \times 10^{-4}$ | $4^{+0.8}_{-1.0}$ | $1.6^{+0.6}_{-0.2}$ | $197^{+105}_{-97}$ | $0.04^{+0.02}_{-0.02}$ | $2.4 \times 10^4$ | 0.05 | $2.4 \times 10^4$ | 0.67–1.46 | 7–51 |
| GW230601_224134 | 11.8 | $1.8 \times 10^{-10}$ | $103^{+17}_{-15}$ | $70^{+16}_{-24}$ | $3423^{+1953}_{-1665}$ | $0.57^{+0.26}_{-0.25}$ | $2.3 \times 10^3$ | 5.5 | 154 | ∼1–∼1 | 100–100 |
| GW230605_065343 | 11.1 | $1.8 \times 10^{-7}$ | $21^{+7}_{-4}$ | $13^{+2.8}_{-3}$ | $1037^{+604}_{-469}$ | $0.20^{+0.10}_{-0.09}$ | 967 | 0.18 | $2.2 \times 10^3$ | ∼1–1.02 | 91–100 |
| GW230606_004305 | 10.7 | $4.1 \times 10^{-4}$ | $54^{+15}_{-10}$ | $38^{+9}_{-12}$ | $2689^{+1434}_{-1370}$ | $0.47^{+0.20}_{-0.21}$ | $1.2 \times 10^3$ | 1.7 | 186 | ∼1–∼1 | 100–100 |
| GW230608_205047 | 10.2 | $1.2 \times 10^{-3}$ | $76^{+14}_{-13}$ | $49^{+17}_{-19}$ | $3389^{+2172}_{-1666}$ | $0.57^{+0.28}_{-0.25}$ | $1.8 \times 10^3$ | 4.5 | 0 | ∼1–∼1 | 100–100 |
| GW230609_064958 | 10.0 | $1.4 \times 10^{-4}$ | $54^{+11}_{-8}$ | $40^{+9}_{-12}$ | $3302^{+1776}_{-1701}$ | $0.56^{+0.23}_{-0.25}$ | $1.3 \times 10^3$ | 2.7 | 41 | ∼1–∼1 | 100–100 |
| GW230624_113103 | 10.0 | $1.8 \times 10^{-4}$ | $35^{+16}_{-7}$ | $22^{+6}_{-6}$ | $1897^{+1226}_{-935}$ | $0.35^{+0.18}_{-0.16}$ | 989 | 0.80 | 841 | ∼1–1.01 | 100–100 |
| GW230627_015337 | 28.3 | $6.8 \times 10^{-39}$ | $9^{+1.9}_{-1.3}$ | $6^{+1.0}_{-0.9}$ | $307^{+62}_{-131}$ | $0.07^{+0.01}_{-0.03}$ | 92 | $3.3 \times 10^{-4}$ | 213 | 0.30–0.77 | 18–56 |
| GW230628_231200 | 15.3 | $8.0 \times 10^{-24}$ | $45^{+7}_{-4}$ | $38^{+5}_{-7}$ | $2286^{+759}_{-1062}$ | $0.41^{+0.11}_{-0.17}$ | 519 | 0.37 | 14 | ∼1–∼1 | 100–100 |
| GW230630_125806 | 9.0 | 0.16 | $95^{+32}_{-19}$ | $62^{+21}_{-27}$ | $5260^{+5031}_{-2921}$ | $0.81^{+0.59}_{-0.40}$ | $3.6 \times 10^3$ | 27 | 0 | ∼1–∼1 | 100–100 |
| GW230630_234532 | 9.9 | $4.2 \times 10^{-4}$ | $12^{+4}_{-2.0}$ | $8^{+1.5}_{-1.8}$ | $1074^{+507}_{-473}$ | $0.21^{+0.09}_{-0.09}$ | $1.1 \times 10^3$ | 0.19 | $4.9 \times 10^3$ | 0.98–∼1 | 89–100 |
| GW230702_185453 | 9.8 | $5.3 \times 10^{-6}$ | $60^{+30}_{-19}$ | $25^{+12}_{-9}$ | $2330^{+1689}_{-1034}$ | $0.41^{+0.24}_{-0.17}$ | $2.1 \times 10^3$ | 2.9 | 298 | ∼1–∼1 | 100–100 |
| GW230704_021211 | 9.4 | 0.21 | $48^{+12}_{-9}$ | $29^{+8}_{-8}$ | $2574^{+1805}_{-1408}$ | $0.45^{+0.25}_{-0.22}$ | $1.4 \times 10^3$ | 2.3 | 411 | ∼1–∼1 | 100–100 |
| GW230706_104333 | 9.2 | 0.23 | $22^{+6}_{-3}$ | $16^{+2.6}_{-3}$ | $1921^{+901}_{-968}$ | $0.35^{+0.14}_{-0.16}$ | $1.3 \times 10^3$ | 0.82 | 685 | ∼1–∼1 | 99–100 |
| GW230707_124047 | 11.9 | $1.1 \times 10^{-3}$ | $78^{+14}_{-9}$ | $63^{+10}_{-16}$ | $4523^{+2180}_{-2299}$ | $0.72^{+0.27}_{-0.32}$ | $2.6 \times 10^3$ | 9.2 | 0 | ∼1–∼1 | 100–100 |
| GW230708_053705 | 8.9 | 0.22 | $45^{+10}_{-6}$ | $36^{+6}_{-7}$ | $3252^{+1953}_{-1657}$ | $0.55^{+0.26}_{-0.25}$ | $1.4 \times 10^3$ | 3.1 | 21 | ∼1–∼1 | 100–100 |
| GW230708_230935 | 9.6 | $3.7 \times 10^{-3}$ | $102^{+26}_{-18}$ | $61^{+23}_{-26}$ | $3364^{+2152}_{-1524}$ | $0.56^{+0.28}_{-0.23}$ | $2.1 \times 10^3$ | 5.2 | 10 | ∼1–∼1 | 100–100 |
| GW230709_122727 | 10.0 | 0.01 | $78^{+19}_{-14}$ | $54^{+16}_{-26}$ | $4590^{+3399}_{-2402}$ | $0.73^{+0.41}_{-0.34}$ | $2.9 \times 10^3$ | 14 | 1 | ∼1–∼1 | 100–100 |
| GW230712_090405 | 9.5 | 0.02 | $43^{+22}_{-14}$ | $16^{+27}_{-7}$ | $1984^{+2244}_{-981}$ | $0.36^{+0.32}_{-0.16}$ | $1.3 \times 10^3$ | 2.0 | 285 | ∼1–∼1 | 100–100 |
| GW230723_101834 | 10.0 | $3.4 \times 10^{-3}$ | $22^{+7}_{-4}$ | $14^{+2.9}_{-3}$ | $1545^{+704}_{-992}$ | $0.29^{+0.11}_{-0.18}$ | 862 | 0.35 | $4.7 \times 10^3$ | 0.98–1.01 | 83–100 |
| GW230726_002940 | 10.5 | $7.8 \times 10^{-6}$ | $49^{+10}_{-6}$ | $39^{+6}_{-8}$ | $2076^{+1177}_{-1048}$ | $0.38^{+0.17}_{-0.17}$ | $2.8 \times 10^4$ | 25 | $5.8 \times 10^3$ | ∼1–∼1 | 100–100 |
| GW230729_082317 | 9.5 | 0.18 | $16^{+11}_{-3}$ | $10^{+2.4}_{-3}$ | $1629^{+793}_{-760}$ | $0.30^{+0.12}_{-0.13}$ | $2.0 \times 10^3$ | 0.90 | $1.9 \times 10^3$ | ∼1–1.01 | 99–100 |
| GW230731_215307 | 12.2 | $3.0 \times 10^{-14}$ | $12^{+3}_{-1.4}$ | $10^{+1.2}_{-1.9}$ | $1113^{+335}_{-453}$ | $0.22^{+0.06}_{-0.08}$ | 629 | 0.08 | $2.8 \times 10^3$ | 0.99–1.01 | 92–100 |
| GW230805_034249 | 9.4 | $3.7 \times 10^{-3}$ | $53^{+17}_{-12}$ | $34^{+12}_{-12}$ | $3145^{+2318}_{-1490}$ | $0.53^{+0.31}_{-0.22}$ | $2.0 \times 10^3$ | 4.9 | 35 | ∼1–∼1 | 100–100 |
| GW230806_204041 | 9.1 | $3.7 \times 10^{-3}$ | $93^{+21}_{-16}$ | $66^{+18}_{-25}$ | $5515^{+3515}_{-2897}$ | $0.84^{+0.41}_{-0.39}$ | $3.8 \times 10^3$ | 22 | 0 | ∼1–∼1 | 100–100 |
| GW230811_032116 | 12.9 | $4.4 \times 10^{-19}$ | $49^{+9}_{-8}$ | $30^{+8}_{-7}$ | $2029^{+1092}_{-1047}$ | $0.37^{+0.16}_{-0.17}$ | 817 | 0.66 | 766 | ∼1–1.01 | 100–100 |
| GW230814_061920 | 10.2 | $6.3 \times 10^{-4}$ | $113^{+16}_{-15}$ | $70^{+23}_{-26}$ | $3774^{+2871}_{-1813}$ | $0.62^{+0.36}_{-0.26}$ | $3.6 \times 10^3$ | 13 | 54 | ∼1–∼1 | 100–100 |
| GW230814_230901 | 42.3 | $5.6 \times 10^{-15}$ | $36^{+2.7}_{-2.0}$ | $30^{+1.9}_{-2.2}$ | $274^{+131}_{-121}$ | $0.06^{+0.03}_{-0.03}$ | $2.5 \times 10^4$ | 0.13 | $5.2 \times 10^4$ | 0.64–0.95 | 17–68 |
| GW230819_171910 | 9.9 | 0.01 | $122^{+85}_{-30}$ | $58^{+31}_{-31}$ | $3865^{+3382}_{-2110}$ | $0.63^{+0.42}_{-0.31}$ | $4.1 \times 10^3$ | 17 | 19 | ∼1–∼1 | 100–100 |
| GW230820_212515 | 9.1 | 0.24 | $105^{+29}_{-23}$ | $53^{+35}_{-31}$ | $3773^{+3000}_{-1905}$ | $0.62^{+0.38}_{-0.28}$ | $1.7 \times 10^3$ | 6.3 | 15 | ∼1–∼1 | 100–100 |
| GW230824_033047 | 10.7 | $4.8 \times 10^{-6}$ | $92^{+17}_{-13}$ | $64^{+16}_{-26}$ | $4621^{+2678}_{-2167}$ | $0.73^{+0.33}_{-0.30}$ | $3.2 \times 10^3$ | 13 | 0 | ∼1–∼1 | 100–100 |
| GW230825_041334 | 8.7 | 0.10 | $76^{+18}_{-14}$ | $51^{+14}_{-17}$ | $5039^{+4172}_{-3043}$ | $0.79^{+0.50}_{-0.42}$ | $3.0 \times 10^3$ | 19 | 1 | ∼1–∼1 | 100–100 |
| GW230904_051013 | 10.5 | $3.9 \times 10^{-5}$ | $13^{+5}_{-2.2}$ | $9^{+1.6}_{-2.1}$ | $1008^{+597}_{-412}$ | $0.20^{+0.10}_{-0.08}$ | $1.7 \times 10^3$ | 0.29 | $5.7 \times 10^3$ | 0.99–1.00 | 90–100 |
| GW230911_195324 | 11.1 | $1.0 \times 10^{-3}$ | $44^{+7}_{-7}$ | $25^{+8}_{-7}$ | $1129^{+971}_{-562}$ | $0.22^{+0.16}_{-0.10}$ | $2.7 \times 10^4$ | 9.4 | $6.5 \times 10^4$ | 0.98–∼1 | 87–100 |
| GW230914_111401 | 15.9 | $5.8 \times 10^{-24}$ | $87^{+11}_{-11}$ | $53^{+19}_{-19}$ | $2588^{+1523}_{-1160}$ | $0.45^{+0.21}_{-0.18}$ | $1.6 \times 10^3$ | 2.2 | 72 | ∼1–∼1 | 100–100 |
| GW230919_215712 | 16.3 | $7.2 \times 10^{-35}$ | $34^{+7}_{-4}$ | $27^{+3}_{-5}$ | $1275^{+734}_{-483}$ | $0.25^{+0.12}_{-0.09}$ | 570 | 0.17 | 716 | 0.99–∼1 | 98–100 |

**Table 8** *continued*

**Table 8** *(continued)*

| Name | SNR | FAR | $m_1^{\rm det}$ | $m_2^{\rm det}$ | $D_{\rm L}$ | $z^*$ | $\Delta\Omega$ | $\Delta V^*$ | $N^*_{\rm gal}$ | Under(over)-density* | Incompleteness* (%) |
|---|---|---|---|---|---|---|---|---|---|---|---|
| – | – | $[{\rm yr}^{-1}]$ | $[M_\odot]$ | $[M_\odot]$ | [Mpc] | – | $[{\rm deg}^2]$ | $[{\rm Gpc}^3]$ | – | – | – |
| GW230920_071124 | 10.1 | $6.4 \times 10^{-6}$ | $48^{+12}_{-7}$ | $36^{+7}_{-10}$ | $2818^{+1691}_{-1330}$ | $0.49^{+0.23}_{-0.20}$ | $1.8 \times 10^3$ | 3.0 | 154 | ∼1–∼1 | 100–100 |
| GW230922_020344 | 12.3 | $8.4 \times 10^{-16}$ | $53^{+12}_{-9}$ | $37^{+7}_{-8}$ | $1453^{+763}_{-609}$ | $0.28^{+0.12}_{-0.11}$ | 270 | 0.10 | 679 | ∼1–1.01 | 98–100 |
| GW230922_040658 | 11.6 | $1.2 \times 10^{-7}$ | $151^{+37}_{-25}$ | $98^{+31}_{-57}$ | $6366^{+4090}_{-3431}$ | $0.95^{+0.47}_{-0.45}$ | $4.4 \times 10^3$ | 32 | 0 | ∼1–∼1 | 100–100 |
| GW230924_124453 | 13.3 | $2.6 \times 10^{-20}$ | $41^{+7}_{-4}$ | $33^{+4}_{-6}$ | $2362^{+982}_{-981}$ | $0.42^{+0.14}_{-0.15}$ | $1.0 \times 10^3$ | 0.88 | 9 | ∼1–∼1 | 100–100 |
| GW230927_043729 | 11.3 | $4.7 \times 10^{-8}$ | $53^{+11}_{-7}$ | $42^{+7}_{-9}$ | $3116^{+1732}_{-1631}$ | $0.53^{+0.23}_{-0.25}$ | $1.1 \times 10^3$ | 2.2 | 188 | ∼1–1.01 | 100–100 |
| GW230927_153832 | 19.8 | $2.7 \times 10^{-36}$ | $27^{+4}_{-3}$ | $20^{+2.8}_{-2.7}$ | $1168^{+385}_{-516}$ | $0.23^{+0.07}_{-0.09}$ | 273 | 0.04 | $1.4 \times 10^3$ | ∼1–1.04 | 94–100 |
| GW230928_215827 | 9.5 | $1.5 \times 10^{-5}$ | $97^{+19}_{-19}$ | $53^{+16}_{-19}$ | $4359^{+3558}_{-2032}$ | $0.70^{+0.44}_{-0.28}$ | $3.0 \times 10^3$ | 14 | 0 | ∼1–∼1 | 100–100 |
| GW230930_110730 | 8.5 | 0.17 | $60^{+17}_{-10}$ | $44^{+10}_{-13}$ | $4857^{+2994}_{-2463}$ | $0.76^{+0.36}_{-0.34}$ | $2.9 \times 10^3$ | 13 | 0 | ∼1–∼1 | 100–100 |
| GW231001_140220 | 10.3 | $1.6 \times 10^{-5}$ | $129^{+23}_{-22}$ | $68^{+31}_{-27}$ | $4173^{+3659}_{-2278}$ | $0.67^{+0.45}_{-0.33}$ | $3.3 \times 10^3$ | 16 | 1 | ∼1–∼1 | 100–100 |
| GW231004_232346 | 8.9 | 0.16 | $111^{+29}_{-22}$ | $58^{+25}_{-24}$ | $4050^{+3308}_{-2009}$ | $0.66^{+0.41}_{-0.29}$ | $2.8 \times 10^3$ | 12 | 0 | ∼1–∼1 | 100–100 |
| GW231005_021030 | 10.4 | 0.01 | $163^{+39}_{-29}$ | $95^{+34}_{-42}$ | $6279^{+4448}_{-3139}$ | $0.94^{+0.51}_{-0.41}$ | $4.9 \times 10^3$ | 38 | 0 | ∼1–∼1 | 100–100 |
| GW231005_091549 | 8.9 | 0.04 | $46^{+13}_{-6}$ | $35^{+6}_{-10}$ | $3585^{+2433}_{-1785}$ | $0.59^{+0.31}_{-0.26}$ | $2.4 \times 10^3$ | 7.1 | 69 | ∼1–∼1 | 100–100 |
| GW231008_142521 | 9.3 | $1.6 \times 10^{-3}$ | $68^{+17}_{-14}$ | $38^{+15}_{-15}$ | $2787^{+2233}_{-1207}$ | $0.48^{+0.30}_{-0.18}$ | $2.6 \times 10^3$ | 5.5 | 87 | ∼1–∼1 | 100–100 |
| GW231014_040532 | 9.0 | 0.21 | $29^{+11}_{-5}$ | $21^{+4}_{-7}$ | $2231^{+1388}_{-1150}$ | $0.40^{+0.20}_{-0.19}$ | $1.6 \times 10^3$ | 1.8 | 484 | ∼1–∼1 | 100–100 |
| GW231020_142947 | 12.0 | $6.6 \times 10^{-10}$ | $15^{+11}_{-3}$ | $9^{+2.2}_{-3}$ | $1229^{+507}_{-643}$ | $0.24^{+0.08}_{-0.12}$ | $1.4 \times 10^3$ | 0.30 | $5.9 \times 10^3$ | 0.98–∼1 | 88–100 |
| GW231028_153006 | 21.0 | $2.8 \times 10^{-24}$ | $157^{+32}_{-24}$ | $110^{+19}_{-47}$ | $4514^{+1178}_{-2032}$ | $0.72^{+0.15}_{-0.28}$ | $1.2 \times 10^3$ | 2.7 | 0 | ∼1–∼1 | 100–100 |
| GW231029_111508 | 10.8 | $5.2 \times 10^{-5}$ | $100^{+17}_{-15}$ | $63^{+22}_{-26}$ | $3015^{+2348}_{-1623}$ | $0.51^{+0.31}_{-0.25}$ | $2.9 \times 10^4$ | 71 | 753 | ∼1–∼1 | 100–100 |
| GW231102_071736 | 13.8 | $8.8 \times 10^{-13}$ | $99^{+13}_{-11}$ | $70^{+16}_{-20}$ | $3619^{+1956}_{-1648}$ | $0.60^{+0.25}_{-0.24}$ | $2.2 \times 10^3$ | 5.4 | 18 | ∼1–∼1 | 100–100 |
| GW231104_133418 | 11.3 | $9.2 \times 10^{-10}$ | $16^{+5}_{-2.3}$ | $11^{+1.8}_{-2.5}$ | $1465^{+525}_{-646}$ | $0.28^{+0.08}_{-0.11}$ | 909 | 0.26 | $1.4 \times 10^3$ | 0.99–∼1 | 98–100 |
| GW231108_125142 | 12.6 | $4.7 \times 10^{-17}$ | $32^{+6}_{-4}$ | $24^{+3}_{-4}$ | $2051^{+702}_{-892}$ | $0.37^{+0.10}_{-0.14}$ | 897 | 0.51 | 302 | ∼1–∼1 | 100–100 |
| GW231110_040320 | 11.4 | $2.9 \times 10^{-11}$ | $26^{+7}_{-5}$ | $16^{+4}_{-3}$ | $1844^{+854}_{-892}$ | $0.34^{+0.13}_{-0.15}$ | 656 | 0.38 | 211 | ∼1–1.01 | 100–100 |
| GW231113_200417 | 10.1 | $3.8 \times 10^{-5}$ | $14^{+5}_{-2.4}$ | $9^{+1.9}_{-2.2}$ | $1146^{+645}_{-524}$ | $0.22^{+0.11}_{-0.10}$ | $1.6 \times 10^3$ | 0.36 | $6.5 \times 10^3$ | 0.99–∼1 | 91–100 |
| GW231114_043211 | 10.0 | $1.3 \times 10^{-4}$ | $29^{+12}_{-7}$ | $10^{+2.8}_{-2.4}$ | $1328^{+830}_{-580}$ | $0.26^{+0.13}_{-0.10}$ | $1.4 \times 10^3$ | 0.51 | $2.8 \times 10^3$ | ∼1–1.01 | 97–100 |
| GW231118_005626 | 10.7 | $1.1 \times 10^{-6}$ | $29^{+10}_{-7}$ | $14^{+5}_{-3}$ | $2083^{+957}_{-917}$ | $0.38^{+0.14}_{-0.15}$ | $1.0 \times 10^3$ | 0.74 | 342 | ∼1–1.02 | 100–100 |
| GW231118_071402 | 9.2 | $2.8 \times 10^{-3}$ | $72^{+20}_{-12}$ | $51^{+14}_{-18}$ | $4074^{+3396}_{-2139}$ | $0.66^{+0.42}_{-0.31}$ | $3.0 \times 10^3$ | 13 | 1 | ∼1–∼1 | 100–100 |
| GW231118_090602 | 11.0 | $9.8 \times 10^{-9}$ | $17^{+20}_{-5}$ | $9^{+3}_{-4}$ | $1369^{+510}_{-620}$ | $0.26^{+0.08}_{-0.11}$ | $1.1 \times 10^3$ | 0.27 | $2.1 \times 10^3$ | ∼1–1.01 | 97–100 |
| GW231119_075248 | 8.3 | 0.02 | $95^{+33}_{-19}$ | $68^{+22}_{-28}$ | $6538^{+5180}_{-3575}$ | $0.97^{+0.59}_{-0.46}$ | $5.1 \times 10^3$ | 46 | 0 | ∼1–∼1 | 100–100 |
| GW231127_165300 | 9.9 | 0.01 | $79^{+21}_{-17}$ | $49^{+18}_{-22}$ | $4312^{+3484}_{-2413}$ | $0.69^{+0.43}_{-0.34}$ | $3.8 \times 10^3$ | 18 | 6 | ∼1–∼1 | 100–100 |
| GW231129_081745 | 9.4 | 0.06 | $73^{+13}_{-13}$ | $39^{+15}_{-14}$ | $3598^{+3306}_{-1945}$ | $0.60^{+0.42}_{-0.29}$ | $3.2 \times 10^3$ | 12 | 3 | ∼1–∼1 | 100–100 |
| GW231206_233134 | 11.9 | $1.4 \times 10^{-14}$ | $54^{+9}_{-7}$ | $43^{+7}_{-10}$ | $3233^{+1739}_{-1815}$ | $0.55^{+0.23}_{-0.28}$ | $2.0 \times 10^3$ | 4.3 | 183 | ∼1–∼1 | 100–100 |
| GW231206_233901 | 20.7 | $1.1 \times 10^{-37}$ | $48^{+9}_{-5}$ | $37^{+6}_{-8}$ | $1483^{+333}_{-492}$ | $0.28^{+0.05}_{-0.08}$ | 292 | 0.06 | 45 | ∼1–∼1 | 99–100 |
| GW231213_111417 | 10.2 | $2.3 \times 10^{-6}$ | $58^{+14}_{-9}$ | $46^{+9}_{-13}$ | $4035^{+2225}_{-2012}$ | $0.66^{+0.28}_{-0.29}$ | $1.9 \times 10^3$ | 5.8 | 9 | ∼1–∼1 | 100–100 |
| GW231223_032836 | 9.4 | $3.8 \times 10^{-4}$ | $77^{+18}_{-12}$ | $54^{+15}_{-28}$ | $3937^{+2960}_{-1945}$ | $0.64^{+0.37}_{-0.28}$ | $3.4 \times 10^3$ | 13 | 2 | ∼1–∼1 | 100–100 |
| GW231223_202619 | 10.0 | $2.0 \times 10^{-3}$ | $13^{+4}_{-1.7}$ | $10^{+1.4}_{-2.3}$ | $884^{+489}_{-434}$ | $0.18^{+0.09}_{-0.08}$ | $2.6 \times 10^4$ | 3.3 | $7.6 \times 10^4$ | 0.98–1.01 | 76–100 |
| GW231224_024321 | 13.0 | $2.0 \times 10^{-16}$ | $11^{+2.5}_{-1.2}$ | $9^{+1.0}_{-1.5}$ | $944^{+283}_{-389}$ | $0.19^{+0.05}_{-0.07}$ | 351 | 0.03 | $1.5 \times 10^3$ | 0.98–1.10 | 88–100 |
| GW231226_101520 | 34.2 | $3.8 \times 10^{-44}$ | $49^{+7}_{-3}$ | $43^{+3}_{-6}$ | $1168^{+222}_{-301}$ | $0.23^{+0.04}_{-0.05}$ | 136 | 0.01 | 141 | ∼1–∼1 | 99–100 |
| GW231231_154016 | 13.4 | $2.3 \times 10^{-8}$ | $27^{+5}_{-3}$ | $21^{+2.7}_{-3}$ | $1058^{+588}_{-527}$ | $0.21^{+0.10}_{-0.10}$ | $2.7 \times 10^4$ | 5.2 | $5.6 \times 10^4$ | 0.97–∼1 | 84–100 |

**Table 8** *continued*

**Table 8** *(continued)*

| Name | SNR | FAR | $m_1^{\rm det}$ | $m_2^{\rm det}$ | $D_{\rm L}$ | $z^*$ | $\Delta\Omega$ | $\Delta V^*$ | $N_{\rm gal}^*$ | Under(over)-density* | Incompleteness* (%) |
|---|---|---|---|---|---|---|---|---|---|---|---|
| – | – | $[{\rm yr}^{-1}]$ | $[M_\odot]$ | $[M_\odot]$ | [Mpc] | – | $[{\rm deg}^2]$ | $[{\rm Gpc}^3]$ | – | – | – |
| GW240104_164932 | 14.8 | $2.1 \times 10^{-10}$ | $57^{+11}_{-7}$ | $44^{+8}_{-11}$ | $1915^{+1073}_{-951}$ | $0.35^{+0.16}_{-0.16}$ | $2.9 \times 10^4$ | 21 | $7.7 \times 10^3$ | ∼1–∼1 | 99–100 |
| GW240107_013215 | 9.1 | 0.03 | $112^{+37}_{-29}$ | $56^{+36}_{-32}$ | $5753^{+4833}_{-3244}$ | $0.88^{+0.56}_{-0.43}$ | $4.0 \times 10^3$ | 31 | 7 | ∼1–∼1 | 100–100 |
| GW240109_050431 | 10.4 | $2.3 \times 10^{-4}$ | $37^{+8}_{-7}$ | $23^{+6}_{-4}$ | $1452^{+967}_{-740}$ | $0.28^{+0.15}_{-0.13}$ | $2.8 \times 10^4$ | 13 | $2.8 \times 10^4$ | ∼1–1.01 | 96–100 |